\documentclass[twocolumn,epj,nopacs]{svjour} 
\RequirePackage[T1]{fontenc}
\RequirePackage{graphicx}
\RequirePackage{mathptmx}      
\RequirePackage{flushend}
\RequirePackage[numbers,sort&compress]{natbib}
\journalname{Eur. Phys. J. A}
\usepackage{amsmath,amssymb,epsfig,bm,mathrsfs}
\usepackage{feynmp,color,graphicx,mathtools}
\usepackage{graphicx,mathptmx,dcolumn,txfonts}
\usepackage{dcolumn,comment,multirow,epstopdf,indentfirst}
\usepackage[section]{placeins}
\usepackage[flushleft]{threeparttable}
\begin{document}
\title{Hypernuclear stars from relativistic Hartree-Fock density functional theory}

\author{Jia Jie Li\inst{1}
        \and
        Wen Hui Long\inst{2} 
        \and 
        Armen Sedrakian\inst{3,1}  
}
\institute{Institute for Theoretical Physics, J. W. Goethe University, D-60438 Frankfurt am Main, Germany
       \and
       School of Nuclear Science and Technology, Lanzhou University, Lanzhou 730000, China
       \and
       Frankfurt Institute for Advanced Studies, D-60438 Frankfurt am Main, Germany
}
\date{Received: date /Accepted: date}
\abstract{
The hypernuclear matter is studied within the relativistic
Hartree-Fock theory employing several parametrizations of the
hypernuclear density functional with density dependent couplings.
The equations of state and compositions of hypernuclear matter are
determined for each parametrization and compact stars are
constructed by solving their structure equations in spherical
symmetry. We quantify the softening effect of Fock terms on the
equation of state, as well as discuss the impact of tensor
interactions, which are absent in the Hartree theories. Starting
from models of density functionals which are fixed in the nuclear
sector to the nuclear phenomenology, we vary the couplings in the
hyperonic sector around the central values which are fitted to the
hyperon potentials in nuclear matter. We use the SU(6) spin-flavor
and SU(3) flavor symmetric quark models to relate the hyperonic
couplings to the nucleonic ones. We find, consistent with previous
Hartree studies, that for the SU(6) model the maximal masses of
compact stars are below the two-solar mass limit. In the SU(3) model
we find sufficiently massive compact stars with cores composed
predominantly of $\Lambda$ and $\Xi$ hyperons and a low fraction of
leptons (mostly electrons). The parameter space of the SU(3) model
is identified where simultaneously hypernuclear compact stars obey
the astrophysical limits on pulsar masses and the empirical
hypernuclear potentials in nuclear matter are reproduced.
}
\maketitle
%
\section{Introduction}

Neutron stars - remnants of luminous stars that are born in
core-collapse supernovas - serve as unique laboratories for the
exploration of dense hadronic matter~\cite{Glenden_book,Weber_book,
Lattimer2004,Weber2007,Sedrakian2007,Oertel2017}.

Observations of the mass and/or the radius of a neutron star can
provide a stringent constraint on the equation of state (EoS) of 
dense matter. The data indicates that the masses of pulsars in 
neutron star binaries are tightly clustered around the canonical 
mass value 1.4~$M_\odot$, as was initially established for the 
pulsar B1913+16 discovered in the Hulse-Taylor binary. More recently,
higher neutron star masses were measured with high precision in 
binaries containing millisecond pulsars, specifically $1.667\pm
0.021$~$M_\odot$ for PSR J1903+0327~\cite{Freire2011}, $1.93 \pm
0.02$~$M_\odot$ for PSR J1614-2230~\cite{Demorest2010,Fonseca2016}
and $2.01 \pm 0.04$~$M_\odot$ for PSR J0348+\\0432~\cite{Antoniadis2013}.
The last two measurements provide observationally reliable lower 
bounds on the maximum mass of a neutron star and thus set astrophysical
constraints on the EoS of dense matter.

The accuracy with which the neutron star radii can be measured is
less precise. These are commonly extracted from the analysis of
X-ray binaries, which depends on the measurements of their distances,
the amount of intervening absorbing material, and their atmospheric
compositions. The values of neutron star radii conjectured in the
literature lie in the range $10 \le R \le 14$~km~\cite{Lattimer2016,
Lattimer2014,Watts2016,Steiner2016,Steiner2017} with uncertainties in the
range $\sim30\%$ or greater. NASA's NICER experiment~\cite{Gendreau2017},
which was recently launched, will allow measurements of neutron star
radii to unprecedented precision with better than 10~percent uncertainty.
Finally, an alternative and independent information on neutron star 
radii can be obtained from the gravitational-wave (GW) signals emitted
in neutron star mergers. The accuracy of a radius measurement in this
case can be comparable to the best X-ray observations~\cite{Bauswein2012,
Lackey2015}.

The recent simultaneous detection of a GW signal by LIGO and
Virgo~\cite{TheLIGOScientific2017} of neutron star binary merger
event (GW170817 event) set bounds on the tidal deformabilities 
of compact stars~\cite{TheLIGOScientific2017}. This information, 
which is complimentary to the mass and radius measurements mentioned
above, can be used to constraint further the EoS of ultra-dense
matter~\cite{Annala2017,Paschalidis2017}.

While medium mass stars may contain purely nucleonic matter in their
centers, the stars close to the maximum mass reach central densities
roughly by an order of magnitude larger than the saturation density. 
At such densities a number of new degrees of freedom such as hyperons, 
deltas, and deconfined quark matter, may be present. The formation of
hyperons which is predicted by modern models of dense matter at densities
of about (2-3)$\rho_0$, where $\rho_0$ is the nuclear saturation density,
leads to sizeable softening of the EoS and, consequently, the maximum 
mass of neutron stars with hyperons decreases to values which are 
incompatible with the  mass-measurements quoted above. This is known 
as the "hyperon puzzle" which is the main focus of this work.

The relativistic density functional theories (DFTs)~\cite{Reinhard1989,
Ring1996,Serot1997,Vretenar2005,Meng2006} provide a convenient tool to 
address the problem of the equation of state (EoS) of dense matter and 
the structure of neutron stars, in particular the problem of hyperonization
at high densities. Several version of the relativistic DFTs have been 
employed for this purpose so far: (a) DFs which incorporate nonlinear (NL)
self-interaction of the mesons~\cite{Bonanno2012,Weissenborn2012a,
Weissenborn2012b,Bednarek2012,Gusakov2014,Chatterjee2016,Maslov2015,Tolos2017,
Spinella2017PhD}; (b) DFs with density-dependent (DD) meson-baryon
couplings~\cite{Long2012,Colucci2013,Dalen2014,Colucci2014,Banik2014,
Gomes2015,Char2015,Oertel2015,Oertel2016,Marques2017,Raduta2017,Fortin2017};
(c) quark-meson coupling models~\cite{Saito2007,Stone2007,Massot2012,
Whittenbury2014,Miyatsu2015}.

In this work we provide a first treatment of the full baryon octet
within the relativistic Hartree-Fock theory based on a DF with DD
couplings, i.e., the class (b) models. To date the hyperonization of
dense matter within the class (b) of DF models was addressed only at
the level of the Hartree approximation~\cite{Colucci2013,Dalen2014,
Colucci2014,Banik2014,Gomes2015,Char2015,Oertel2015,Oertel2016,Marques2017,
Raduta2017,Fortin2017}, the only exception being the previous treatment 
which included the $\Lambda$ hyperon~\cite{Long2012}. In doing so, 
we adopt the strategy of using the SU(3) symmetric quark model for 
fixing the DD coupling constants of hyperons to the nucleonic ones, 
previously used in the Hartree theories, see e.g.~\cite{Weissenborn2012a,
Weissenborn2012b,Colucci2013,Dalen2014,Colucci2014}, combined with 
constraints placed by the depth of the hyperonic potentials in nuclear matter.

Thus, we take the previous studies based on Hartree approximation to a
new level of many body theory, which has a number of advantages: (a)
the contributions of the Fock terms to the hyperonic self-energies are
explicit; (b) the tensor interactions arising for example from pion
exchanges, which are absent in the Hartree theories, are fully taken
into account. (c) the space component of vector self-energy arising 
from Fock terms could be important beyond the saturation density; 
(d) the Fock contributions ensure the proper antisymmetrization of 
baryon wave function in matter. We take into account two constraints 
available to tune the hypernuclear interactions: (i) the depth of 
hyperonic potentials in nuclear matter are used as a guide to fix 
the range of hyperonic couplings; (ii) the measured masses of neutron
stars are used to select the viable models.

In closing this section, we note that the case of non-strange nuclear
matter within the framework adopted in this work has been extensively
studied and gauged to reproduce the nuclear phenomenology~\cite{Long2006a,
Long2007,Liang2008,Long2010a,Long2010b,Li2014,Li2015}. In particular 
the role of the Fock terms and the tensor interactions in determining
the nuclear structure properties has been discussed in Refs.~\cite{Long2008,
Long2009,Li2016b,Jiang2015,Li2016a}. The non-strange models produce heavy
enough neutron stars with the maximum in the range 2.4 $M_{\odot}$~\cite{Sun2008}, 
which guarantees that moderate softening due to introduction of hyperons
can still produce large enough masses for hypernuclear compact stars.

This paper is organized as follows. In Sect.~\ref{section2}, we describe
in some detail the formalism of DFT in relativistic Hartree-Fock (RHF) 
approximation and its extension to the baryon octet. In Sect.~\ref{section3}
we discuss the EoS of hypernuclear matter. Our numerical results are presented
in Sect.~\ref{section4}. Section~\ref{section5} contains our conclusions and
the perspectives of this research. Some details of our calculations and 
the input physics are relegated to the Appendices.

\section{Theoretical formalism}
\label{section2}

In this section we outline the theoretical framework of the present
work, which is based on the covariant DFT for hypernuclear matter. The
functional is generated by evaluating the baryon self-energies in the
Hartree-Fock approximation. The interaction part of the Lagrangian of
hypernuclear matter contains the couplings between the baryons and
mesons which, as usual, have to be fitted to the experimental
(empirical) data. Table ~\ref{tab:Properties of baryon octet} lists
the key properties of the spin-1/2 baryon octet considered in this
work. The meson degrees of freedom included in our treatment are
listed in Table~\ref{tab:Quantum numbers of mesons} and are arranged
according to their quantum number ($J^\pi, T$), where $J$ is the spin,
$\pi$ is the intrinsic parity and $T$ is the isospin. In addition to
the usual set of meson acting in nuclear matter, we have added two
additional hidden-strangeness mesons, $\sigma^\ast$ and $\phi$. Note
that the mass of the $\sigma$ meson, which is supposed to represent
the two $\pi$-exchange contribution to the nuclear force, is not known
with precision and lies around 500~MeV.

\begin{table}[tb]
\caption{The properties of baryon octet with a spin $1/2$. The mass (in MeV),
spin $j$, isospin $\tau$ and its third component $\tau_3$, charge $q$, and strangeness $s$.}
\setlength{\tabcolsep}{10.2pt}
\label{tab:Properties of baryon octet}
\begin{tabular}{cccrrrr}
\hline
 Octet      &   Mass   &  $j$  & $\tau$ & $\tau_3$ & $q$ & $s$ \\
\hline
$n$         &  939.57  &  1/2  &   1/2  &   -1/2   &  0  &  0  \\
$p$         &  938.27  &       &        &    1/2   &  1  &     \\
\\
$\Lambda$   & 1115.68  &  1/2  &    0   &     0    &  0  &  -1 \\
\\
$\Sigma^-$  & 1197.45  &  1/2  &    1   &    -1    & -1  &  -1 \\
$\Sigma^0$  & 1192.64  &       &        &     0    &  0  &     \\
$\Sigma^+$  & 1189.37  &       &        &     1    &  1  &     \\
\\
$\Xi^-$     & 1321.71  &  1/2  &   1/2  &   -1/2   & -1  & -2  \\
$\Xi^0$     & 1314.86  &       &        &    1/2   &  0  &     \\
\hline
\end{tabular}
\end{table}
%
\begin{table}[tb]
\caption{Quantum numbers and mass (in MeV) of mesons.}\setlength{\tabcolsep}{8.8pt}
\label{tab:Quantum numbers of mesons}
\begin{tabular}{ccccccc}
\hline
 Meson   &  $\pi$  &  $\rho$ & $\sigma$ & $\omega$ & $\sigma^\ast$ & $\phi$ \\
\hline
 $J^\pi$ &  $0^-$  &  $1^-$  &   $0^+$  &   $1^-$  &     $0^+$     &  $1^-$ \\
   $T$   &    1    &    1    &     0    &     0    &       0       &    0   \\
  Mass   &   138   &   769   &$\sim 500$&    783   &      975      &  1020  \\
\hline
\end{tabular}
\end{table}

\subsection{Computation of the self-energies}

The Lagrangian of the hypernuclear matter is the sum of the free
baryonic and mesonic Lagrangians, $\mathscr{L}_B$ and $\mathscr{L}_M$
and an interaction Lagrangian $\mathscr{L}_{\text{int}}$ which
describes the coupling between baryon fields via meson exchange
\begin{align}\label{eq:Lagrangian density}
\mathscr{L}=\mathscr{L}_B+\mathscr{L}_M+\mathscr{L}_{\text{int}}.
\end{align}

The free baryonic Lagrangian is given by the sum of Dirac Lagrangians 
of individual baryons of mass $M_B$
\begin{align}\label{eq:Lagrangian density of baryon}
\mathscr{L}_B = \sum_B \bar{\psi}_B(i\gamma_\mu\partial^\mu - M_B)\psi_B,
\end{align}
where index $B$ sums over the baryonic octet. The mesonic Lagrangian has the form
\begin{align}
\mathscr{L}_M = & + \frac{1}{2}\partial^\mu\sigma\partial_\mu \sigma -\frac{1}{2} m^2_\sigma \sigma^2
                  + \frac{1}{2}\partial^\mu\sigma^\ast \partial_\mu \sigma^\ast - \frac{1}{2} m^2_{\sigma^\ast} \sigma^{\ast 2} \nonumber \\
                & - \frac{1}{4}\Omega^{\mu\nu}\Omega_{\mu\nu} + \frac{1}{2} m^2_\omega \omega^\mu\omega_\mu
                  - \frac{1}{4}\Phi^{\mu\nu}\Phi_{\mu\nu} + \frac{1}{2} m^2_\phi\phi^\mu\phi_\mu \nonumber \nonumber \\
                & - \frac{1}{4}\vec{R}^{\mu\nu}\vec{R}_{\mu\nu} + \frac{1}{2} m^2_\rho\vec{\rho}^\mu\vec{\rho}_\mu
                  + \frac{1}{2}\partial^\mu\vec{\pi}\partial_\mu\vec{\pi} -\frac{1}{2} m^2_\pi\vec{\pi}^2,
\end{align}
with $\Omega^{\mu\nu}, \Phi^{\mu\nu}$ and $\vec{R}^{\mu\nu}$ being the field strength
tensors for the vector mesons. The interaction Lagrangian is given by
\begin{align}\label{eq:interaction Lagrangian}
\mathscr{L}_{\text{int}} = & \sum_B \bar{\psi}_B\Big(-g_{\sigma B}\sigma-g_{\sigma^\ast B}\sigma^\ast \nonumber \\
                           & -g_{\omega B}\gamma^\mu\omega_\mu-g_{\phi B}\gamma^\mu\phi_\mu -g_{\rho B}\gamma^\mu\vec{\rho}_\mu\cdot\vec{\tau}_B \nonumber \\
                           & +\frac{f_{\omega B}}{2M_B}\sigma^{\mu\nu}\partial_\mu\omega^\nu +\frac{f_{\phi B}}{2M_B}\sigma^{\mu\nu}\partial_\mu\phi^\nu
                             +\frac{f_{\rho B}}{2M_B}\sigma^{\mu\nu}\partial_\mu\vec{\rho}^\nu\cdot\vec{\tau}_B \nonumber \\
                           & -\frac{f_{\pi B}}{m_\pi}\gamma_5\gamma^\mu\partial_\mu\vec{\pi}\cdot\vec{\tau}_B \Big)\psi_B,
\end{align}
where $\sigma^{\mu\nu} =\frac{i}{2}[\gamma^\mu, \gamma^\nu]$, $\vec{\tau}_B$ is 
the vector of isospin Pauli matrices, with $\tau_{3,B}$ being its third component. 
The mesons couple to the baryons with the strengths determined by the coupling 
constants $g_{\alpha B}$ and $f_{\alpha B}$.

The full baryon propagator is defined through Dyson's equation
\begin{align}
G(k) = G^0(k) + G^0(k) \Sigma(k) G(k),
\end{align}
where $G^0$ is the Green's function in free space, $k$ is the four momentum of baryon, 
and $\Sigma$ is the baryon self-energy. Because of the requirement of translational and
rotational invariance in the rest frame of infinite matter, the most general form of the
decomposition of the self-energy in the Dirac space is given by
\begin{align}\label{eq:Lorentz_decomp}
\Sigma(k)=\Sigma_S(k)+\gamma_0\Sigma_0(k)+\bm{\gamma}\cdot\bm{\hat{k}}\Sigma_V(k),
\end{align}
where $\bm{\hat{k}}$ is the unit vector along $\bm{k}$, with $\Sigma_S$, $\Sigma_0$ 
and $\Sigma_V$ being the scalar, time and space components of the vector self-energies.

By introducing the following auxiliary quantities
\begin{subequations}
\begin{align}
\bm{k}^\ast&=\bm{k}+\hat{\bm{k}}\Sigma_V,\\
M^\ast&=M+\Sigma_S,\\
E^\ast&=E-\Sigma_0,
\end{align}
\end{subequations}
the Dirac equation in infinite matter can be written in a form resembling the free-space
Dirac equation
\begin{align}
(\bm{\gamma}\cdot\bm{k}^\ast + M^\ast )u(k,s,\tau)= \gamma^0 E^\ast u(k,s,\tau),
\end{align}
where $u(k,s,\tau)$ have the meaning of Dirac spinors. In addition we define the quantities
$\hat{P}$ and $\hat{M}$
\begin{align}
\hat{P}\equiv\frac{\bm{k}^\ast}{E^\ast},\qquad\hat{M}\equiv\frac{M^\ast}{E^\ast},
\end{align}
which we will use in the expressions of self-energies.

As well known, the ground state of {\it interacting} fermionic matter is obtained by filling
energy levels with spin-isospin degeneracy $\gamma$ up to the Fermi momentum $k_F$. A 
straightforward computation of the Hartree (i.e. direct interaction) contribution to 
the components of Lorentz decomposition~\eqref{eq:Lorentz_decomp} of the self-energy gives
\begin{subequations}\label{eq:Hartree self-energies}
\begin{align}
\Sigma_{S, B}^H & = - g_{\sigma B}\bar{\sigma} - g_{\sigma^\ast B}\bar{\sigma^\ast}, \\
\Sigma_{0, B}^H & = + g_{\omega B}\bar{\omega} + g_{\phi B}\bar{\phi} + g_{\rho B} \tau_{3, B} \bar{\rho}, \\
\Sigma_{V, B}^H & = 0.
\end{align}
\end{subequations}
In the present calculation, we ignore the retardation effects which amount to rather small
contributions (at most a few percent) to the self-energies~\cite{Gotz1989}. The contributions
of the Fock terms (corresponding to exchange interactions) are therefore given by
\begin{subequations}\label{eq:Fock self-energies}
\begin{align}
\Sigma^F_{S, B}(k, \tau) = &
\frac{1}{(4\pi)^2k}\sum_{\alpha,B^\prime}\tau^2_\alpha
\int^{k_{F,B^\prime}}_0 k^\prime dk^\prime\Big[\hat{M}(k^\prime)B_\alpha(k,k^\prime) \nonumber \\
& +\frac{1}{2}\hat{P}(k^\prime)D_\alpha(k,k^\prime)\Big],\\
\Sigma^F_{0, B}(k, \tau) = &
\frac{1}{(4\pi)^2k}\sum_{\alpha,B^\prime}\tau^2_\alpha \int^{k_{F,B^\prime}}_0 k^\prime dk^\prime A_\alpha(k,k^\prime),\\
\Sigma^F_{V, B}(k, \tau) = &
\frac{1}{(4\pi)^2k}\sum_{\alpha,B^\prime}\tau^2_\alpha
\int^{k_{F,B^\prime}}_0 k^\prime dk^\prime\Big[\hat{P}(k^\prime)C_\alpha(k,k^\prime) \nonumber \\
& +\frac{1}{2}\hat{M}(k^\prime)D_\alpha(k,k^\prime)\Big],
\end{align}
\end{subequations}
where $\alpha$ sums over mesons, $k_{F, B}$ is the baryon Fermi momentum and $\tau_\alpha$ 
is the isospin factor at the meson-baryon vertex in the Fock diagram. The explicit expression
for the functions $A_{\alpha}$, $B_{\alpha}$, $C_{\alpha}$ and $D_{\alpha}$ in Eqs.~(\ref{eq:Fock self-energies})
are given in Appendix~A.

The meson fields obey in general inhomogeneous Klein-Gordon equations for scalar mesons and Proca equations
for vector mesons. As well known, the current conservation implies that the Proca equations can be further
reduced to Klein-Gordon equations. In the mean-field approximation, the meson fields are replaced by their
respective mean-field expectation values,
\begin{subequations}
\begin{align}
\bar{\sigma} &= \frac{1}{m^2_\sigma}\sum_B g_{\sigma B} \rho_{s, B}, \qquad \bar{\sigma^\ast} 
              = \frac{1}{m^2_{\sigma^\ast} } \sum_B g_{\sigma^\ast B} \rho_{s, B}, \\
\bar{\omega} &= \frac{1}{m^2_\omega } \sum_B g_{\omega B} \rho_{v, B}, \qquad \bar{\phi} 
              = \frac{1}{m^2_{\phi}}\sum_B g_{\phi B} \rho_{v, B}, \\
              & \qquad \quad \bar{\rho} = \frac{1}{m^2_{\rho}} \sum_B g_{\rho B} \tau_{3, B}\rho_{v, B},
\end{align}
\end{subequations}
where the scalar and vector densities are given by
\begin{subequations}\label{eq:source densities}
\begin{align}
\rho_{s,B} &\equiv \langle\bar{\psi}\psi\rangle = \frac{1}{\pi^2}\int^{k_{F,B}}_0k^2dk\hat{M}_B,\\
\rho_{v,B} &\equiv \langle \psi^\dag\psi\rangle = \frac{1}{\pi^2}\int^{k_{F,B}}_0k^2dk = \frac{1}{3\pi^2}k^3_{F,B}.
\end{align}
\end{subequations}
In addition, in infinite nuclear matter the spatial gradients of the fields can be neglected.

Because of the density dependence of meson-baryon couplings, we need to take into account the 
rearrangement in the interaction, which amounts to additional $\Sigma_R$ contribution to the 
self-energy component $\Sigma_0(k)$
\begin{align}\label{eq:rearrangement}
\Sigma_R =
\sum_\alpha\frac{\partial g_{\alpha B}}{\partial\rho_B}\sum_B\frac{1}{\pi^2}\int k^2d k\Big[\hat{M}_B(k)\Sigma^\alpha_{S,B}(k) \nonumber \\
+\Sigma^\alpha_{0,B}(k)+\hat{P}_B(k)\Sigma^\alpha_{V,B}(k)\Big].
\end{align}
Including rearrangement is mandatory to ensure thermodynamic consistency of our model. 
Note that the rearrangement does not contribute to the energy density but modifies 
the pressure. Attempts have been made to include phenomenologically the effect of 
short-range correlations in the Hartree-Fock studies of nuclear matter and nuclei~\cite{Marcos1996},
which should mimic the effect that arises from the resummations of ladder diagrams
in the non-perturbative approaches. We do not include any phenomenological corrections 
that should account for short-range correlations.

\subsection{Thermodynamics relations}

Once the Hartree-Fock self-energies are determined, the computation of the energy density and
the pressure of hypernuclear matter at zero temperature is standard. One starts with the field-theoretical
expression for the energy-momentum tensor $T^{\mu\nu}$ in the Lagrangian formalism
\begin{align}
T^{\mu\nu} = \frac{\partial \mathscr{L}}{\partial (\partial_\mu \varphi_i)}\partial^\nu \varphi_i - \eta^{\mu\nu}\mathscr{L},
\end{align}
where $\varphi_i$ stands generically for a boson or fermion field. The energy density is given by
\begin{align}
\mathscr{E} \equiv \langle T^{00}\rangle,
\end{align}
which includes sum over all baryons and mesons in the model and $\langle \dots \rangle$ refers 
to the statistical average. The energy density is then obtained as
\begin{align}
\mathscr{E}_B = \frac{\gamma_B}{2\pi^2}\int^{k_{F, B}}_0 k^2dk[T_B(k) + \frac{1}{2}V_B(k)]
\end{align}
with
\begin{subequations}
\begin{align}
T_B(k) = \hat{P}_Bk_B & + \hat{M}_BM_B, \\
V_B(k) = \hat{M}_B\Sigma_{S, B}(k) & + \hat{P}_B\Sigma_{V, B}(k) - \Sigma_{0, B}(k).
\end{align}
\end{subequations}
In a similar way, one finds the pressure as the statistical average of the trace of the spatial
component $T^{ij}$ of energy-momentum tensor
\begin{align}
\mathscr{P}  \equiv \frac{1}{3}\sum_i \langle T^{ii} \rangle.
\end{align}
Thermodynamic consistency implies that the same can be obtained from the thermodynamic relation,
\begin{align}
\mathscr{P}_B = \rho^2_B\frac{\partial}{\partial \rho_B}\frac{\mathscr{E}_B}{\rho_B}.
\end{align}
%

\subsection{Stellar matter and structure}

We now complete the discussion of thermodynamics of hypernuclear matter by pointing out 
the additional conditions of weak equilibrium and change neutrality that prevail in 
neutron stars. These conditions imply that the stellar matter consists of not only 
baryon octet but also leptons $l$ ($l= e^-, \mu^-$). The Lagrangian density for 
noninteracting leptons is given by the standard Dirac Lagrangian and their energy 
density and pressure at zero temperature read
\begin{subequations}
\begin{align}
\mathscr{E}_l &= \frac{1}{\pi^2}\int^{k_{F,l}}_0 dk \, k^2 (k^2+m^2_l)^{1/2}, \\
\mathscr{P}_l &= \frac{1}{3 \pi^2 }\int^{k_{F,l}}_0 dk \, k^4 (k^2+m^2_l)^{-1/2},
\end{align}
\end{subequations}
where $k_{F,l}$ is the lepton Fermi momentum, $m_l$ is the lepton bare mass. 
The contribution of leptons should be added to the energy density and pressure 
of hadronic matter once the chemical potentials of the baryon-lepton mixture 
are determined, i.e.,
\begin{align}
\mathscr{E}_H = \sum_B \mathscr{E}_B + \sum_l \mathscr{E}_l, \qquad \mathscr{P}_H= \sum_B \mathscr{P}_B + \sum_l \mathscr{P}_l.
\end{align}
In $\beta$ equilibrium the chemical potentials of the particles are related to each other by
\begin{equation}
\mu_B = b_B \mu_n - q_B \mu_e,
\end{equation}
where $b_B$ and $q_B$ denote the baryon number and electric charge of baryon species $B$. 
This condition guarantees that all reactions which conserve charge and baryon number are allowed.
Explicitly, the condition of $\beta$ equilibrium is expressed as
\begin{subequations}\label{eq:Chemical equilibrium}
\begin{align}
\mu_n = \mu_\Lambda & = \mu_{\Sigma^0} = \mu_{\Xi^0}, \\
\mu_n + \mu_e & = \mu_{\Sigma^-} = \mu_{\Xi^-}, \\
\mu_n - \mu_e & = \mu_p = \mu_{\Sigma^+}, \\
\mu_e & = \mu_\mu,
\end{align}
\end{subequations}
where the chemical potentials for baryons and leptons are given by
\begin{subequations}
\begin{align}
\mu_B &= \Sigma_0(k_{F,B}) + E^\ast(k_{F,B}), \\
\mu_l &= \big(k^2_{F,l} + m^2_l\big)^{1/2}.
\end{align}
\end{subequations}
In the expressions above, $\Sigma_0$ contains Hartree [$\Sigma^H_0$, Eq.~\eqref{eq:Hartree self-energies}]
and Fock [$\Sigma^F_0$, Eq.~\eqref{eq:Fock self-energies}] contributions as well as the rearrangement 
term $\Sigma_R$ which is given by Eq.~\eqref{eq:rearrangement}. Furthermore, charge neutrality is imposed as
\begin{align}
\sum_B q_B\rho_B + \sum_l q_l\rho_l = 0.
\end{align}
The conservation laws imply that there are only two independent chemical potentials related to 
baryon number density and total charge density.

The above conditions, together with the field equations for baryons and mesons, allow one to 
determine the equilibrium composition $\rho_B$ and $\rho_l$ at a given baryon number density 
and determine the EoS of matter.

The spherically symmetric solutions of Einstein's equations for self-gravitating fluids are
given by the Tolman-Oppenheimer-Volkoff (TOV) equations~\cite{Tolman1939, Oppenheimer1939}. 
In the geometrized units $c = G = 1$, the TOV equations read
\begin{subequations}
\begin{align}
\frac{d P (r)}{d r} &= - \frac{[P(r) + \varepsilon (r)][M(r) + 4\pi r^3 P(r)]}{r [r - 2 M(r)]}, \\
\frac{d M (r)}{d r} &= 4\pi r^2 \varepsilon(r),
\end{align}
\end{subequations}
where $P(r)$ is the pressure of the star at radius $r$, and $M(r)$ is the total star mass inside
a sphere of radius $r$. To construct equilibrium models of compact stars we supplement the EoS of infinite
hypernuclear matter with the EoS of inhomogeneous low-density matter in the crusts at the transition
density $\rho_0/2$. Specifically, for the inner and outer crust we use the EoS of Refs.~\cite{Baym1971a}
and~\cite{Baym1971b}. It is worthwhile to mention that the crust-core matching procedure can affect
the value of the radius of less massive stars~\cite{Fortin2016}. However, in the present work, we 
concentrate mainly on the maximum-mass compact stars, for which the radius is not very sensitive to 
the matching procedures~\cite{Fortin2016}.

\section{Equations of state}
\label{section3}

The wealth of nuclear data allows one to constrain the nucleon-nucleon ($NN$) interaction within reasonable 
accuracy, whereas this is not the case for hyperon-nucleon ($YN$) and hyperon-hyperon ($YY$) interactions, 
where data are scarce. We try to construct an effective $NY$ or $YY$ interaction for use in the many-body 
environment, starting from a nucleonic RHF density functional (hereafter DF) which quantitatively fits the
nuclear data.

\subsection{Meson-nucleon couplings}
%
\begin{table*}[tb]
\caption{Bulk properties of symmetric nuclear matter at the saturation point: density $\rho_0$ (fm$^{-3}$), 
binding energy $E_{B}$ (MeV), compression modulus $K$ (MeV), symmetry energy $J$ (MeV) and its slope $L$ (MeV),
Dirac mass $M^\ast_D$ ($M$), where $M$ is the bare mass, and nonrelativistic effective mass $M^\ast_{NR}$ ($M$)
predicted by selected RHF and RH DFs. The nonrelativistic effective masses of neutrons $M^\ast_{NR}(n)$ and 
protons $M^\ast_{NR}(p)$ in neutron matter are also listed.}
\setlength{\tabcolsep}{10.2pt}
\label{tab:NMP}
\begin{tabular}{cccccccccccc}
\hline
\multirow{2}*{DF} & \multirow{2}*{Interaction} & \multicolumn{7}{c}{Symmetric matter} & & \multicolumn{2}{c}{Neutron matter}  \\
\cline{3-9}\cline{11-12} &  & $\rho_0$ & $E_{B}$ & $K$ & $J$ & $L$& $M^\ast_D$ & $M^\ast_{NR}$ & & $M^\ast_{NR}(n)$ & $M^\ast_{NR}(p)$ \\
\hline
RHF & PKA1    & 0.160 & $-15.83$ &  229.96 &  36.02 & 103.50 & 0.55 & 0.68 &   & 0.68 & 0.70  \\
    & PKO1    & 0.152 & $-16.00$ &  250.28 &  34.37 &  97.71 & 0.59 & 0.75 &   & 0.73 & 0.76  \\
    & PKO2    & 0.151 & $-16.03$ &  249.53 &  32.49 &  75.92 & 0.60 & 0.76 &   & 0.75 & 0.77  \\
    & PKO3    & 0.153 & $-16.04$ &  262.44 &  32.99 &  82.99 & 0.59 & 0.74 &   & 0.74 & 0.76  \\
\\
RH  & DD-ME2  & 0.152 & $-16.14$ &  251.15 &  32.31 &  51.27 & 0.57 & 0.65 &   & 0.64 & 0.70  \\
    & GM1     & 0.153 & $-16.33$ &  300.22 &  32.51 &  93.96 & 0.70 & 0.77 &   & 0.73 & 0.81  \\
\hline
\end{tabular}
\end{table*}

\begin{figure}[tb]
\centering
\ifpdf
\includegraphics[width = 0.40\textwidth]{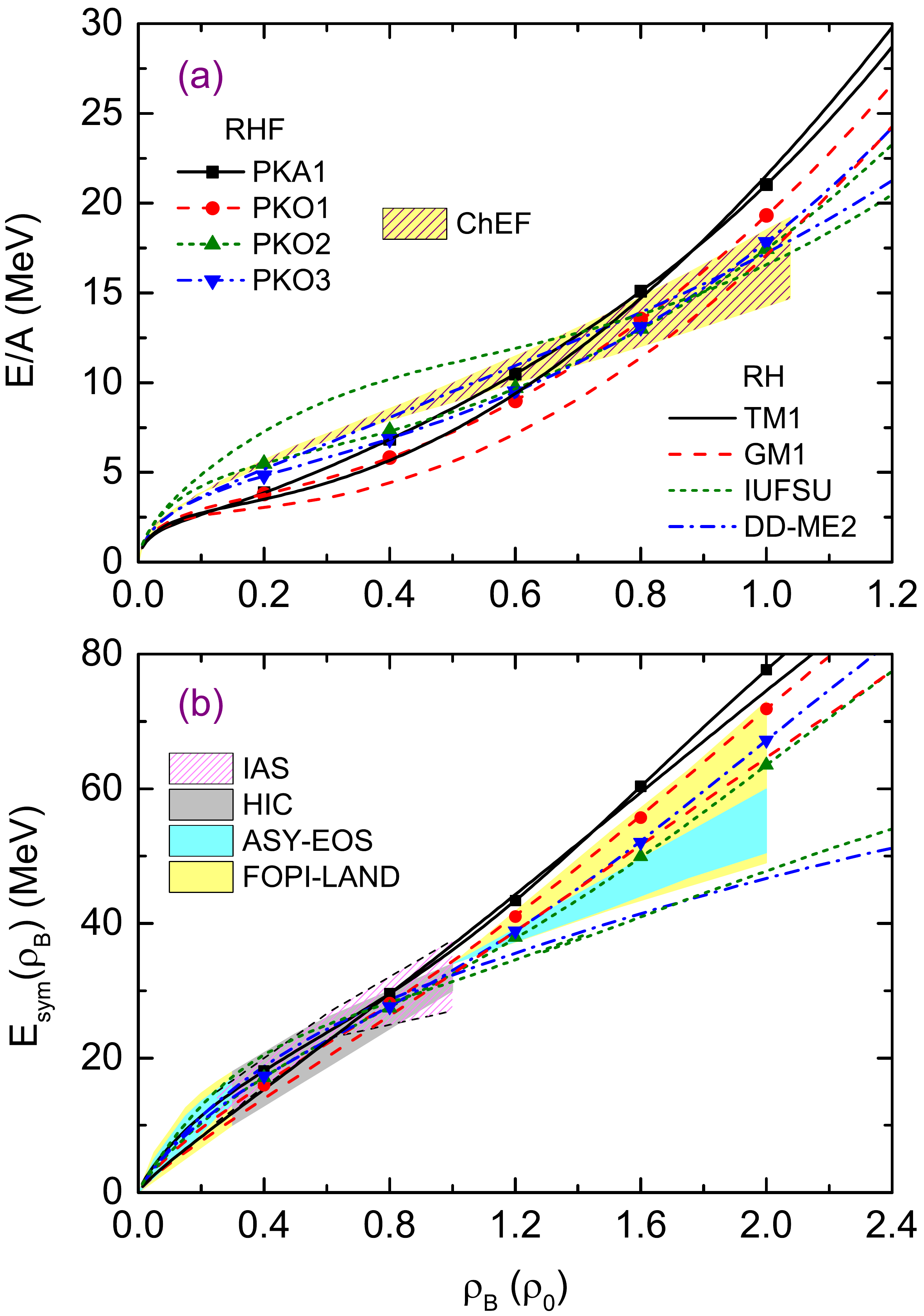}
\else
\includegraphics[width = 0.40\textwidth]{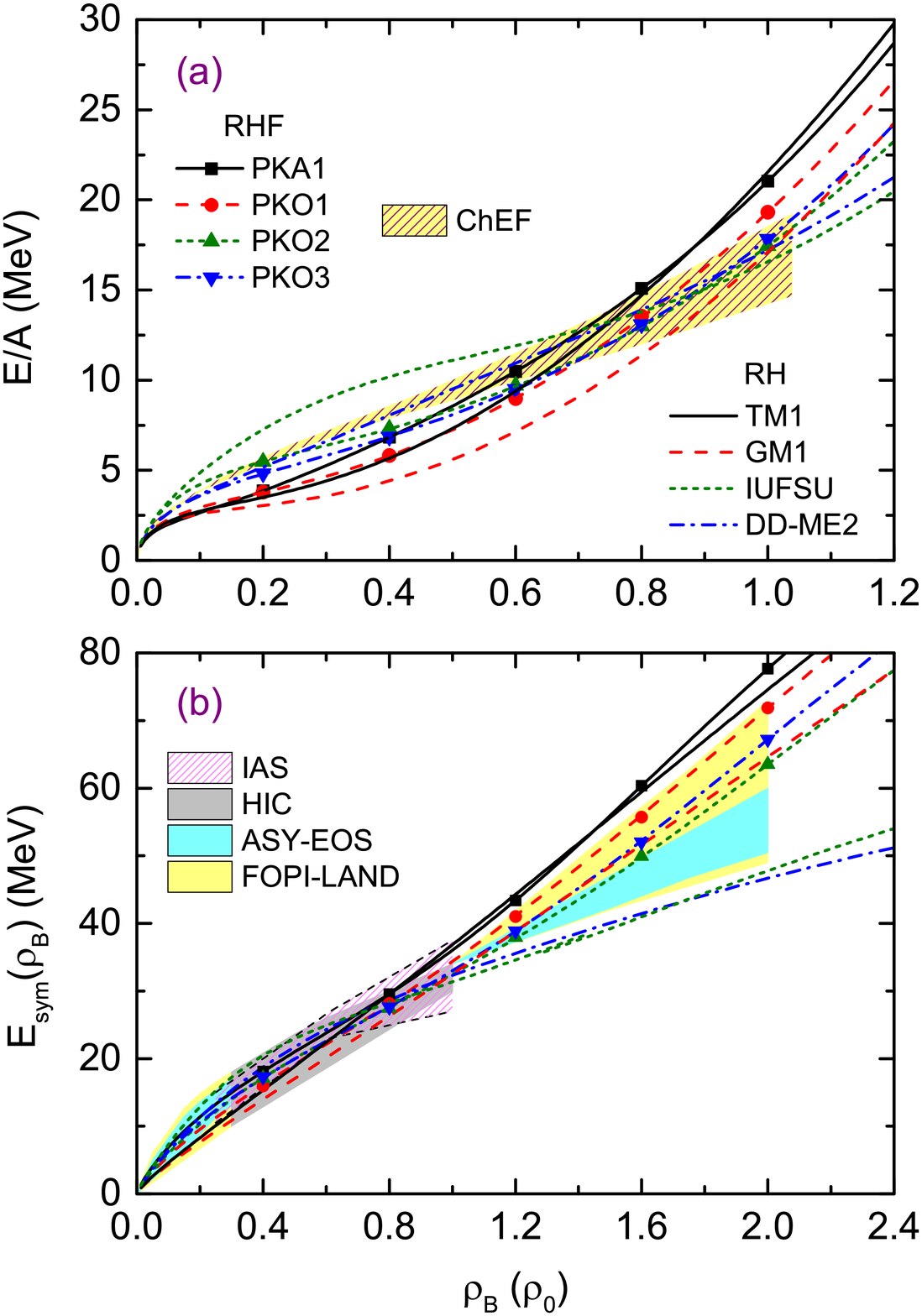}
\fi
\caption{ (a) The energy per nucleon in pure neutron matter as functions of baryon density within
different DFs compared with the chiral effective field (ChEF) theory of Ref.~\cite{Hebeler2013}.
(b) The symmetry energy as a function of baryon density. The shaded regions represent the experimental
constraints deduced from HIC (Sn+Sn)~\cite{Tsang2009}, FOPI-LAND~\cite{Russotto2011}, 
ASY-EOS~\cite{Russotto2016} and IAS~\cite{Danielewicz2014}.}
\label{fig:LAS}
\end{figure}

In this work we will use density-dependent meson-baryon couplings, which are designed to account
in an economical manner for the many-body corrections that arise beyond the mean-field approximation. 
Such density dependencies of the coupling constants are designed to account for the influence of the 
medium on the scattering amplitude of baryons, as explicitly accounted for in the DBHF theories of 
nuclear matter. Below we consider four parametrizations designed for RHF computations, namely 
PKO1-3~\cite{Long2006a,Long2008} and PKA1~\cite{Long2007}, as well as two standard parametrizations
of RH DF, specifically DD-ME2~\cite{Lalazissis2005} and GM1~\cite{Glendenning1991}. Some details of
the RHF parametrizations are given in Appendix~B. The predictions of these parametrizations for 
nuclear matter characteristics are compared in Table~\ref{tab:NMP}. Note also that the RH-GM1 DF is 
fitted only to the properties of bulk nuclear matter.

In Fig.~\ref{fig:LAS} (a) we show the energy per nucleon for pure neutron matter at and below the
saturation density. The band represents the results based on chiral effective field theory (ChEFT), 
which includes an estimate of the uncertainties related to the three-body force~\cite{Hebeler2013}. 
In the relevant density region $0.5\le \rho_B/\rho_0\le 1.2$, i.e., the region where one deals
with homogeneous neutron matter in neutron stars, the DFs PKO2 and PKO3 are fully consistent with
the results of ChEFT. The DFs PKA1 and PKO1 overshoot the ChEFT band by several MeV at close to
saturation density. Among the RH DFs the DD-ME2 is seen to be consistent with the ChEFT results, 
whereas the remaining DFs show considerable deviation in the low-density (GM1, IUFSU) and 
close-to-saturation (TM1) regime.

The symmetry energy and its derivatives are important  characteristics of any model of nuclear EoS.
Intensive efforts, both theoretical and experimental, have been made to constrain symmetry energy 
and its density dependence. Figure~\ref{fig:LAS}(b) shows the symmetry energy as a function of baryon
density for the models discussed above. We present also the data from simulations of heavy ion 
collisions~\cite{Tsang2009,Russotto2011,Russotto2016} and nuclear structure studies, which are
based on excitation to isobaric analog states~\cite{Danielewicz2014}. As can be seen, the symmetry
energy at low densities ($\rho_B \leq \rho_0$) predicted by our collection of models is consistent
with the experimental data. However, sizeable deviations are seen for densities beyond the saturation
density. In particular, the symmetry energy of the RHF DFs above saturation is higher than the one
suggested by the ASY-EOS experiment~\cite{Russotto2016} and the one predicted by PKA1 DF lies above
the band of FOPI-LAND experiment~\cite{Russotto2011}.

\begin{figure}[tb]
\centering
\ifpdf
\includegraphics[width = 0.40\textwidth]{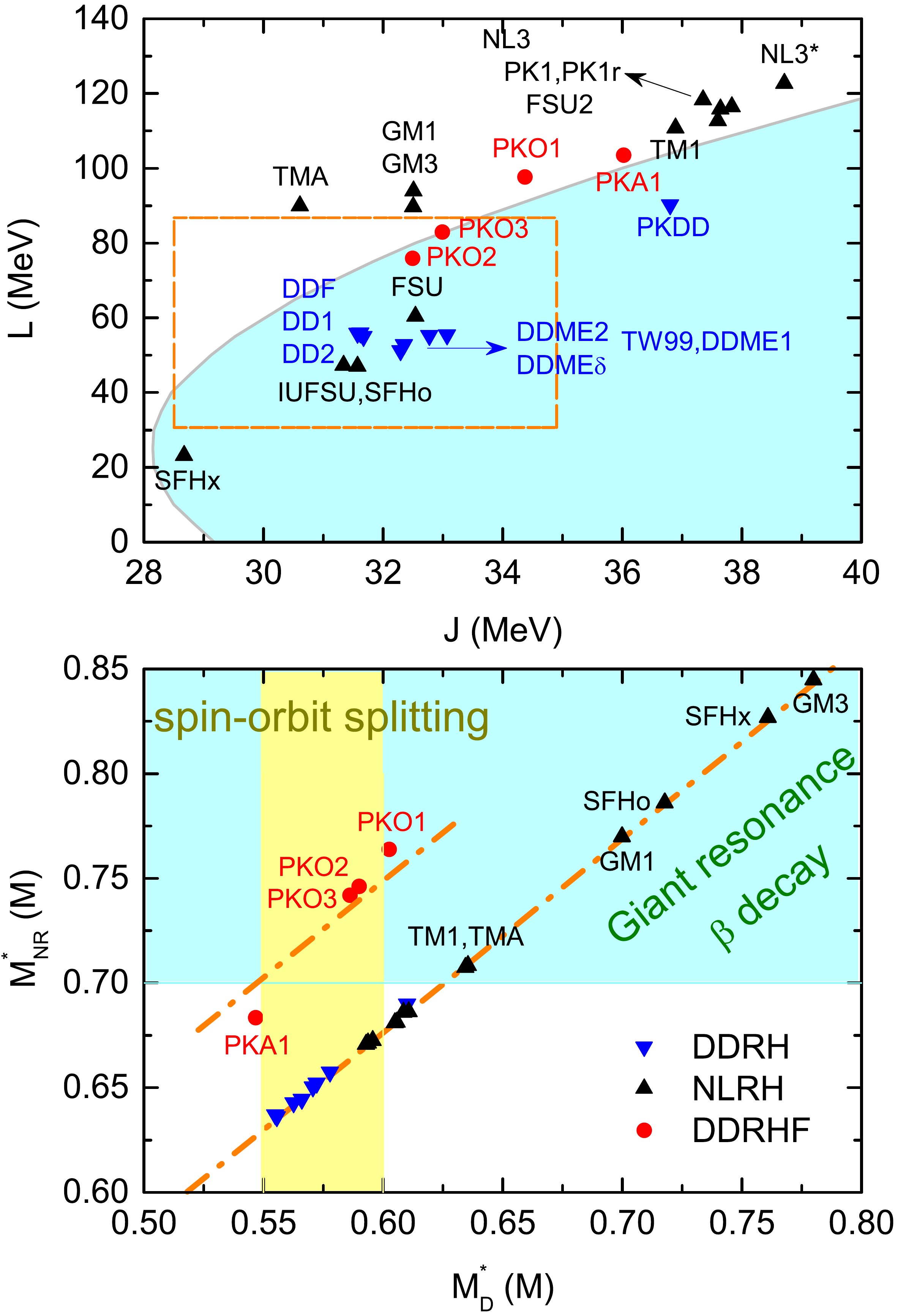}
\else
\includegraphics[width = 0.40\textwidth]{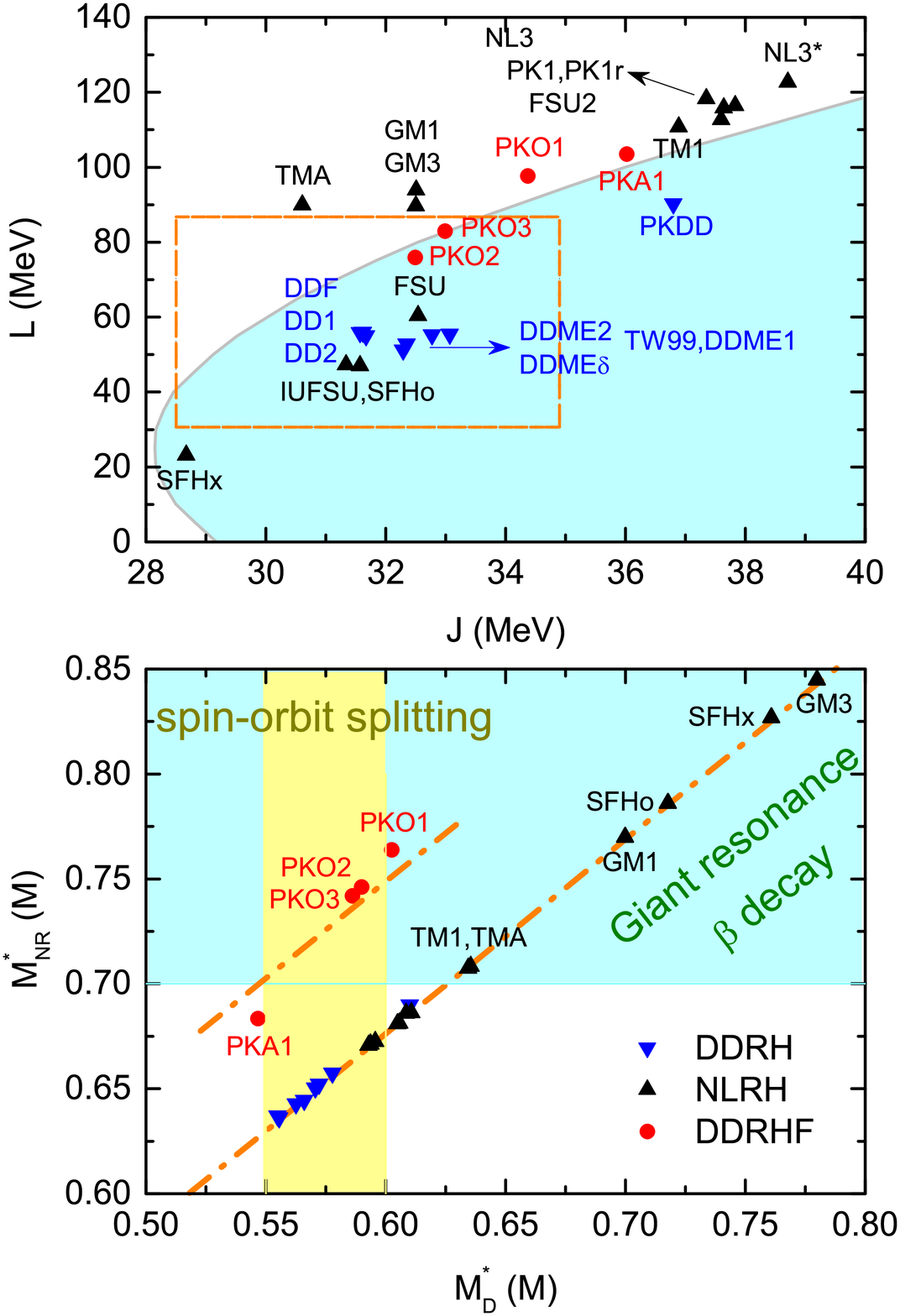}
\fi
\caption{ (a) The location of different relativistic nuclear DFs in the plane spanned by the symmetry 
energy $J$ and its slope $L$. The rectangle shows the bounds on the most probable values of symmetry 
energy $J = 31.7\pm 3.2$~MeV and the slope $L = 58.7\pm 28.1$~MeV obtained from the combined analysis
of astrophysical constrains and terrestrial experiments~\cite{Oertel2017}. The shaded region is the 
one allowed by the unitary gas bounds~\cite{Tews2017}. (b) The nonrelativistic effective masses 
$M^\ast_{NR}$ versus Dirac masses $M^\ast_D$ predicted by the DFs defined in panel (a). The dashed
lines are shown to guide the eye. The realistic description of the spin-orbit splitting requires 
$M^\ast_D \in [0.55, 0.60]$~\cite{Typel2005}, and the giant resonances require $M^\ast_{NR} \in [0.70, 0.90]$
~\cite{Chabanat1998, Reinhard1999, Niksic2005, Marketin2007}.}
\label{fig:SNM}
\end{figure}

Figure~\ref{fig:SNM}(a) summarizes the symmetry energy at the saturation density $J$ versus its slope
$L$ for RHF DFs used in this work along with some other frequently used RH DFs. The collection of 
shown DFs include the DD-RHF parametrizations PKO1-3~\cite{Long2006a, Long2008}, PKA1~\cite{Long2007}, 
the DD-RH parametrizations include TW99~\cite{Typel1999}, DD1~\cite{Typel2005}, DD2~\cite{Typel2010}, 
DDF~\cite{Klahn2006}, PKDD~\cite{Long2004}, DD-ME1~\cite{Niksic2002}, DD-ME2~\cite{Lalazissis2005}, 
DD-ME$\delta$~\cite{Roca2011}, and the NL-RH parametrizations GM1, GM3~\cite{Glendenning1991},
TM1~\cite{Sugahara1994}, TMA~\cite{Toki1995}, NL3~\cite{Lalazissis1997}, NL3$^\ast$~\cite{Lalazissis2009} 
PK1, PK1r~\cite{Long2004}, FSU~\cite{Todd2005}, FSU2~\cite{Chenwj2014}, IU-FSU~\cite{Fattoyev2010}, SFHo,
SFHx~\cite{Steiner2013b}. Extensive, independent studies have been performed to constrain the values of
these quantities, but the uncertainties are still large especially for the slope $L$. In this figure, 
the soft EoSs are located at the lower left corner where the values of $J$ and $L$ are small, whereas
the hard EoSs are located at the upper right corner where the values of these parameters are large. 
It is clearly seen that the softest ones are those based on the DD-RH DFs, which are followed by 
moderately soft DD-RHF DFs, then by the hard non-linear RH DFs.

The effective mass, which characterizes the quasiparticle properties in a strongly interacting medium, 
is another important characteristic of a model; one distinguishes the Dirac mass, which is a genuine 
relativistic quantity without a nonrelativistic counterpart and the nonrelativistic mass defined in 
the context of Fermi-liquid theories~\cite{Dalen2005b}. Figure~\ref{fig:SNM}(b) shows the Dirac mass 
$M^\ast_D$ and the nonrelativistic effective mass $M^\ast_{NR}$ for a number of DFs. The realistic 
description of the spin-orbit splitting in finite nuclei places the constraint $ 0.55 \le M^\ast_D \le 0.60$
on the Dirac mass~\cite{Typel2005}. The excitation energies of quadrupole giant resonances in nuclei 
have shown that a realistic choice for the non-relativistic effective mass should be in the range
$0.7 \le M^\ast_{NR} \le 0.90$~\cite{Chabanat1998, Reinhard1999, Niksic2005, Marketin2007}.

\begin{table}[tb]
\caption{
Astrophysical characteristics of the nucleonic EoS models: maximum mass $M_{\text{max}}$ ($M_\odot$), 
the corresponding central densities  $\rho_c$ (fm$^{-3}$) and radii $R_{\text{max}}$ (km), the radii
$R_{1.4}$ (km) for the canonical mass 1.4$M_\odot$ neutron stars, the density $\rho_{DU}$ (fm$^{-3}$) 
and mass $M_{DU}$ ($M_\odot$) threshold for the onset of direct Urca process (no entry means that the 
process is forbidden), calculated using RHF and RH DFs.}\setlength{\tabcolsep}{6.6pt}
\label{tab:nucleonic neutron star}
\begin{tabular}{ccccccc}
\hline
 DF &  $M_{\text{max}}$  &  $R_{\text{max}}$  & $\rho_c$ &  $R_{1.4}$ &  $\rho_{DU}$ & $M_{DU}$\\
\hline
 PKA1  &  2.42  &    12.34    &    0.810  &   13.99  &   0.252  &   0.99  \\
 PKO1  &  2.44  &    12.41    &    0.801  &   14.13  &   0.251  &   1.01  \\
 PKO2  &  2.45  &    12.30    &    0.804  &   13.79  &   0.294  &   1.25  \\
 PKO3  &  2.49  &    12.49    &    0.780  &   13.96  &   0.282  &   1.23  \\
\\
DD-ME2 &  2.48  &    12.07    &    0.817  &   13.22  &     -    &    -    \\
 GM1   &  2.36  &    11.97    &    0.862  &   13.81  &   0.278  &   1.10  \\
\hline
\end{tabular}
\end{table}

To complete the survey of models with non-strange baryons, we list in Table~\ref{tab:nucleonic neutron star}
the maximal mass of spherically-symmetrical, non-rotating and non-magnetized configuration supported by each model,
its radius and central density. The maximum masses quoted are compatible with the current lower bound on the
maximum mass of a neutron star. Clearly, the values of maximal masses leave some room for softening of the EoS
through the onset of hyperons, which will be discussed later on. Table~\ref{tab:nucleonic neutron star} also
lists the radii predicted by the models for a star with the canonical mass 1.4~$M_\odot$; these lie close to the
upper range of radii inferred from the analysis of X-ray data~\cite{Lattimer2014, Steiner2016, Steiner2017}.
Finally, the density and mass threshold for the onset of direct Urca (DU) process in purely nucleonic
matter are shown. For RHF DFs the DU threshold is quite low, which means that the stars will cool rapidly
by this process if nucleonic pairing does not slow down the cooling rate significantly.

\subsection{Including strange mesons: $\phi$-$B$ and $\sigma^*$-$B$ couplings}

\begin{figure}[b]
\centering
\ifpdf
\includegraphics[width = 0.40\textwidth]{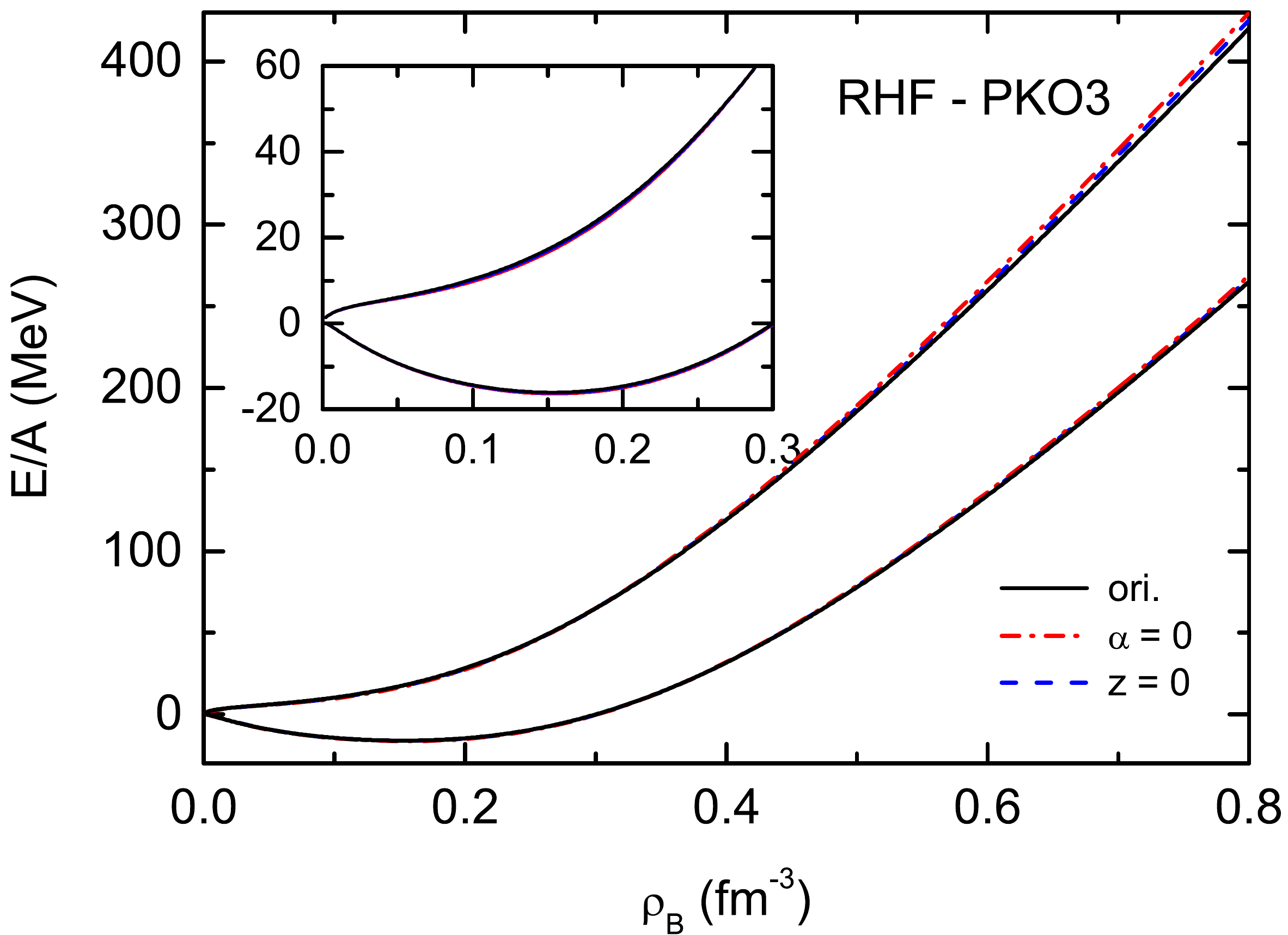}
\else
\includegraphics[width = 0.40\textwidth]{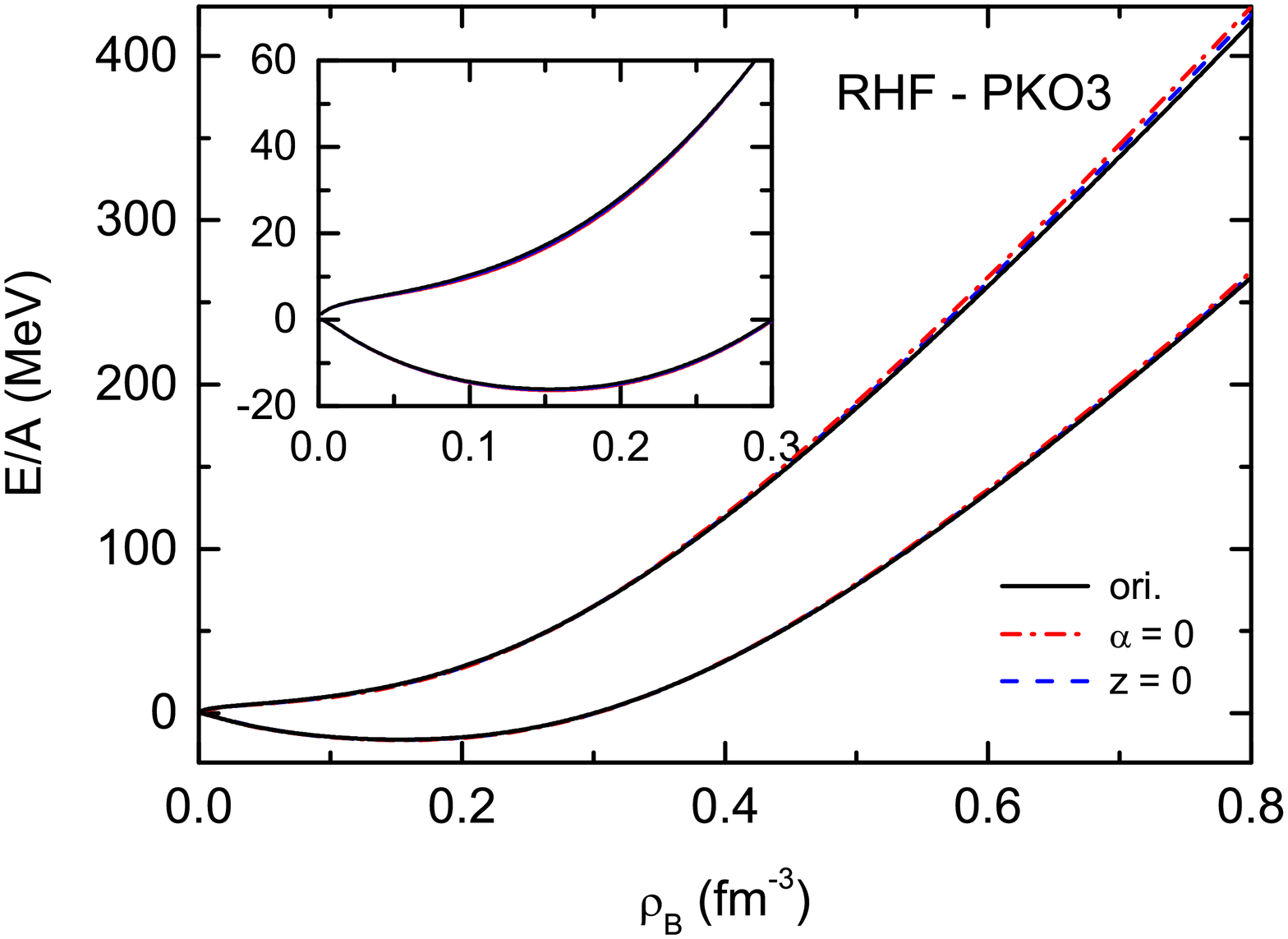}
\fi
\caption{ Dependence of binding energies per nucleon in symmetric nuclear and pure neutron matter (lower and
upper curves, respectively) on the baryonic density. The results are calculated with RHF theory with the original
PKO3 parameterization as well as its modifications within the SU(3) symmetric model. The parameters of this model
are chosen as $z = 0, \alpha_v= 1$ and $z = 1/\sqrt{6}, \alpha_v= 0$ which correspond to the cases where the $\phi$
mesons plays an important role. The inset shows the low-density region on a smaller scale.}
\label{fig:EoS_NM}
\end{figure}

The baryonic interactions involving strangeness will be treated below within either the SU(6) or SU(3) symmetric
quark models~\cite{Swart1963, Dover1984, Schaffner1994}, see Appendix~C. In the SU(6) symmetric model 
the $\phi$ meson has a vanishing $\phi$-$N$ coupling, whereas it does couple to nucleon in SU(3) symmetric model
through the relation~\eqref{eq:phiome relation} of Appendix~C. As commonly assumed, in what follows 
we will take $\tan \theta = 1/\sqrt{2}$, corresponding to ideal mixing; then one has
\begin{align}
\frac{g_{\phi N}}{g_{\omega N}} & = -
\frac{\sqrt{3} + \sqrt{2}z(1-4\alpha_v)}{\sqrt{6}-z(1-4\alpha_v)}.
\end{align}
However, we need to insure that this new coupling in the SU(3) model does not spoil the fits to the purely nuclear
data. To do so, we can make, consistent with Fierz transformation~\cite{Maruhn2001, Liang2012}, the replacement
\begin{align}
\frac{\tilde{g}^2_{\omega N}}{m^2_\omega}
=  \frac{g^2_{\omega N}}{m^2_\omega} + \frac{g^2_{\phi N}}{m^2_\phi},
\end{align}
where the $\tilde{g}_{\omega N}$ denotes the coupling for the case of $g_{\phi N} = 0$.

In Fig.~\ref{fig:EoS_NM} we show the EoSs of symmetric nuclear matter and pure neutron matter calculated using RHF DF with
PKO3 parameterization for two typical cases in the SU(3) model, $z = 0, \alpha_v= 1$ and $z = 1/\sqrt{6}, \alpha_v= 0$, where
the $\phi$ mesons plays an important role. It is clear seen that the EoSs of pure nuclear matter is almost independent of
the variations in $\alpha_v$ and $z$. Similarly, one could introduce $\sigma^\ast$-meson within the SU(3) symmetric model
without destroying the properties of nuclear matter. In this work, we assume that the $\sigma^\ast$ meson does not couple
to nucleons ($g_{\sigma^\ast N} = 0$). In other words, we shall use $\sigma$ and $\sigma^\ast$ mesons to constrain the
nucleon-hyperon ($NY$) and hyperon-hyperon ($YY$) interactions, respectively. A non-zero $g_{\sigma^\ast N}$ will lead to
a global readjustment of $g_{\sigma N}$ and $g_{\sigma^\ast Y}$, without any new information introduced.

\subsection{Meson-hyperon couplings}

The available experimental information on nucleon-hyperon ($NY$) and hyperon-hyperon ($YY$) interactions is scarce.
Hypernuclei provide experimental information on the depths of hyperonic potential-wells in symmetric nuclear matter
at saturation density. Additional information is obtained from nuclear reactions involving strangeness. Specifically,
the reactions producing $\Lambda$-hypernuclei in the $(\pi^+, K^+)$ associated production reactions~\cite{Hasegawa1996,
Hotchi2001} and $(K^-, \pi^-)$ strangeness exchange reactions, show that the $\Lambda N$ interaction is definitely
attractive. The binding energies of light  to heavy $\Lambda$-hypernuclei are well reproduced by mean-field models
~\cite{Millener1988, Glendenning1993, Tsushima1997, Keil2000, Cugnon2000, Vidana2001, Dalen2014, Fortin2017}. The
presently accepted value of the potential of the $\Lambda$ particle in nuclear matter is $U^{(N)}_\Lambda (\rho_0) \approx-30$~MeV.
The information for $\Xi N$ interaction is more uncertain, the few current experimental data indicating that the corresponding
potential depth $U^{(N)}_\Xi (\rho_0)$ is negative too, i.e., the interaction is attractive. The analysis of the missing-mass
spectra of production reactions $^{12}$C~$(K^-, K^+)$~$^{12}_{\Xi}$Be suggests that the attractive potential for $\Xi$ is smaller,
$U^{(N)}_\Xi (\rho_0)\approx-14$~MeV~\cite{Khaustov2000}. Finally, the situation with the $\Sigma N$ interactions is ambiguous.
The analysis of $(\pi^-, K^+)$ reactions on medium to heavy nuclei revealed a repulsive potential of the order of 40 MeV or
less~\cite{Noumi2002, Kohno2006}, while the observation of a $^4_\Sigma$He bound state in the $^4$He$(K^-, \pi^-)$ reaction
seems to be in favor of an attractive potential~\cite{Nagae1998}. Furthermore, the fits of $\Sigma$-atomic data indicate a
transition from an attractive $\Sigma$ potential at the surface to a repulsive one in the interior of a nucleus~\cite{Friedman2007}.
Therefore, we shall adopt below a repulsive potential for $\Sigma$ hyperon in nuclear matter $U^{(N)}_\Sigma (\rho_0) \simeq 30~\text{MeV}$.

Motivated by the above considerations, we determine the coupling constants, $g_{\sigma Y}$, using the following 
values of hyperon potentials in symmetric nuclear matter at saturation:
\begin{align}
\label{eq:hyp_potentials}
U^{(N)}_\Lambda (\rho_0) = & -30~\text{MeV}, \nonumber \\
U^{(N)}_\Xi (\rho_0) = -14~\text{MeV}, &\quad U^{(N)}_\Sigma (\rho_0) = 30~\text{MeV}.
\end{align}
Note that the above potentials need to be considered as isoscalar hyperon potentials.

Using the baryon self-energies given in Eqs.~(\ref{eq:Hartree self-energies}) and~(\ref{eq:Fock self-energies}), 
the potential for a single hyperon $Y$ embedded in the nucleonic matter can be written as
\begin{align}
\label{eq:UN}
U^{(N)}_Y(\bm{k}) = \Sigma_{S,Y}(\bm{k}) + \Sigma_{0,Y}(\bm{k}).
\end{align}

In the RHF theory, the self-energies are momentum dependent, therefore we consider their value at $\bm{k}=0$,
which corresponds to zero-momentum hyperon. Furthermore, the contribution of isovector mesons is small, 
even in symmetric nuclear matter, therefore we consider only the $\sigma$-scalar and $\omega(\phi)$-vector
couplings in the above expression. Note that rearrangement terms enter Eq.~\eqref{eq:UN} via the $\Sigma_{0,Y}$ terms.

The existing few data on multi-hyperon nuclei comes from the measurements on light double-$\Lambda$ nuclei. This provides the
means of extracting the $\Lambda\Lambda$ interaction between $\Lambda$ hyperons from the binding energy difference between 
double-$\Lambda$ and single-$\Lambda$ hypernuclei. However, the measured bond energies are subject to large uncertainty. 
The data on $^{10}_{\Lambda\Lambda}$Be and $^{13}_{\Lambda\Lambda}$B suggests that this energy is $\Delta B_{\Lambda\Lambda} \approx 5$
~MeV~\cite{Dover1991, Aoki1991}. The data on $^{6}_{\Lambda\Lambda}$He suggests a lower value $\Delta B_{\Lambda\Lambda} \approx
0.67\pm0.17$~MeV~\cite{Takahashi2001, Ahn2013}. Physically, one can interpret the bond energy as a rough estimate of the
$U^{(\Lambda)}_\Lambda$ potential at the average $\Lambda$ density $(\approx \rho_0/5)$ inside the hypernucleus~\cite{Vidana2001}.
We adopt, therefore, the value
\begin{align}
\label{eq:hyphyp_potentials}
U^{(\Lambda)}_\Lambda (\rho_0/5) = - 0.67~\text{MeV},
\end{align}
which reproduces the most accurate experimental data to date corresponding to the Nagara event~\cite{Takahashi2001, Ahn2013}.
Note that this value represents the potential well of a zero-momentum of $\Lambda$ particle in $\Lambda$ matter. 
This information we use to fix the value of the coupling $g_{\sigma^\ast \Lambda}$. However, we would like to 
note that the value given by Eq.~\eqref{eq:hyphyp_potentials} is suggestive and is somewhat lower than the one obtained
from the studies of $^{6}_{\Lambda\Lambda}$He nucleus~\cite{Fortin2017}. This study uses various RH DFs and calibrates
the DF parameters to obtain the bond energy quoted above. From this procedure it find that
$-3 \le U^{(\Lambda)}_\Lambda (\rho_0/5)\le -8$~MeV~\cite{Fortin2017}. Also, the nonrelativistic DF studies have shown 
that one could obtain precise bond energy by optimizing the ratio of the average $\Lambda$ density to the saturation 
density in He~\cite{Khan2015,Margueron2017}. Note, however, that the extrapolation of information derived from few-body
physics to statistical systems with large number of particles (here the infinite hypernuclear matter) is associated with
large uncertainties, which translate into uncertainties in the couplings among hyperons.

The coupling of remaining hyperons $\Xi$ and $\Sigma$ to the $\sigma^*$ is constrained by the relations
\begin{align}\label{eq:sigmastar relations}
\frac{g_{\sigma^\ast Y}}{g_{\phi Y}}
= \frac{g_{\sigma^\ast\Lambda}}{g_{\phi\Lambda}}, \quad Y \in (\Xi, \Sigma),
\end{align}
where $\phi$ refers to the $\phi$-meson.

\section{Numerical results and discussions}
\label{section4}

We now study the hyperonic matter within the RHF theory using several
parametrizations of the meson-baryon couplings implied by the SU(6)
and SU(3) flavor symmetric quark models and empirical hypernuclear
data. The baryon-exchange and baryon-transition processes are beyond
the scope of the present work and are disregarded below. We will use 
instead of $g_{\alpha Y}$ the ratio $R_{\alpha Y} = g_{\alpha Y}/g_{\alpha N}$.
The density dependence of the meson-hyperon couplings is the same as 
the meson-nucleon ones.

The parametrizations PKO1-3 and PKA1 have been carefully tuned in the nuclear sector, therefore 
we prefer not to modify the parameters of these CDFs. At the same time, as a guide, we will use 
the SU(3)/SU(6) quark model relations in the hyperonic sector in order to reduce the number of
the unknown coupling constants. By this, we do not imply a consistent SU(3)/SU(6) treatment of 
the full baryonic octet, since as mentioned above the nuclear sector remains fixed throughout 
our computation. Consequently, the true parameters are the ratios of the hyperonic to nucleonic
couplings and not the parameters of the SU(3)/SU(6) model.

\subsection{Stellar matter within the SU(6) symmetry}

We start our discussion with the case where the coupling constants of vector meson-hyperon interactions 
are fixed by the SU(6) symmetric model, see Table~\ref{tab:Couplings in SU6} of~\ref{app:C}. Although not
entirely realistic, this model is instructive because it allows us to study the effects of the Fock terms, 
including the Lorentz tensor couplings associated with terms $\propto\sigma^{\mu\nu}$ in 
Eq.~\eqref{eq:interaction Lagrangian}, on the EoS for neutron star matter.

\subsubsection{Hartree approximation vs Hartree-Fock approximation}

\begin{figure}[tb]
\centering
\ifpdf
\includegraphics[width = 0.40\textwidth]{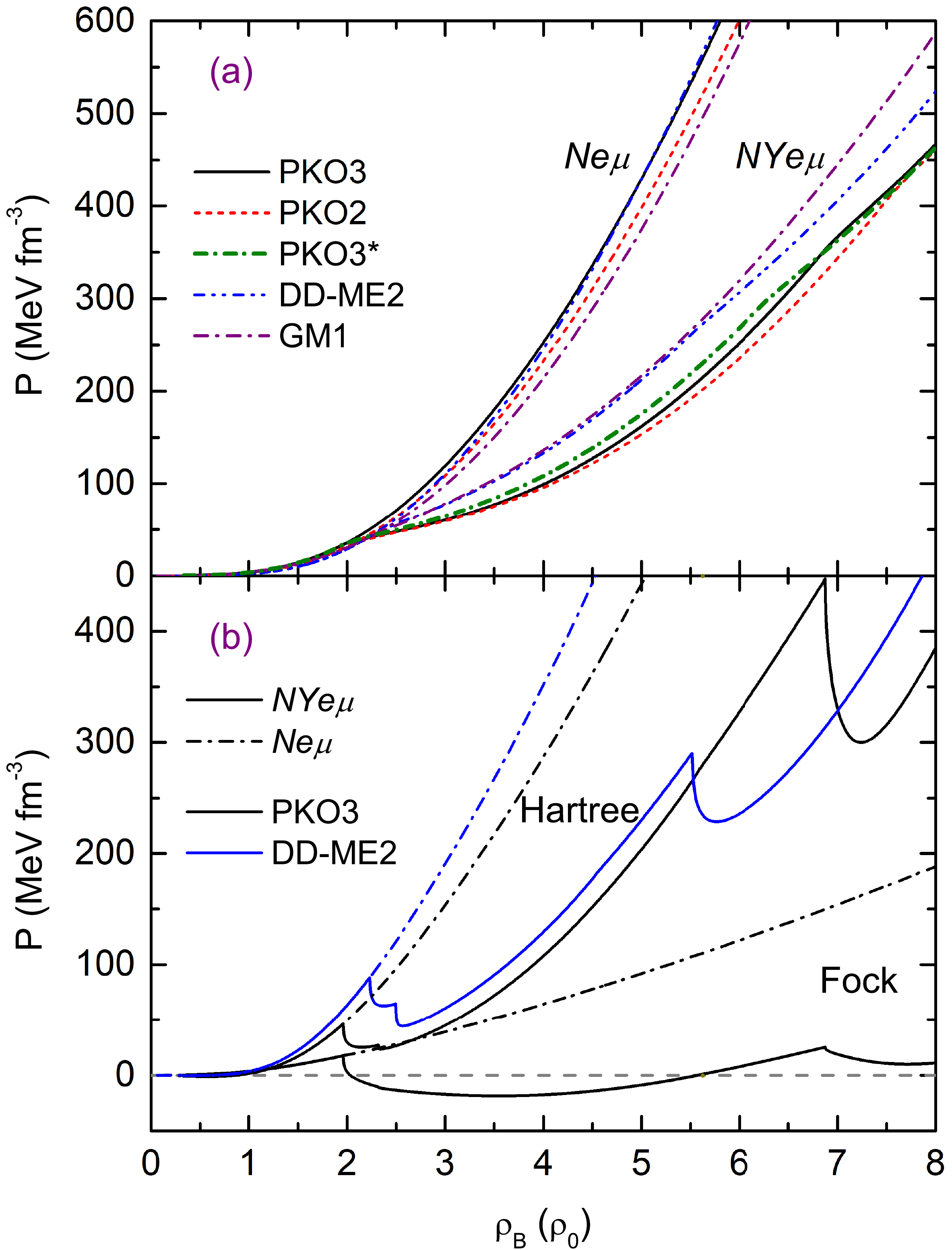}
\else
\includegraphics[width = 0.40\textwidth]{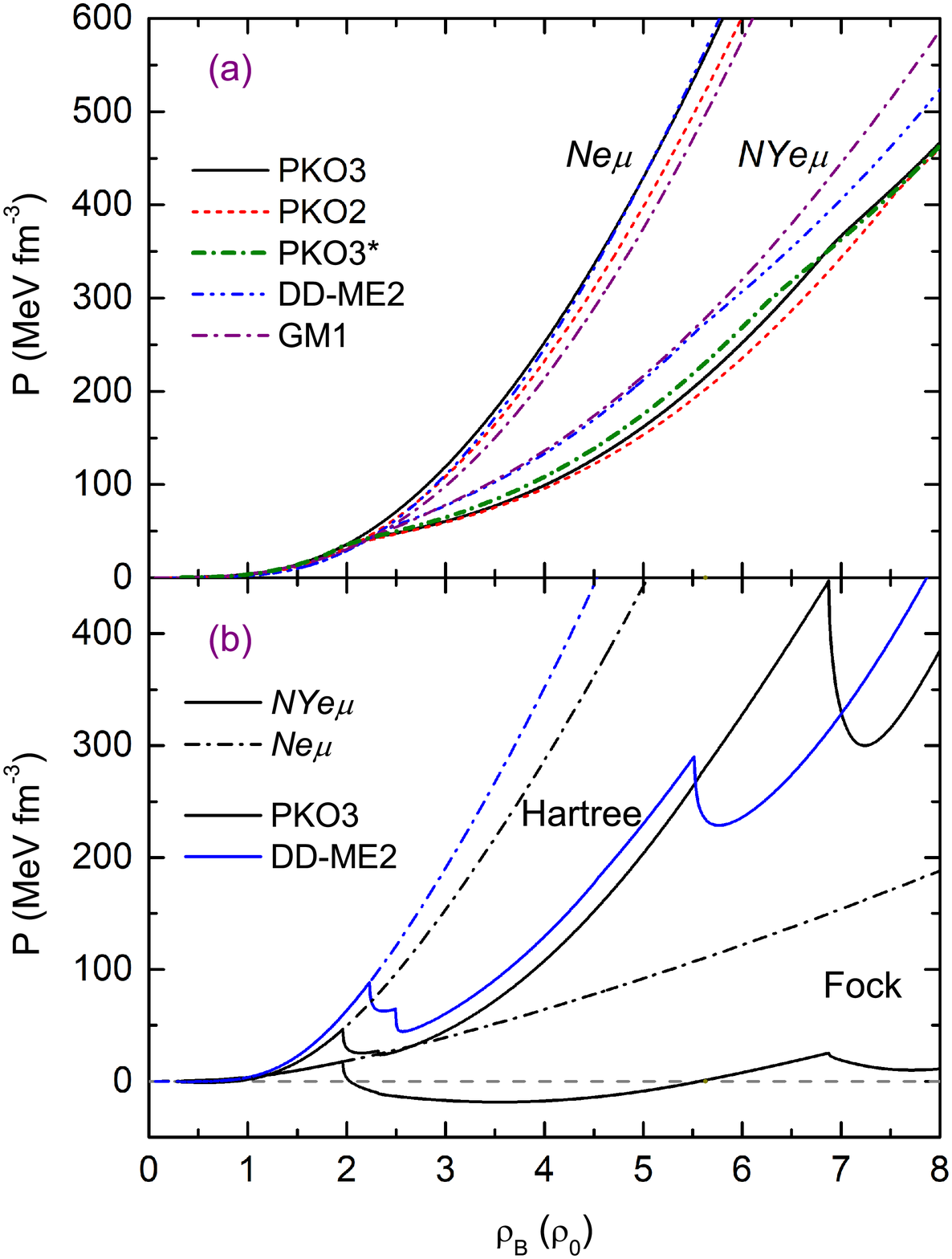}
\fi
\caption{(a) Dependence of the pressure of stellar matter with purely nucleonic and hyperonic compositions (labeled $Ne\mu$ and
$NYe\mu$, with $e, \mu$ indicating electron and muon) on the normalized baryonic density. Calculations shown include the
RHF parametrizations PKO2, PKO3 and the RH parametrizations DD-ME2, GM1 which have similar symmetry energy at the saturation
point. The PKO3$^\ast$ result represents the EoS where the $\pi$-$Y$ and the vector meson-$Y$ tensor couplings are neglected.
(b) Decomposition of the pressure resulting from the potential part of the interaction into Hartree and Fock channels. The
results are shown in the cases of PKO3 and DD-ME2 parametrizations. The vector meson-hyperon coupling constants are fixed
according to the SU(6) symmetric model.}
\label{fig:EoS_SET}
\end{figure}

Our detailed numerical results were obtained using four parametrizations. Two of these are based on the RHF DFT and
correspond to the PKO2 and PKO3 parametrizations; another two are based on more commonly used RH DFT, specifically 
on DD-ME2 and GM1 parametrizations. All these DFs predict similar EoS for purely nucleonic (i.e. $Ne\mu$) matter, 
see Fig.~\ref{fig:EoS_SET}(a). As expected, the appearance of hyperons ($NYe\mu$) softens the EoS, which is reflected
in the sudden change in the slope of the pressure at baryon density $\rho_B \simeq 2.5\rho_0$, which corresponds to the
threshold of the hyperon production. It is seen that quite generally the hyperonic EoSs based on RHF DFs are softer
than those based on the RH DFs, as the Fock terms contribute with an opposite sign to the self-energy of fermions 
in general.

To elucidate this argument we show the separate contributions from the Hartree and Fock channels to the pressure in $Ne\mu$ matter
and $NYe\mu$ matter in Fig.~\ref{fig:EoS_SET}(b). For both compositions the pressures are dominated by the Hartree channels.
In $Ne\mu$ matter both the Hartree and Fock contributions are smoothly increasing functions of the baryonic density, whereas
in $NYe\mu$ matter these are discontinuous at the onset of hyperons. At the onset of hyperons the slope of the pressure of the
Hartree and Fock contributions changes in a manner corresponding to a less repulsive interaction. We find that the Fock contribution
at intermediate densities is decreasing and can become negative. One may thus conclude that the softening on the EoS in the RHF DFs
is largely due to a pressure reduction induced by the Fock contribution.

The tensor interactions, originate from the $\pi$- and $\rho(\omega)$-meson exchanges are clearly important in the $NN$-scattering
and binding of deuteron. Since in the RH DFT the Fock diagrams are simply dropped, the RHF DFT becomes the only relativistic model
which generates a tensor force. To show explicitly the impact of the tensor interactions on the EoS of dense matter we show in
Fig.~\ref{fig:EoS_SET}(a) a representative EoS PKO3$^\ast$ in which the $\pi$-$Y$ and vector meson-$Y$ tensor couplings are excluded.
It is seen that within the SU(6) symmetric parameterization the inclusion of the tensor couplings tends to soften the EoS. Note that
the tensor effects are mainly due to the isoscalar-vector mesons, since $\omega(\phi)$-$Y$ tensor couplings are much larger than
$\rho(\pi)$-$Y$ ones. Furthermore, in the present models, the isovector-meson couplings depend exponentially on
density, therefore, the isovector fields are largely suppressed at high densities important for neutron stars.

\begin{figure}[tb]
\centering
\ifpdf
\includegraphics[width = 0.40\textwidth]{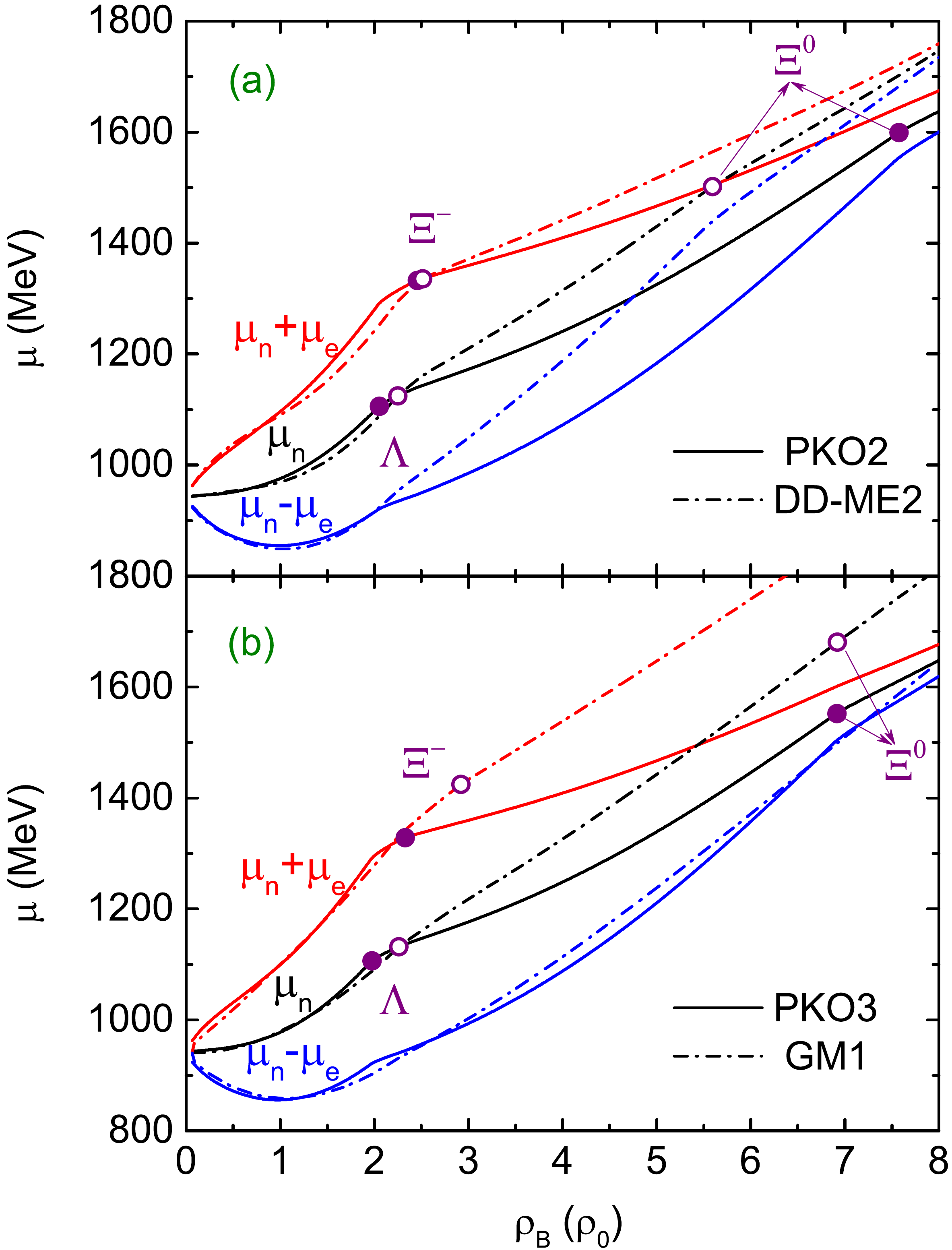}
\else
\includegraphics[width = 0.40\textwidth]{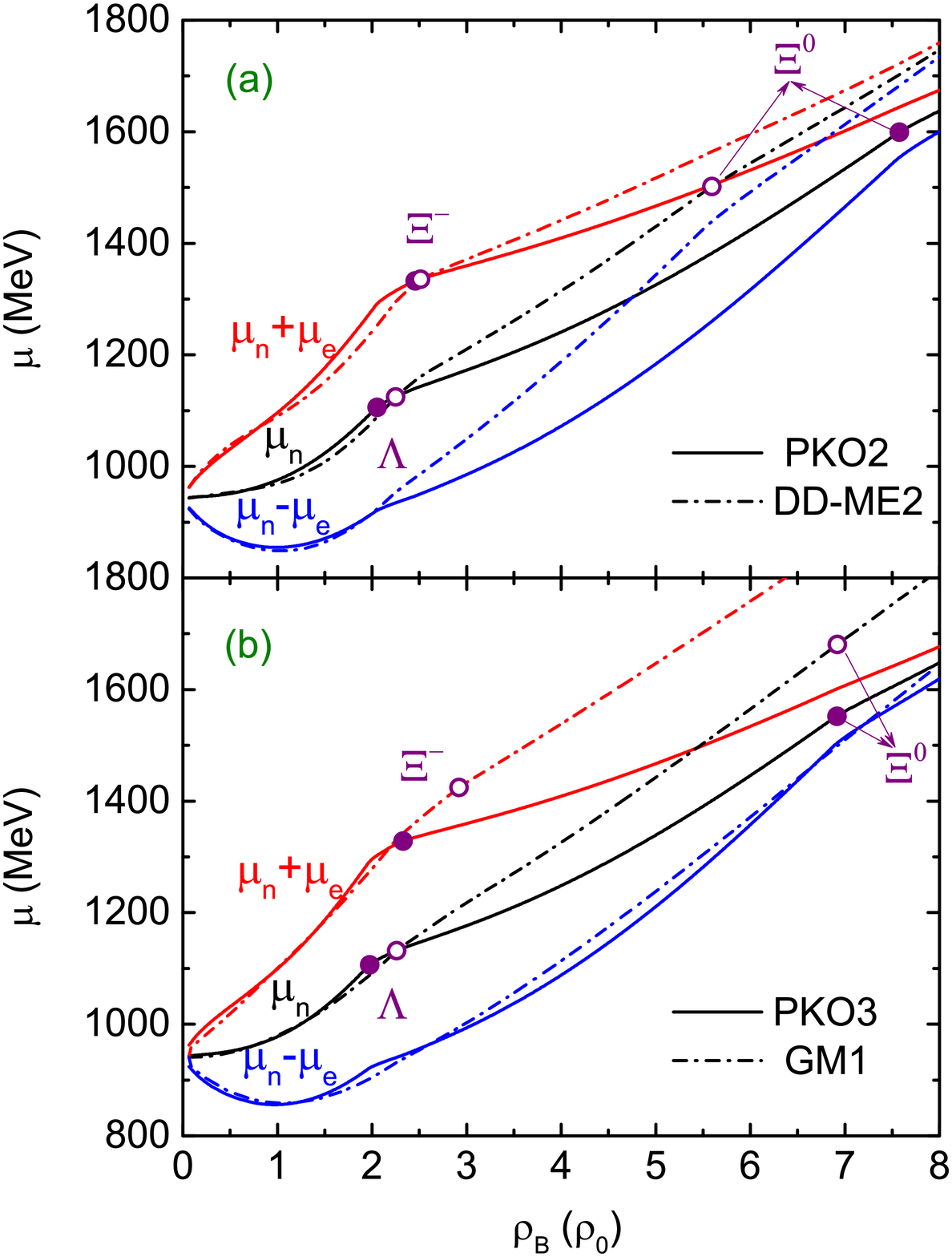}
\fi
\caption{Chemical potential for neutral, positively charged, and negatively charged baryons in $\beta$ equilibrium,
as a function of the normalized baryonic density. The results are calculated in the RHF-PKO2 (PKO3) DFs and are
compared to the RH-DD-ME2 (GM1) DFs. The onsets of hyperons in the RHF DFs are shown by filled circles, in RH
DFs by open ones. The vector meson-hyperon coupling constants are fixed according to SU(6) symmetric model.}
\label{fig:Ferm_SET}
\end{figure}

The chemical equilibrium conditions listed in Eq.~(\ref{eq:Chemical equilibrium}) determine the onset of hyperons, according to the
criterion that hyperons appear in dense stellar matter when the maximal energy of the neutrons (Fermi energy) becomes comparable to
the rest mass difference between hyperons and nucleons. In Fig.~\ref{fig:Ferm_SET} we show the chemical potentials for neutral,
positively charged, and negatively charged baryons in $\beta$ equilibrium, as well as the points where the onset of hyperons occur. 
It is seen that the types of hyperons that are being populated are the same for the four selected parametrizations. The neutral
baryon $\Lambda$ appears first due to its small rest mass and larger attractive potential, with all the models predicting similar
threshold densities ($2\rho_0$). In particular, the chemical potentials in RHF DFT with PKO2 and PKO3 parametrizations are lower
than that those in RH DFT with DD-ME2 and GM1 parametrizations above the threshold density of $\Lambda$ hyperon.

In $Ne\mu$ matter, the values of symmetry energy $J$ and its slope $L$ directly affect the chemical equilibrium of matter, 
in particular the neutron chemical potential. In general, for larger values of $J$ the neutron chemical potential $\mu_n$ 
increases more rapidly with the density. As a result, $\mu_n$, that defines the chemical potential of charge neutral baryons
(e.g., $\Lambda$) is larger for models with large $J$. A larger value of $L$ implies that the proton and electron chemical 
potentials (and their fractions) increase faster with the density. Because the sum $\mu_n + \mu_e$ defines the chemical
potential of negatively charged baryons (e.g., $\Xi^-$), their onset densities is lower for larger values of $L$.

These arguments are useful for understanding the onset densities of $\Lambda$ and $\Xi^-$ hyperons in $NYe\mu$ matter.
Indeed, to give a concrete example, we compare the RHF-PKO2 (PKO3) and RH-DD-ME2 DFs which differ mainly in the value of $L$.
As seen from Fig.~\ref{fig:Ferm_SET}(a), below the threshold density of $\Lambda$, the neutron chemical potential for 
RHF-PKO2 DF (stiff symmetry energy) is slightly larger than the one of RH-DD-ME2 DF (soft symmetry energy), which 
leads to a smaller $\Lambda$ onset density. However, the situation is reversed for the DFs PKO3 and GM1, used in 
Fig.~\ref{fig:Ferm_SET}(b), because in addition to the factors mentioned above isoscalar parameters of the DFs, 
such as the incompressibility $K$ and Dirac mass $M^\ast_D$ play a role.

\begin{figure*}[tb]
\centering
\ifpdf
\includegraphics[width = 0.92\textwidth]{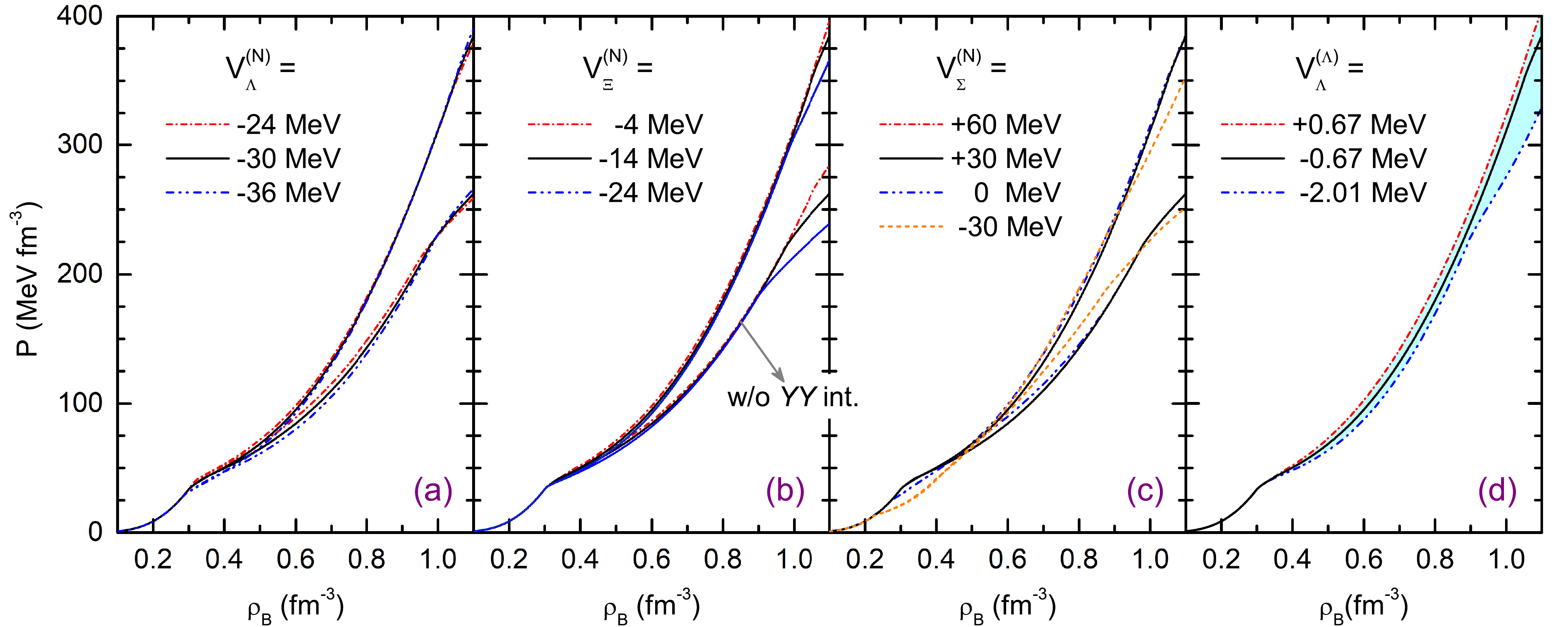}
\else
\includegraphics[width = 0.92\textwidth]{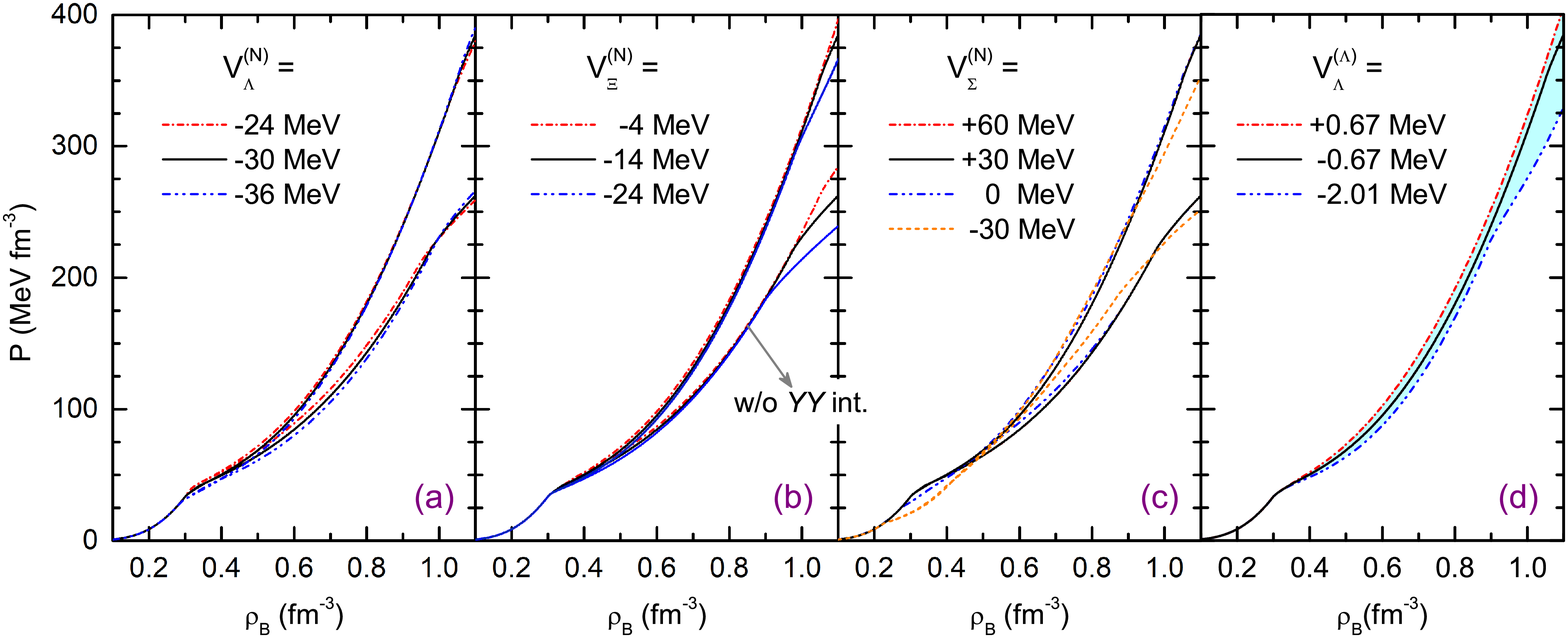}
\fi
\caption{Equations of state of hyperonic matter for different values of isoscalar hypernuclear potentials in nuclear matter
$V^{(N)}_Y (\rho_0)$ (panels a to c) and hyperonic potentials in hyperonic matter $V^{(\Lambda)}_\Lambda (\rho_0/5)$ (panel d).
We tune the coupling constants $\sigma(\sigma^\ast)$-$Y$ while the others are fixed assuming SU(6) symmetry.
(a) $V_Y \equiv ( V^{(N)}_\Lambda, V^{(N)}_\Xi, V^{(N)}_\Sigma, V^{(\Lambda)}_\Lambda )= (-30\pm6, -14, +30, -0.67)$~MeV,
the ordering of the EoSs according to their stiffness depends on the density interval; (b) $V_Y = (-24, -14\pm10, +30, -0.67)$ MeV;
(c) $V_Y = (-30, -4, +30\pm30, -0.67)$ MeV, and in addition the case where $V^{(N)}_\Sigma = -30$ MeV; the EoSs are not much altered
for positive $V^{(N)}_\Sigma$; (d) $V_Y = (-30, -14, +30, -0.67\pm1.34)$ MeV. The lower set of lines in panels (a-c) shows the results
in the case where $YY$ interactions are neglected. These results were calculated using the RHF DF with PKO3 parameterization.}
\label{fig:EoS_SU6}
\end{figure*}
\begin{figure*}[tb]
\centering
\ifpdf
\includegraphics[width = 0.92\textwidth]{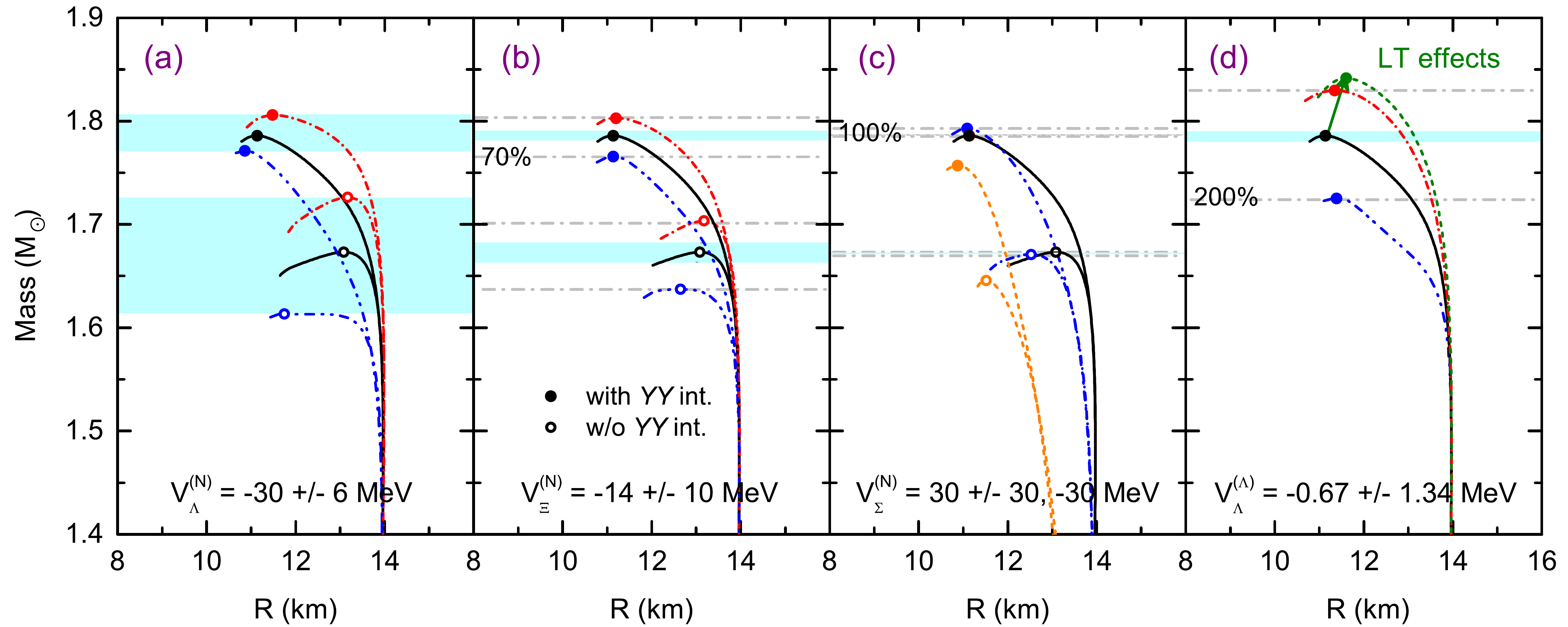}
\else
\includegraphics[width = 0.92\textwidth]{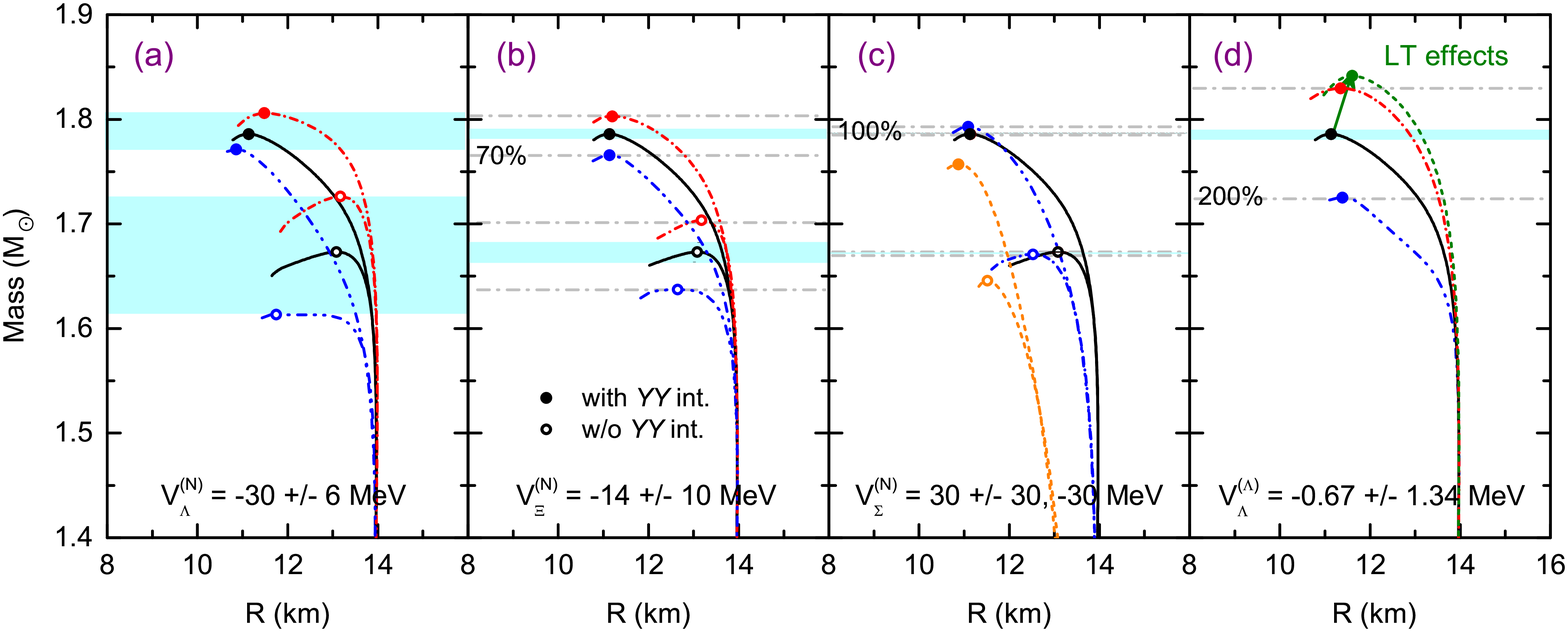}
\fi
\caption{Mass-radius relations for neutron stars for different values of isoscalar hypernuclear potentials. The circles show the point at
which a neutron star reaches the maximum mass. The notations are the same as in Fig.~\ref{fig:EoS_SU6}. The shaded bands indicate the
influences of varying $V_Y$ by 20\%, and those horizontal lines denote the influences of larger variations: for $V^{(N)}_\Xi$ by 70\%
(b), for $V^{(N)}_\Sigma$ by 100\% (c), for $V^{(\Lambda)}_\Lambda$ by 200\% (d). The effects of Lorentz tensor (LT) couplings are also
indicated in panel (d).}
\label{fig:ToV_SU6}
\end{figure*}

It is also seen in these figures that as soon as hyperons appear, there are large differences between the values of 
the chemical potentials of the RHF and RH DFs. It is not easy to understand these differences in terms of the characters
of the purely nucleonic matter. As seen from Fig.~\ref{fig:Ferm_SET}, in general, the density-dependence of the 
chemical potentials for RHF DFs are weaker than in RH DFs. We can speculate that the reason lies in the fact that 
Fock terms contribute to the chemical potentials with a sign which is opposite to the Hartree ones.

In this subsection, we illustrated the important features introduced by the Fock terms: they make the hyperonic EoS
rather soft with their contribution to the pressure of matter becoming even negative. In the following, we shall restrict
our attention to the RHF DFs.

\subsubsection{Varying hyperon potentials}

We now discuss the uncertainty in the hyperon potentials in nucleonic and hyperonic matter. The potential depth for $\Lambda$ has been
determined more reliably than those for $\Xi$ and $\Sigma$. Our strategy will be to keep fixed all hyperonic potentials but one and
vary the latter in a certain reasonable range by readjusting the $\sigma(\sigma^\ast)$-$Y$ coupling of the model. In Fig.~\ref{fig:EoS_SU6},
we show the EoSs of hyperonic matter for different values of isoscalar hypernuclear potentials $V^{(N)}_Y (\rho_0)$ in nuclear matter
[panels (a-c)] and the value of hyperonic $V^{(\Lambda)}_\Lambda (\rho_0/5)$ potential in hyperonic matter [panel (d)]. The corresponding
mass-radius relations of the compact star models derived from these EoSs are shown in Fig.~\ref{fig:ToV_SU6}. In both figures we also
show the results where the $YY$ interactions are omitted, i.e., $\sigma^*$-$Y$ and $\phi$-$Y$ couplings are set to zero. It is clearly
seen that the inclusion of strange $\sigma^\ast$ and $\phi$ mesons has an overall repulsive effect and stiffens the EoS. It also suppresses
the influence of uncertainty of hyperon potentials on the EoS.

We start our discussion with the potential depth of $\Lambda$ by changing its value in the range $-36\le V_\Lambda^{(N)} \le -24$~MeV 
which corresponds to $20\%$ variations around its accepted value $-30$ MeV. We do this by readjusting the couplings $g_{\sigma\Lambda}$ 
and keeping all other potential depth fixed at values given in Eqs.~\eqref{eq:hyp_potentials} and~\eqref{eq:hyphyp_potentials}. In the 
case where $YY$ interactions are included there is no sizeable change in the EoS of hypernuclear matter and only slight variations when
$YY$ interactions are set to zero. The variations in the maximum mass of corresponding compact stars are in the range of 0.05 $M_{\odot}$
in the first case and 0.1~$M_\odot$ in the second case and the variations in the radii are of the order of 0.5~km.

In the case of $\Xi$ potential we vary it in the range $-24 \le V_\Xi^{(N)}\le -4$~MeV around its accepted value $-14$ MeV. This corresponds
to a variation of $70\%$ around the central value. (The $20\%$ variation in the magnitude of this potential has no effect on the EoS and
stellar mass, see panel (b) of Fig.~\ref{fig:ToV_SU6}.) This leads to small variations in the EoS, which are reflected in the
uncertainty band in the panel (b) of Fig.~\ref{fig:ToV_SU6} of the order of 0.07 $M_\odot$ in the maximum mass.

In the case of $\Sigma$ hyperons, changing $\Sigma$ potential $V^{(N)}_\Sigma(\rho_0)$ by $20\%$ from the experimentally 
motivated value $30$~MeV, we find no changes at all. In this case the $\Sigma$s are not present up to density $\rho_B = 8\rho_0$. 
Next we consider variations of this potential in the range $0 \le V_\Sigma^{(N)} \le 60$ MeV (which correspond to variation of
$100\%$ around the central value $30$~MeV). We find that the EoS and the mass-radius relation are only slightly altered in this case. 
In addition we find that a weak repulsive potential, for example $V^{(N)}_\Sigma(\rho_0)$ ($\sim 5$ MeV), implies that $\Sigma^-$
appears at lower density than the lightest hyperon $\Lambda$. This is a straightforward consequence of the conditions
~(\ref{eq:Chemical equilibrium}a) and~(\ref{eq:Chemical equilibrium}b) according to which introducing $\Sigma^-$ and removing
an electron becomes energetically more favorable than adding a $\Lambda$ particle. In addition, we study the effect of strongly
attractive potential value $V_\Sigma^{(N)} = -30$~MeV. A strong attractive $V^{(N)}_\Sigma(\rho_0)$ leads to strong modifications
of the EoS (in the case of interacting hyperons -- to softening of the EoS), see Fig.~\ref{fig:EoS_SU6}(c), and to smaller masses
and radii of the stars, see Fig.~\ref{fig:ToV_SU6}(c).

For $YY$ interactions, represented by the potential $V^{(\Lambda)}_\Lambda(\rho_0/5)$, the situation is rather clear: deeper potentials yield
softer EoSs. Note that the variation of $V^{(\Lambda)}_\Lambda(\rho_0/5)$ do not affect the onset of $\Lambda$ which is essentially determined
by $V^{(N)}_\Lambda(\rho_0)$ but they do affect the abundance of these species. Varying the value of $V^{(\Lambda)}_\Lambda(\rho_0/5)$ by $20\%$
changes the maximum mass of corresponding stars by only about $0.01M_\odot$. For a larger variation of $200\%$ we find that the maximum mass
varies in the range $1.72 \le M/M_\odot \le 1.82$, see panels (d) of Figs.~\ref{fig:EoS_SU6} and~\ref{fig:ToV_SU6}.

Finally, we would like to illustrate the effects of Lorentz tensor couplings on the mass-radius relations of neutron stars. In
Fig.~\ref{fig:ToV_SU6}(d) we show also the case where the vector meson-$Y$ tensor couplings are set to zero. In this case we observe
that the maximum mass shifts up by about $0.06M_\odot$.

\begin{figure}[htbp]
\centering
\ifpdf
\includegraphics[width = 0.44\textwidth]{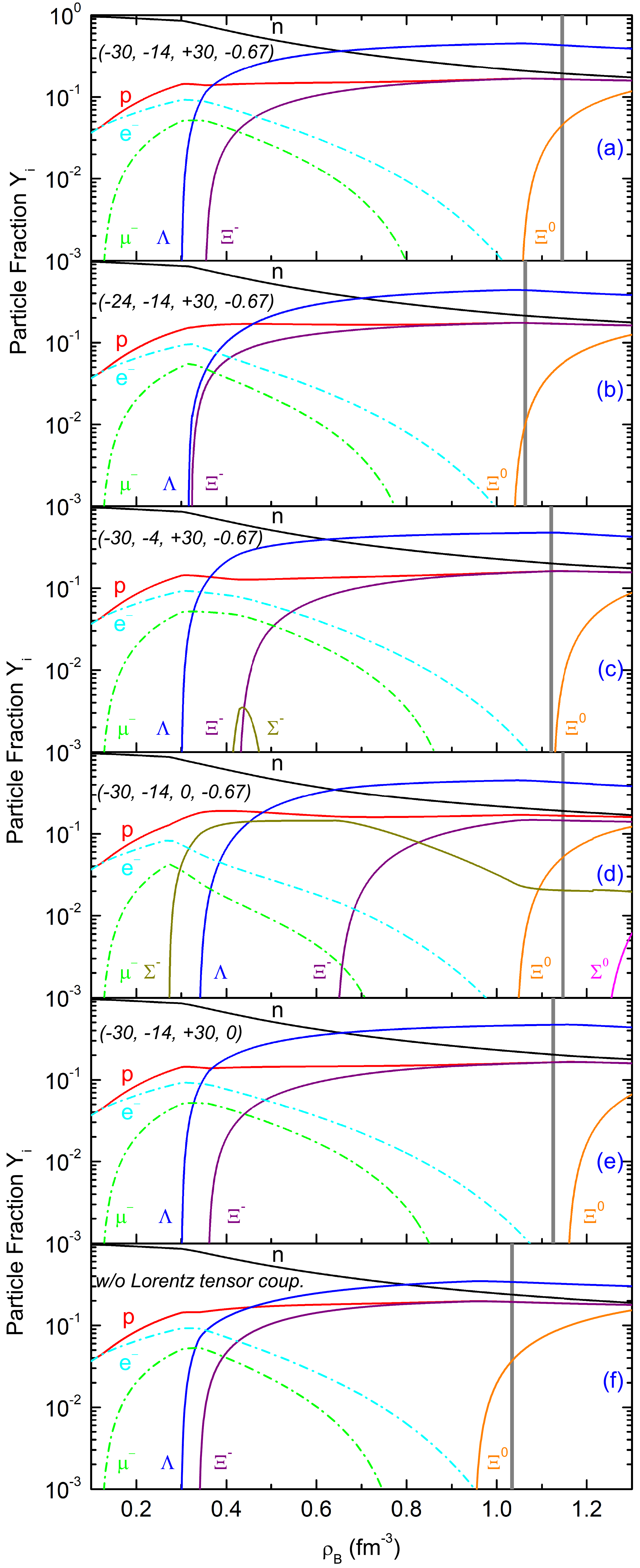}
\else
\includegraphics[width = 0.44\textwidth]{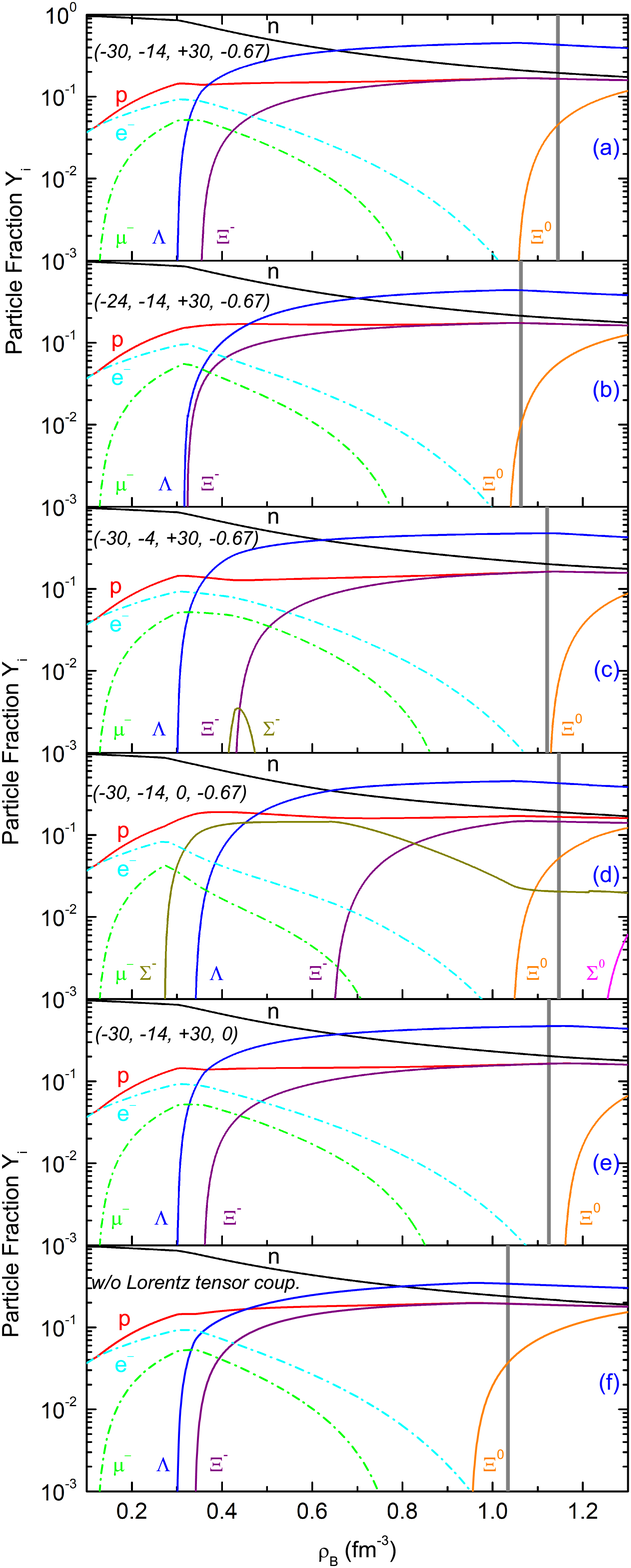}
\fi
\caption{ Particle fractions of hyperonic matter for different values of isoscalar hypernuclear potentials in nuclear matter
$V^{(N)}_Y (\rho_0)$ and hyperonic potentials in hyperonic matter $V^{(\Lambda)}_\Lambda (\rho_0/5)$. We tune the coupling constants
$\sigma(\sigma^\ast)$-$Y$ while the others are fixed according to the SU(6) symmetric model.
(a) $V_Y \equiv (V^{(N)}_\Lambda, V^{(N)}_\Xi, V^{(N)}_\Sigma, V^{(\Lambda)}_\Lambda) = (-30, -14, +30, -0.67)$~MeV;
(b)$V_Y = (-24, -14, 30, -0.67)$ MeV; (c)$V_Y = (-30, -4, 30, -0.67)$~MeV; (d)$V_Y = (-30, -14, 0, -0.67)$~MeV;
(e)$V_Y = (-30, -14, +30, 0)$ MeV; (f) same as (a) but switch off the Lorentz tensor (LT) couplings. The thick vertical lines 
indicate the central density of the respective maximum mass configurations. The results are calculated using the RHF DF with 
PKO3 parameterization.}
\label{fig:Frac_SU6}
\end{figure}

Because EoS represents the sum of contributions from each baryons (and leptons), we varied so far their potentials in turn, to see the
amount of changes associated with each baryon. These variations can be better understood if one examines the particle fractions 
corresponding to those EoSs, which are shown in Fig.~\ref{fig:Frac_SU6} for selected values of potential depths.

Firstly, comparing panels (a) and (b) of Fig.~\ref{fig:Frac_SU6}, we find that a less deep potential $V^{(N)}_\Lambda(\rho_0)$ pushes up
the threshold density of $\Lambda$, while the onsets of $\Xi^{-, 0}$ shift down in density and the overall fraction of particles vary to
some extent. Similarly, a less deep potential $V^{(N)}_\Xi(\rho_0)$ pushes up the threshold densities of $\Xi^{-, 0}$, while new specie
$\Sigma^-$ could arise, see panel (c). We see that if one type of hyperon is pushed up in density by the change of their potential, a
different hyperon takes its place; due to this compensation mechanism the variations in the $\Lambda (\Xi)$ potentials do not affect the
EoS substantially. For example, as discussed above, the $20\%$ variation in the $V^{(N)}_\Lambda(\rho_0)$ results in the variation in the
maximum mass and the radius of hyperonic stars by $0.04M_\odot$ and 0.50~km respectively.

Secondly, comparing panels (a) and (d) of Fig.~\ref{fig:Frac_SU6} we notice that the change in the $\Sigma^-$ potential can drastically
change their onset density and matter composition. Indeed in panel (a) where the $\Sigma^-$ potential is large and positive it is completely
absent from matter composition, whereas in panel (d) where its potential is zero it is the first hyperon to appear. In this case $\Sigma^-$
populates matter up to densities of about $4\rho_0$ where $\Xi^-$ sets in at a similar density as in the case of repulsive
$V^{(N)}_\Sigma(\rho_0) = 30$~MeV. At higher density, $\Xi^-$ essentially replaces $\Sigma^-$. Thus, we conclude that $\Sigma^-$, if it
appears, it will affect the intermediate density region ($3\sim5 \rho_0$) of the EoS.

Next, comparing the panels (a) and (e) of Fig.~\ref{fig:Frac_SU6} we find that the variation of $YY$ interaction for $\Lambda$ particles
does not change their onset density, but modify their fraction. We note that according to Eq.~(\ref{eq:sigmastar relations}) any reduction
$g_{\sigma^\ast\Lambda}$ leads to a reduction of $g_{\sigma^\ast\Xi}$ and $g_{\sigma^\ast\Sigma}$. Therefore, the threshold densities for
$\Xi$s are higher (which is more notable for $\Xi^0$s) and their abundances are overall decreased. (The same would apply to $\Sigma^-$ if
the parameter choice allows for their appearance).

Finally we study the effect of the Lorentz tensor coupling (present only in the HF computations) by comparing panel (a) to panel (e) of
Fig.~\ref{fig:Frac_SU6}, in which this coupling has been set to zero. We find that the overall attraction provided by the vector meson-hyperon
tensor couplings shifts down the onsets of $\Xi^-$s, and reduces the abundances of hyperons. It is thus clear that the tensor couplings,
which contribute only through the Fock diagrams, have an important effect on the particle populations in dense matter.

In conclusion, we find that the variations of hyperon potentials change the threshold densities at which the hyperons appear. We also
observe that the particle fractions vary depending on the magnitude of the potentials, especially the appearance of $\Sigma^-$ crucially
depends on the value of its potential. In addition we find a compensation mechanism which shows that when the $\Lambda$ particles
are disfavored, then the $\Xi^-$ hyperons replace them, thus keeping the total amount of strangeness in matter at the same level. As a
consequence, the variations of particle fractions do not affect the global properties of stars, i.e. their maximum mass and the corresponding
radius. Note however, that the inclusion of $YY$ interactions has an important effect on the maximum mass.

The maximum masses obtained from our models so far do not satisfy the lower bound on the maximum mass $M \sim 2 M_\odot$ of compact stars
set by the observations of PSR J0348+0432 and PSR J1614-2230. The ways to overcome this difficulty suggested in the literature include:
(a) reduction of the strength of the hyperon coupling to meson~\cite{Colucci2013, Dalen2014}, (b) increase in the hyperon potentials
$U^{(N)}_Y$~\cite{Weissenborn2012a, Long2012}, or (c) modifying the values of the hyperon couplings away from the values implied SU(6)
symmetry~\cite{Weissenborn2012b, Katayama2012}. We will now follow the prescription (c) at the same time keeping fixed the
experiment estimates of $U^{(N)}_Y(\rho_0)$ and $U^{(\Lambda)}_\Lambda(\rho_0/5)$ discussed above.

\subsection{Stellar matter within the SU(3) symmetry}

In this section, we fix the coupling constants of vector meson-hyperon interactions according to the more general SU(3) flavor symmetry in
order to allow for stellar sequences which contain members with masses of the order of observed 2$M_\odot$. For the vector meson-$Y$ tensor
couplings, represented by the ratios $\kappa_{\alpha Y}$, we will again use their SU(6) values which are listed in Table~\ref{tab:Couplings in SU6}
of Appendix~C.

\subsubsection{Varying parameters $z$ and $\alpha_v$}

\begin{figure}[tb]
\centering
\ifpdf
\includegraphics[width = 0.40\textwidth]{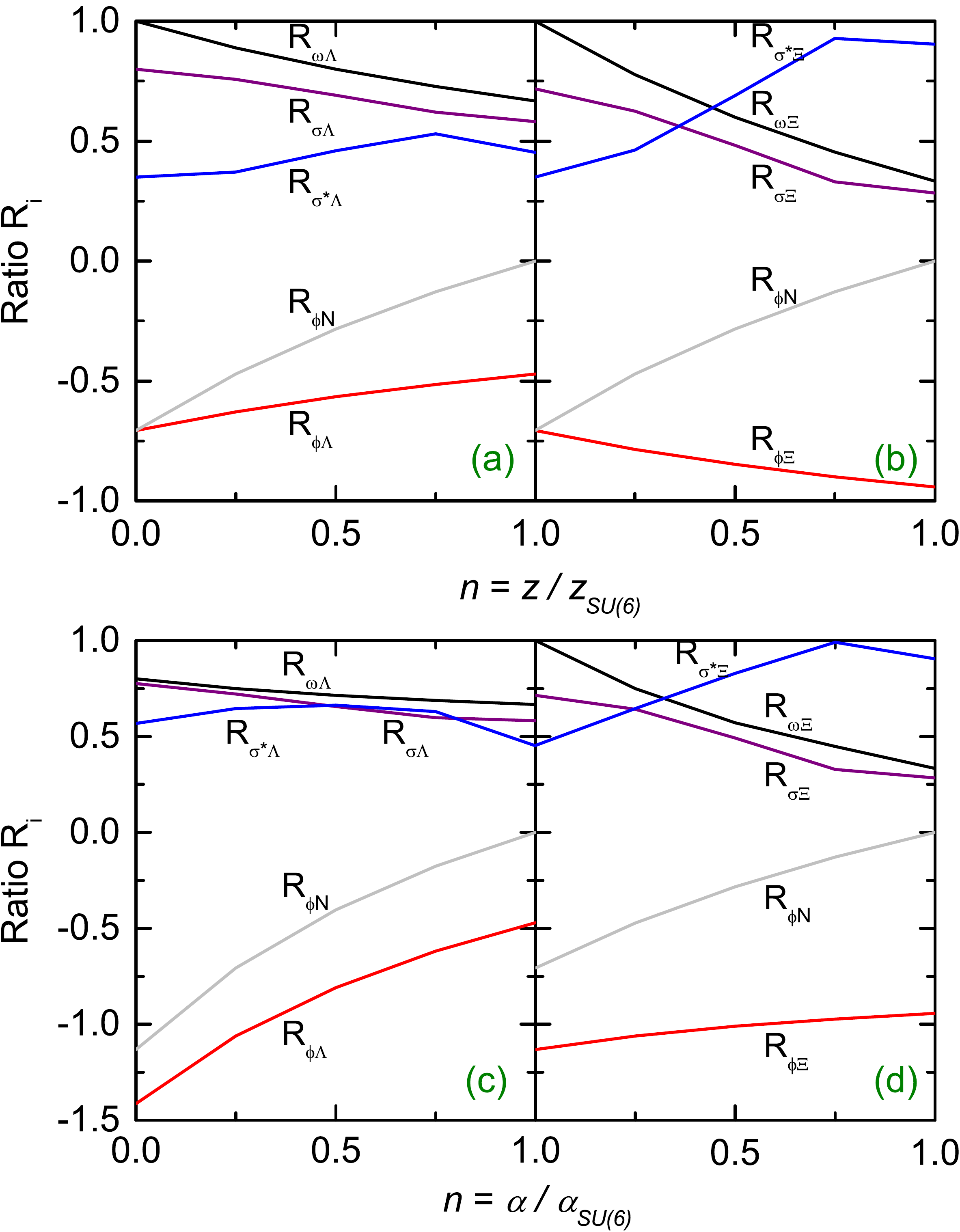}
\else
\includegraphics[width = 0.40\textwidth]{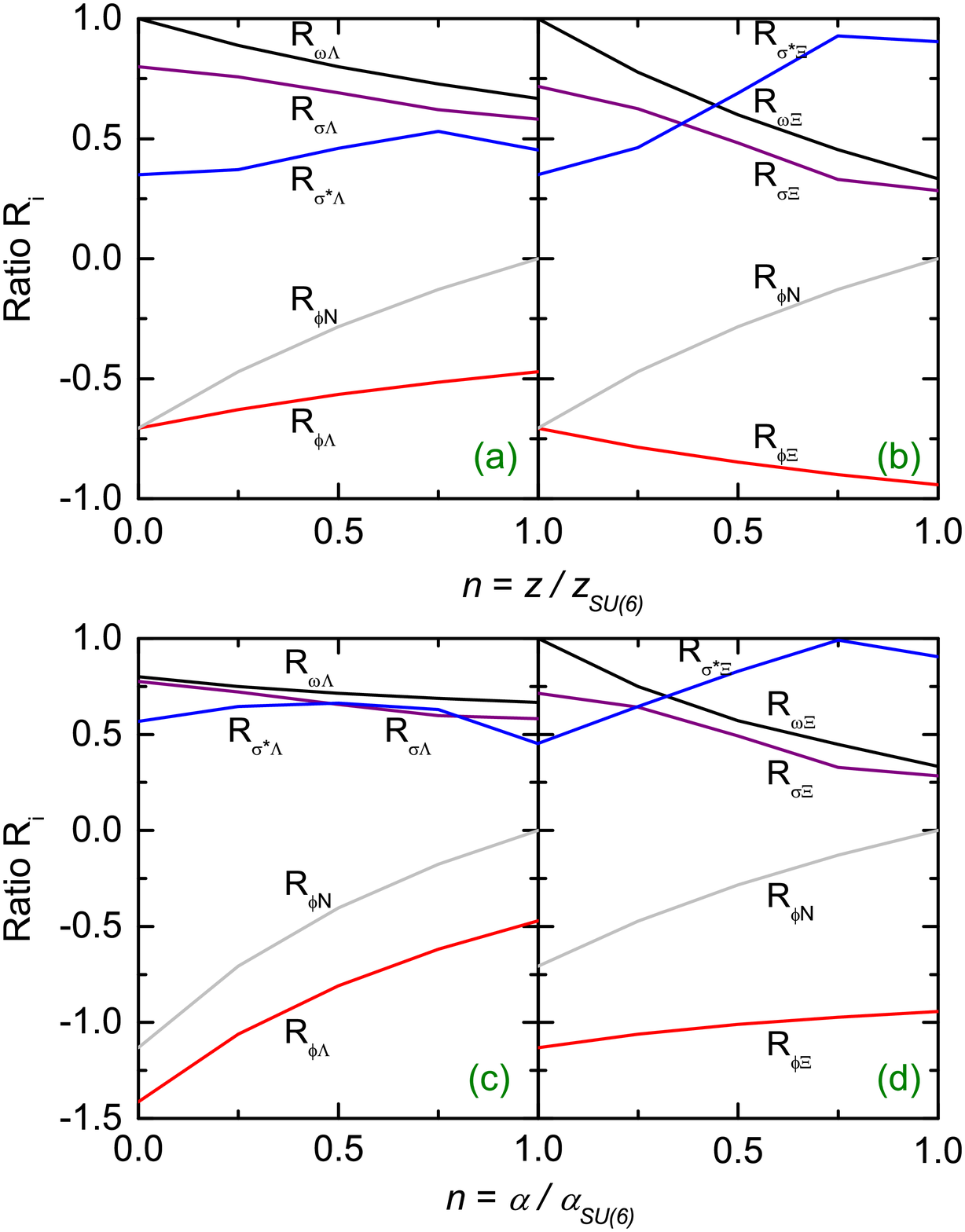}
\fi
\caption{Relative coupling strengths for $\Lambda$ hyperon (panels a, c) and $\Xi$ hyperon (panels b, d) in the isoscalar channels as a
function of $z (\alpha)$ and normalized to their values in SU(6) model for fixed $\alpha = 1 (z = 1/\sqrt{6})$.}
\label{fig:CouZA}
\end{figure}
%
\begin{figure*}[tb]
\centering
\ifpdf
\includegraphics[width = 0.90\textwidth]{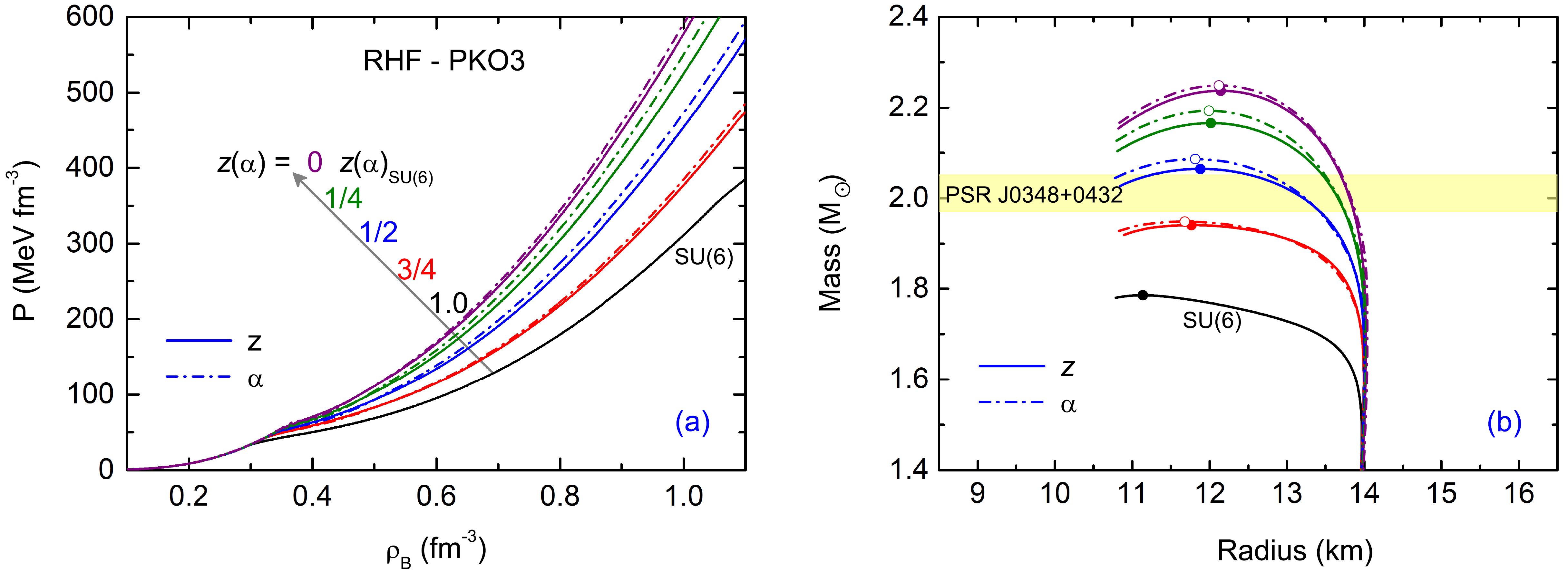}
\else
\includegraphics[width = 0.90\textwidth]{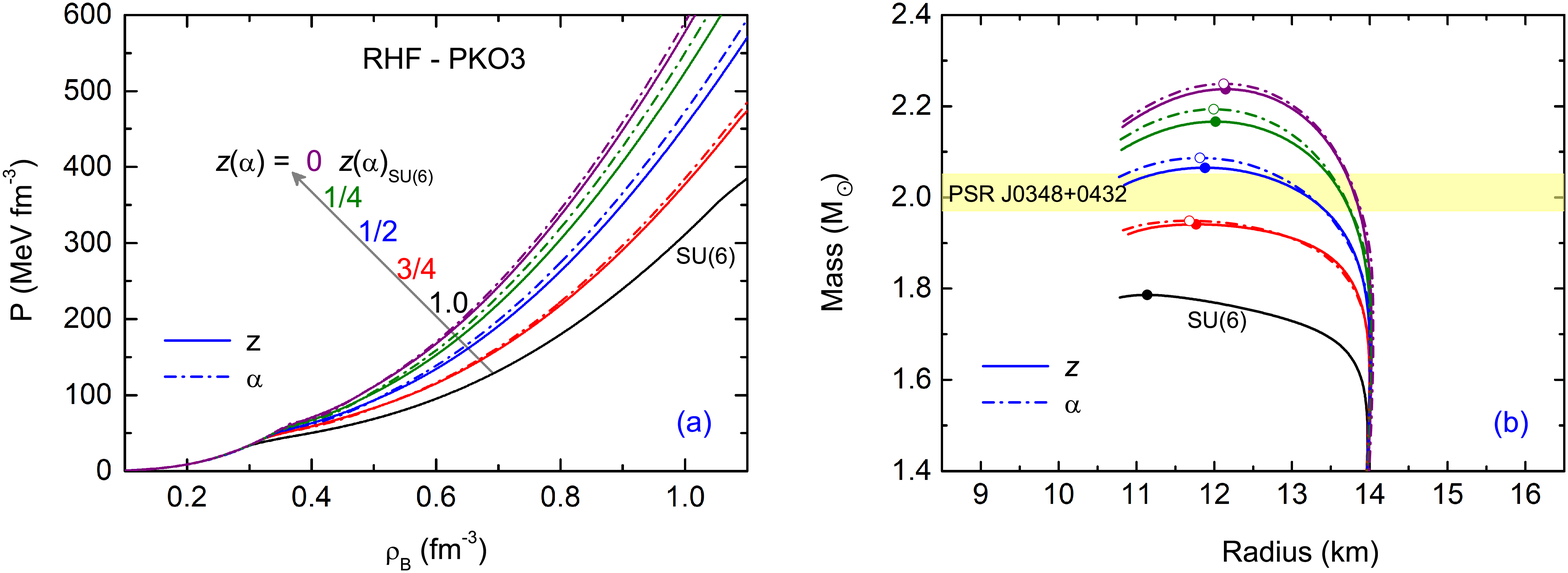}
\fi
\caption{Equation of state of hyperonic matter (a) and mass-radius relation (b) within different value of $z (\alpha)$ under the SU(3)
flavor symmetry. The coupling constants $\sigma(\sigma^\ast)$-$Y$ for each case are fitted by the isoscalar hypernuclear potentials
$V_Y (\rho_0) \equiv (V^{(N)}_\Lambda, V^{(N)}_\Xi, V^{(N)}_\Sigma) = (-30, -14, +30)$~MeV and $V^{(\Lambda)}_\Lambda (\rho_0/5) = -0.67$ MeV.}
\label{fig:EoS_SU3}
\end{figure*}

Consider the ideal mixing case where $\theta_v = \tan^{-1}(1/\sqrt{2})$ for vector mesons. The two choices which allow us to go beyond SU(6)
symmetry are the tuning of the parameters $z$ or $\alpha_v$, see Appendix~C. Below we drop the subscript $v$ for simplicity.

We first probe the effects of variation of the ratio $z$ (which corresponds to variation of the coupling of baryons to the meson octet
$g_8$) on the stiffness of the hadronic EoS and corresponding mass-radius relation. We restrict $z$ to the interval $z\in[0, 1/\sqrt{6}]$,
where the upper bound corresponds to the SU(6) value. We do not consider values larger than this upper bound because the larger the $z$ value
the softer is the EoS.

\begin{figure*}[tb]
\centering
\ifpdf
\includegraphics[width = 0.90\textwidth]{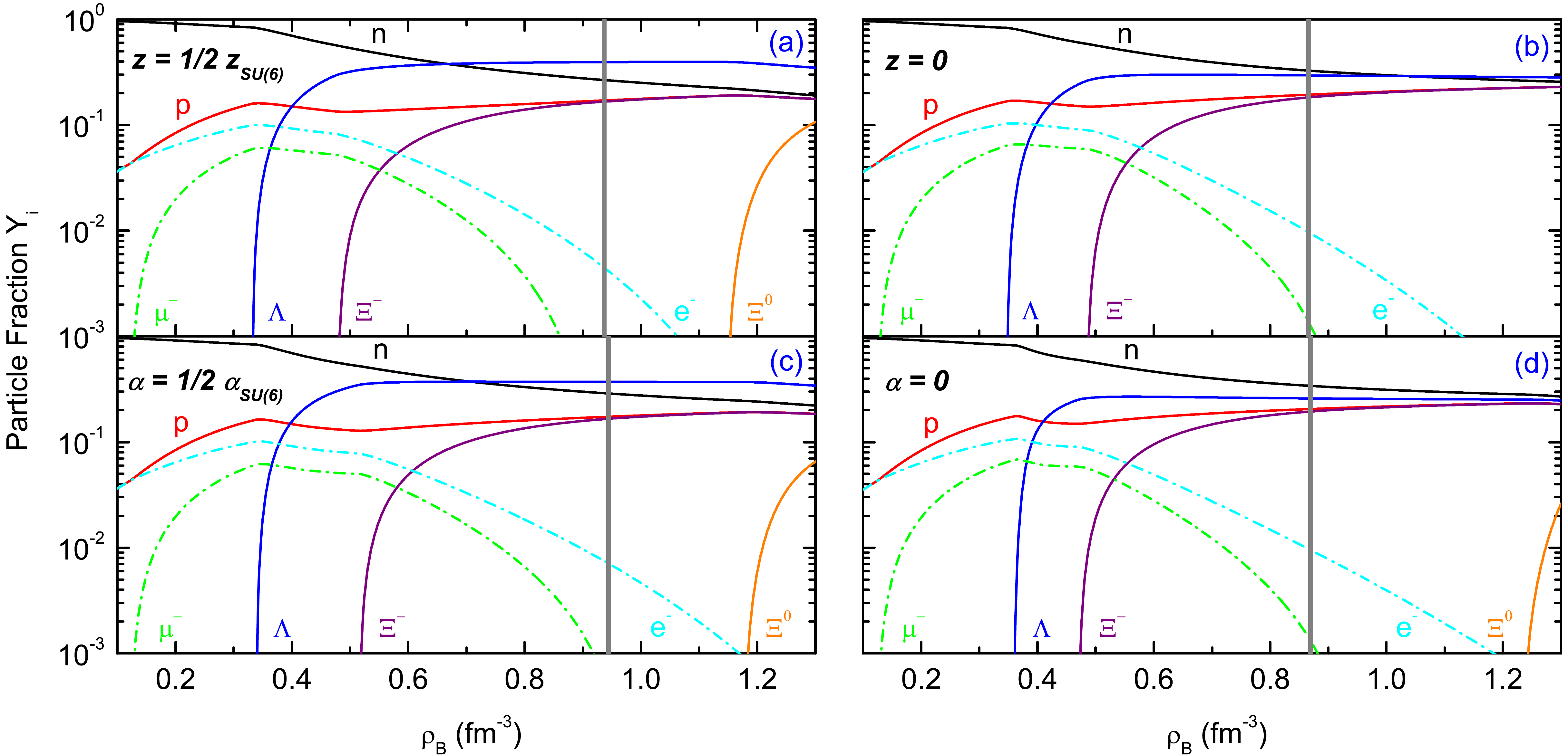}
\else
\includegraphics[width = 0.90\textwidth]{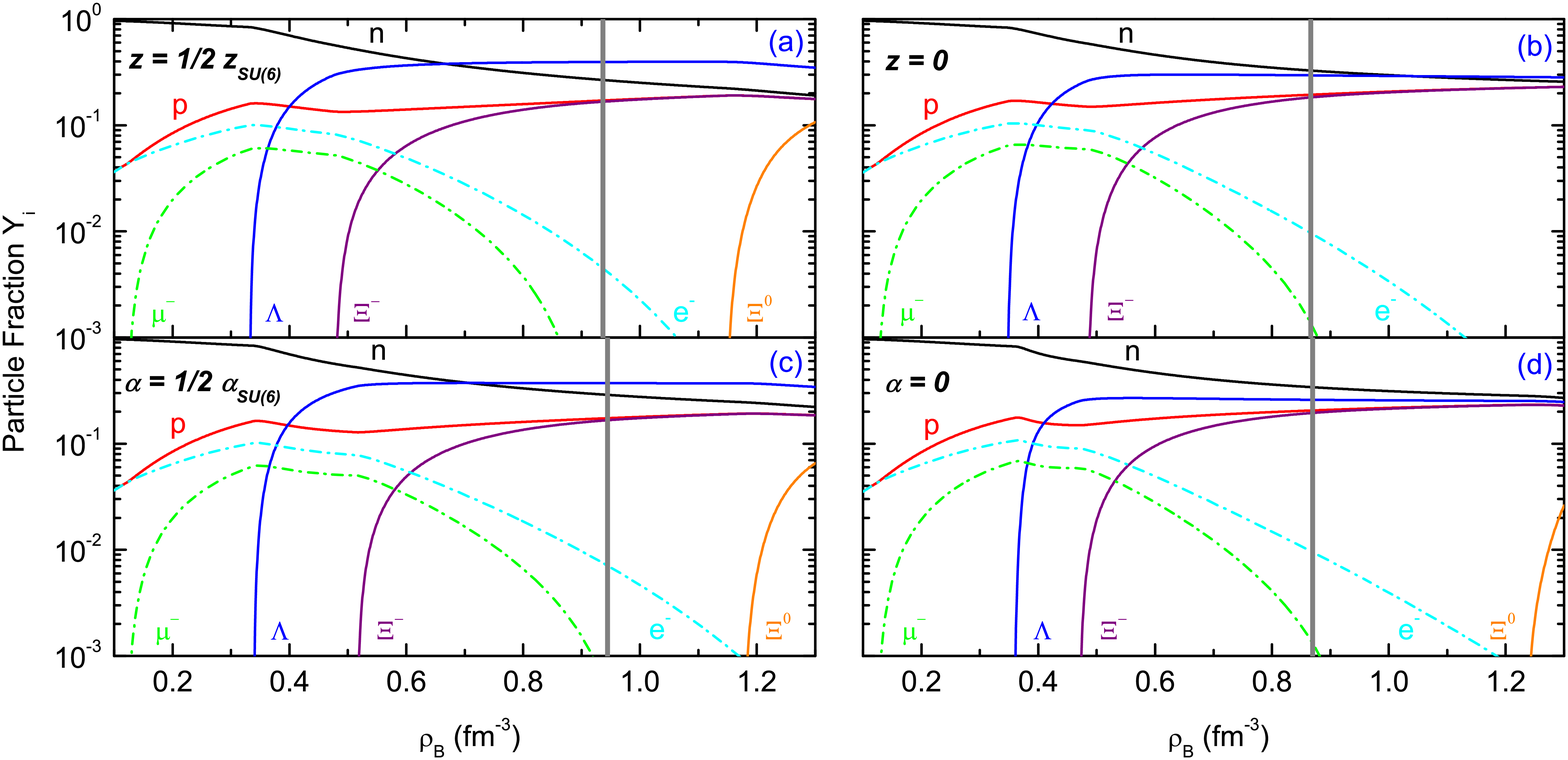}
\fi
\caption{Particle fractions of hyperonic matter in the SU(3) flavor symmetric model for (a) $z = 0.5z_{SU(6)}$, (b) $z = 0$,
where $\alpha$ is fixed to its SU(6) value, (c) $\alpha= 0.5\alpha_{SU(6)}$, (d) $\alpha = 0$, where $z$ is fixed to its SU(6) value. 
The SU(6) case is given by $z_{SU(6)} = 1/\sqrt{6}$ and $\alpha_{SU(6)} = 1$.
The thick vertical lines indicate the central density of the respective maximum mass configurations. The results are calculated
using RHF approach with PKO3 parameterization.}
\label{fig:Frac_SU3}
\end{figure*}

\begin{figure*}[tb]
\centering
\ifpdf
\includegraphics[width = 0.70\textwidth]{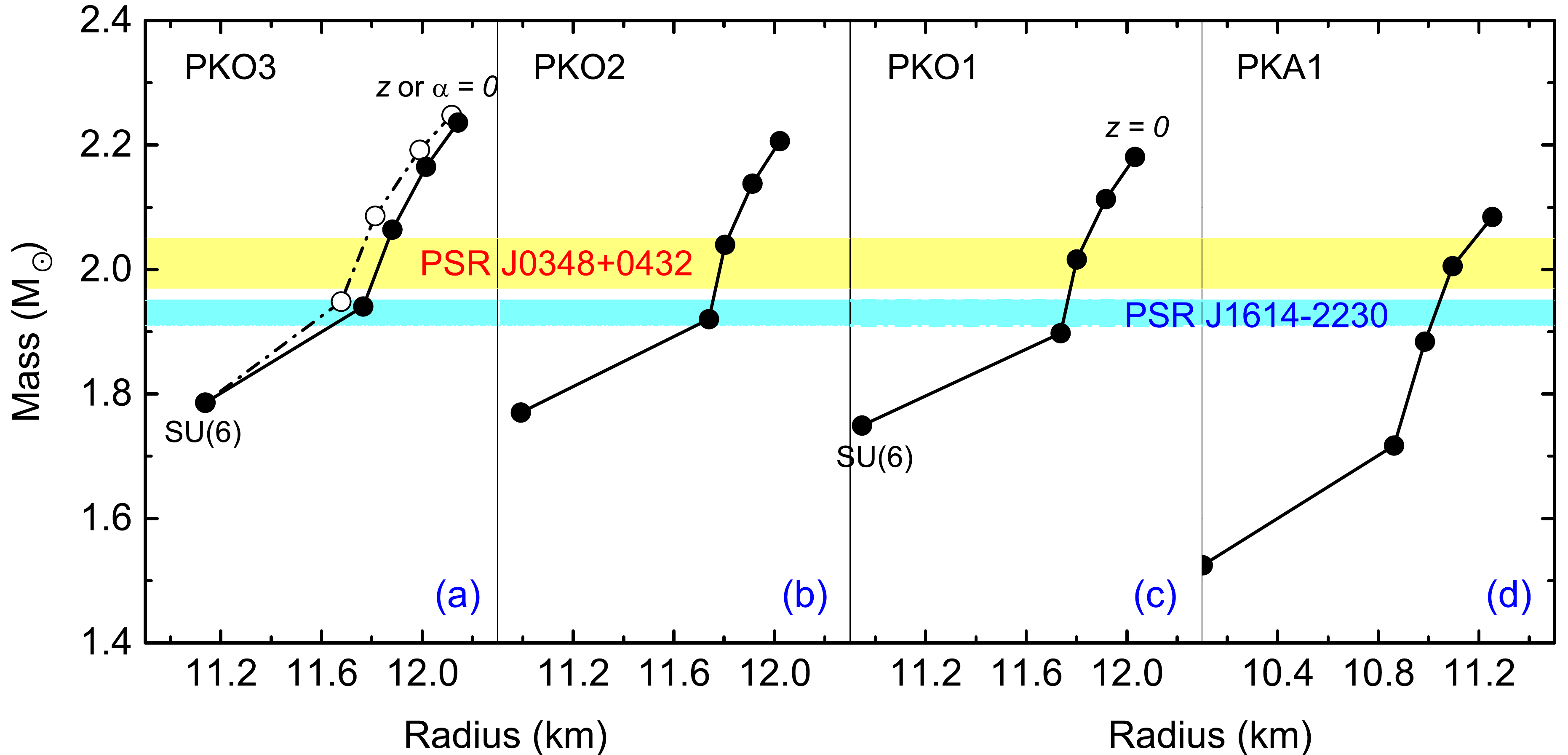}
\else
\includegraphics[width = 0.70\textwidth]{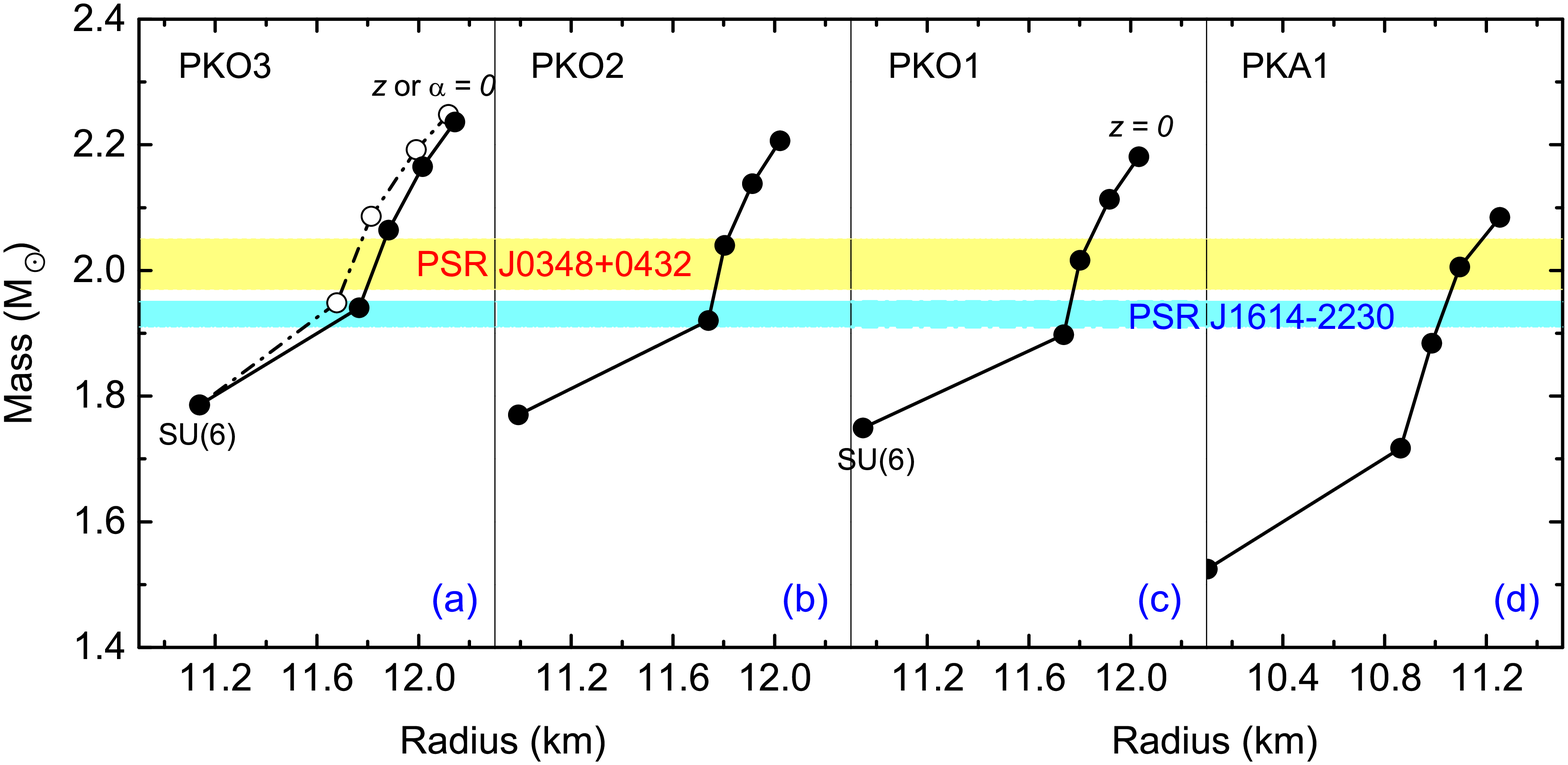}
\fi
\caption{The maximum masses versus radii for the hyperonic stars obtained from different RHF DF and for the range of SU(3) symmetric
model values $0 \le z\le 1/\sqrt{6}$ (solid lines). In the case of PKO3 parameterization (panel a) we also show the same results, but
for $0\le \alpha \le 1$. The shaded areas show the masses of two massive pulsars, PSR J0348+0432 with $M=2.01\pm 0.04 M_\odot$, and
PSR J1614-2230 with $M=1.93\pm 0.02 M_\odot$.}
\label{fig:Sum}
\end{figure*}

We now consider the relative couplings ratios defined by the relation $R_{\alpha Y} = g_{\alpha Y}/g_{\alpha N}$; the corresponding values
in the case of SU(6) symmetric model are given in the Appendix~C and Table~\ref{tab:Couplings in SU6}. These ratios for $\Lambda$
and $\Xi$ as functions of $z$ under SU(3) symmetry are shown in Fig.~\ref{fig:CouZA}. We see that with decreasing $z$ the $\omega$-$\Lambda$
and $\sigma$-$\Lambda$ couplings increase, the $\phi$-$\Lambda$ couplings decrease and $\sigma^*$-$\Lambda$ stays approximately constant.
In the case of $\Xi$ hyperons the coupling behave in the same manner, except the strange meson coupling $\sigma^\ast$-$\Xi$ is now decreasing for
$z/z_{SU(6)}\le 0.75$ and that for $\phi$-$\Xi$ is increasing; here $z_{SU(6)} =1/\sqrt{6} $ is the SU(6) value of the $z$. We see that
with decreasing $z$ the combined effect of repulsive vector meson-baryon interactions and attractive scalar meson-baryon interactions becomes
more repulsive. In the case of $\Lambda$ particle this is also due to the increase in the difference $R_{\omega\Lambda}-R_{\sigma\Lambda}$.
The same can be said for $\Xi$ mesons, where in addition there is a reduction in the attractive $\sigma^*$-$\Xi$ interaction.

The EoS for parameter values $z/z_{SU(6)} = 0, $ 0.25, 0.50, 0.75, 1.0 with $\alpha = 1$ are plotted in Fig.~\ref{fig:EoS_SU3}(a) (in solid lines). 
It is seen that the EoS stiffens with decreasing $z$. In Fig.~\ref{fig:EoS_SU3}(b) we plot the corresponding mass-radius relations. As expected
from the influence of $z$ on the stiffness of the EoS, the maximum mass grows up from the value $M = 1.79M_\odot$ for $z/z_{SU(6)} = 1$ to
$M = 2.24M_\odot$ for $z/z_{SU(6)} = 0$.

Now we examine at the particle fractions which are shown in Fig.~\ref{fig:Frac_SU3}(a, b) for $z/z_{SU(6)}= 0.5$ and 0, and compare
with the SU(6) case (shown in Fig.~\ref{fig:Frac_SU6}(a)). In the case $z/z_{SU(6)}= 0.5$, we find that the particle composition of
the core matter is the same as in the SU(6) case. On decreasing $z/z_{SU(6)}$ to 0, we find that the onset densities of hyperons are
pushed upward in density and, at the same time, their fractions decrease. We note that the onset density of $\Lambda$'s is not shifted
much, since before the appearance of $\Lambda$s in every case considered we have the same $Ne\mu$ properties together with the same
constraint $V^{(N)}_\Lambda(\rho_0) = -30$~MeV.

We turn now to the examination of the effects of the ratio $\alpha$ on the stiffness of the hadronic EoS and mass-radius relation.
We vary $\alpha$ in the interval $\alpha\in[0, 1]$. The relative coupling ratios for $\Lambda$ and $\Xi$ hyperons as functions of $\alpha$
in the SU(3) symmetric model are shown in Fig.~\ref{fig:CouZA}. Repeating the arguments we used in the case of variations of $z$ we
see that the decrease in $\alpha$ has an overall repulsive effect, as expected.

The stiffness of the EoS depends monotonously on $\alpha$ and, again, we obtain the softest EoS for the SU(6) case (i.e., a pure $F$-type
coupling), while the stiffest EoS is obtained for $\alpha = 0$, which corresponds to the pure $D$-type coupling of the meson-baryon multiples.
The resulting neutron star maximum mass grows up to $M = 2.25M_\odot$, almost coinciding with the $z = 0$ case studied above, see
Fig.~\ref{fig:EoS_SU3}. The particle fractions for $\alpha/\alpha_{SU(6)}= 0.5$ and 0 are shown in Fig.~\ref{fig:Frac_SU3}(c, b).

It is interesting to observe that, working in the case of ideal mixing $\theta^{\text{id.}}$ for vector mesons, tuning $z$ or $\alpha$
down from their SU(6) value, yield comparable results, both for the EoS and the particle abundances.

\begin{table*}[tb]
\caption{Isoscalar meson-hyperon coupling constants for the RHF models of dense matter which produce sequences of compact stars with
massive 2$M_\odot$ stars. We list the model the parameters specifying the SU(3) parameterization, i.e., $z$ and $\alpha_v$, the values
of $R_{\sigma Y} = g_{\sigma Y}/g_{\sigma N}$ and $R_{\sigma^\ast Y} = g_{\sigma^\ast N}/g_{\sigma N}$.}
\setlength{\tabcolsep}{17pt}
\label{tab:SU3 parameters}
\begin{tabular}{ccccccccc}
\hline
  DF   & $z$ & $\alpha_v$ & $R_{\sigma\Lambda}$ & $R_{\sigma\Xi}$ & $R_{\sigma\Sigma}$ & $R_{\sigma^\ast\Lambda}$ &
 $R_{\sigma^\ast\Xi}$ & $R_{\sigma^\ast\Sigma}$ \\
\hline
 PKA1  & $\frac{1}{4\sqrt{6}}$  &  1            & 0.7650 &  0.6410  &  0.5138 & 0.6134 & 0.7668 & 0.6134 \\
 PKO1  & $\frac{1}{2\sqrt{6}}$  &  1            & 0.7029 &  0.4863  &  0.4157 & 0.3802 & 0.5703 & 0.3802 \\
 PKO2  & $\frac{1}{2\sqrt{6}}$  &  1            & 0.6913 &  0.4790  &  0.4053 & 0.4161 & 0.6242 & 0.4161 \\
 PKO3  & $\frac{5}{8\sqrt{6}}$  &  1            & 0.6561 &  0.4059  &  0.3906 & 0.4992 & 0.8112 & 0.4992 \\
 PKO3  & $\frac{1}{\sqrt{6}} $  & $\frac{5}{8}$ & 0.6255 &  0.4083  &  0.4711 & 0.6518 & 0.9125 & 0.3911 \\
\hline
\end{tabular}
\end{table*}

\subsubsection{Alternative parametrizations of the RHF DF}

To explore the dependence of our results on the choice of the DFs, we repeated the above procedures for the remaining three sets
of RHF DFT parameterization. Figure~\ref{fig:Sum} shows the maximum masses and the corresponding radii of neutron stars when the $z$
parameter is varied in the range~$z\in[0, 1/\sqrt{6}]$. One can clearly see that the four parametrizations show similar trends and
differ only in details. The maximum mass increases by about 0.5$M_\odot$ in the cases of PKOi ($i=1$-3) parameterization and by 0.6$M_\odot$
in the case of PKA1 parameterization when $z$ is varied in the range indicated above. In any event, all four parameterization predict
heavy enough ($M> 2.0 M_\odot$) compact star once SU(3) symmetry is assumed.

One may notice that the PKA1 DF is special in our collection as it predicts smaller radii than the PKOi ($i=1$-3)
DFs. The main difference between these DFs are the values of symmetry energy and slope parameter, see Table~\ref{tab:NMP}. 
As a general trend, it has been established that the radius of a compact star increase with the slope parameter $L$, 
which is opposite to the trend we see~\cite{Horowitz2001,Cavagnoli2011}. However, the EoS derived from the covariant
DFT is very sensitive to both the isoscalar and isovector channels. If we only focus on the symmetry energy $J$ and 
the slope $L$ of nucleonic EoS, its effect on the hyperonic EoS could be split into two pieces: on one hand, the 
larger $L$ (or $J$) the stiffer the nucleonic sector of the EoS; on the other hand, the larger $L$ (or $J$) the 
lower the $\Lambda$ onset density and the larger the $\Lambda$ fraction. This latter effect makes the EoS softer
for our models in which $\Lambda$ is the dominant hyperon. Therefore, in our case, it is the interplay between
the above two effects that determines the properties of hyperonic EoS. Furthermore, the density dependence of 
the couplings in the two isoscalar channels for PKA1 parameterization is quite different from the PKOi ($i=1$-3)
parametrizations, all of which feature similar density dependences of couplings, see Refs.~\cite{Long2007, Sun2008}. 
The EoS derived from the covariant DFT is very sensitive to balance between the two isoscalar channels, therefore
the resulting differences are expected.

\begin{figure*}[tb]
\centering
\ifpdf
\includegraphics[width = 0.90\textwidth]{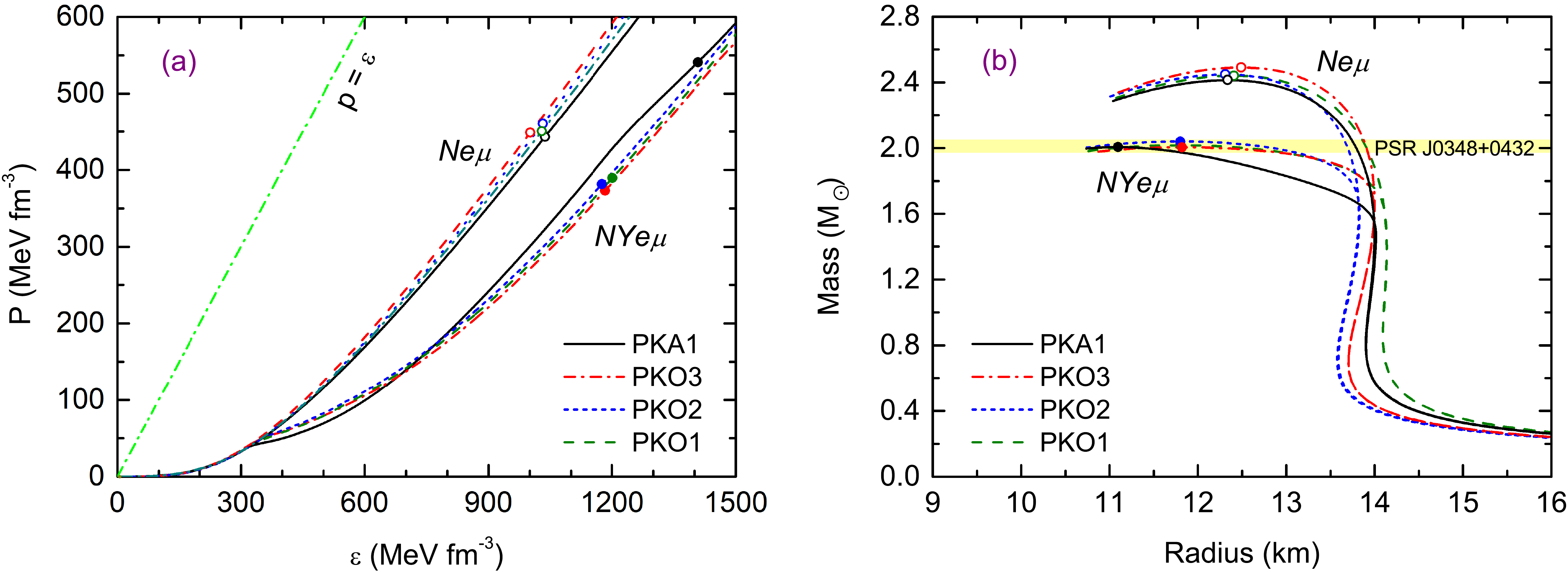}
\else
\includegraphics[width = 0.90\textwidth]{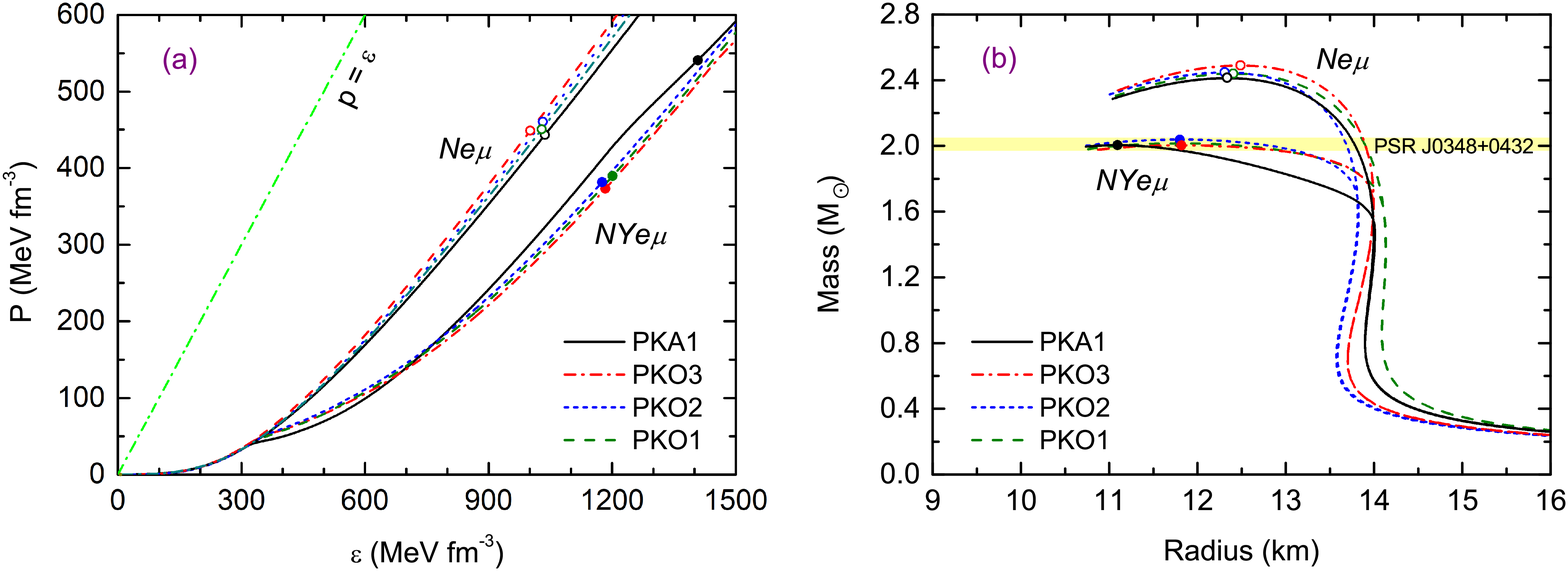}
\fi
\caption{Equation of state (a) and mass-radius relation (b) of hyperonic matter ($NYe\mu$) predicted by the models that favor a maximum gravitational
mass of 2.0$M_\odot$ under SU(3) flavor symmetry. The nucleonic-only ($Ne\mu$) results are also shown. The circles shows the point at which a neutron
star reaches the maximum mass. The shaded area portrays the mass of pulsar PSR J0348+0432, $2.01\pm 0.04 M_\odot$.}
\label{fig:EoS_2M}
\end{figure*}

From our detailed analysis above, we conclude that massive hyperonic neutron stars can be obtained within RHF 
DFT when the couplings of the SU(3) flavor symmetric model are tuned appropriately.

In closing, it is worthwhile to remark here that the canonical $1.4 M_{\odot}$ mass neutron stars do not contain
hyperons in all models described above and their properties are entirely determined by the nuclear covariant DFT. 
Indeed, hyperons appear earliest in the case of soft EoS, which occurs in the SU(6) symmetric model. Even in
this limiting case we find that the hyperons appear in compact stars with masses with $M > 1.5 M_\odot$, i.e., 
masses larger than the canonical pulsar mass.

\subsection{Stellar sequences with maximum mass 2$M_\odot$ and their EoS}

Next we concentrate on the models of dense matter which produce sequences of compact stars with massive 2$M_\odot$ stars and their properties.
The scalar meson-hyperon coupling constants (the ratios $R_{\sigma Y} = g_{\sigma Y}/g_{\sigma N}$ and $R_{\sigma^\ast Y} = g_{\sigma^\ast N}/g_{\sigma N}$)
for the RHF DFs which were used in the following calculations are presented in Table~\ref{tab:SU3 parameters}. The remaining constants for vector
meson-hyperon couplings can be obtained by the SU(3) relations~\eqref{eq:SU3Zr} and \eqref{eq:SU3Z} in Appendix~C. These models are of
practical importance as these can be used in modeling dynamics of neutron stars with hyperons on the basis of EoS which are constrained by the pulsar
mass measurements.

\subsubsection{The mass vs radius relations}

\begin{figure*}[tb]
\centering
\ifpdf
\includegraphics[width = 0.90\textwidth]{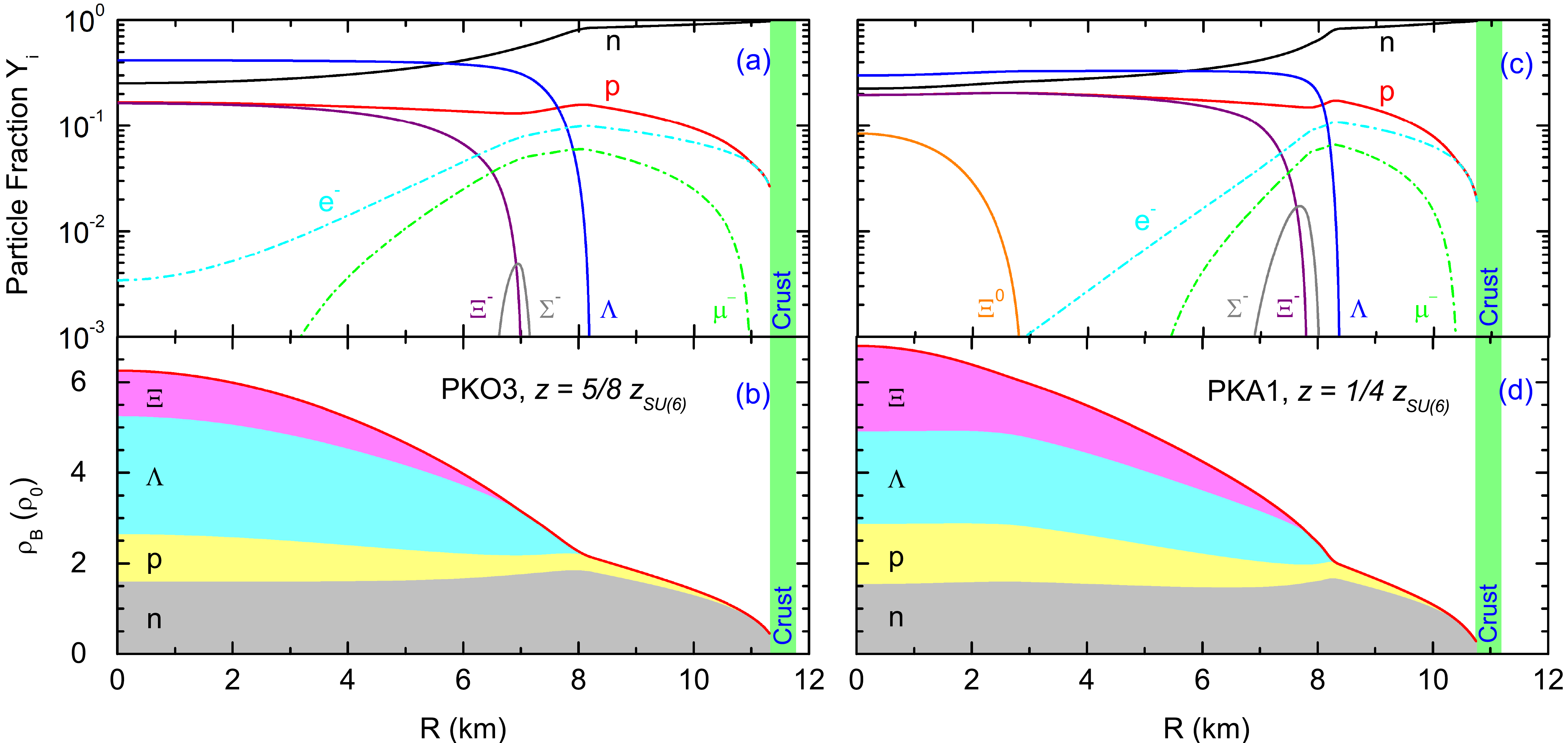}
\else
\includegraphics[width = 0.90\textwidth]{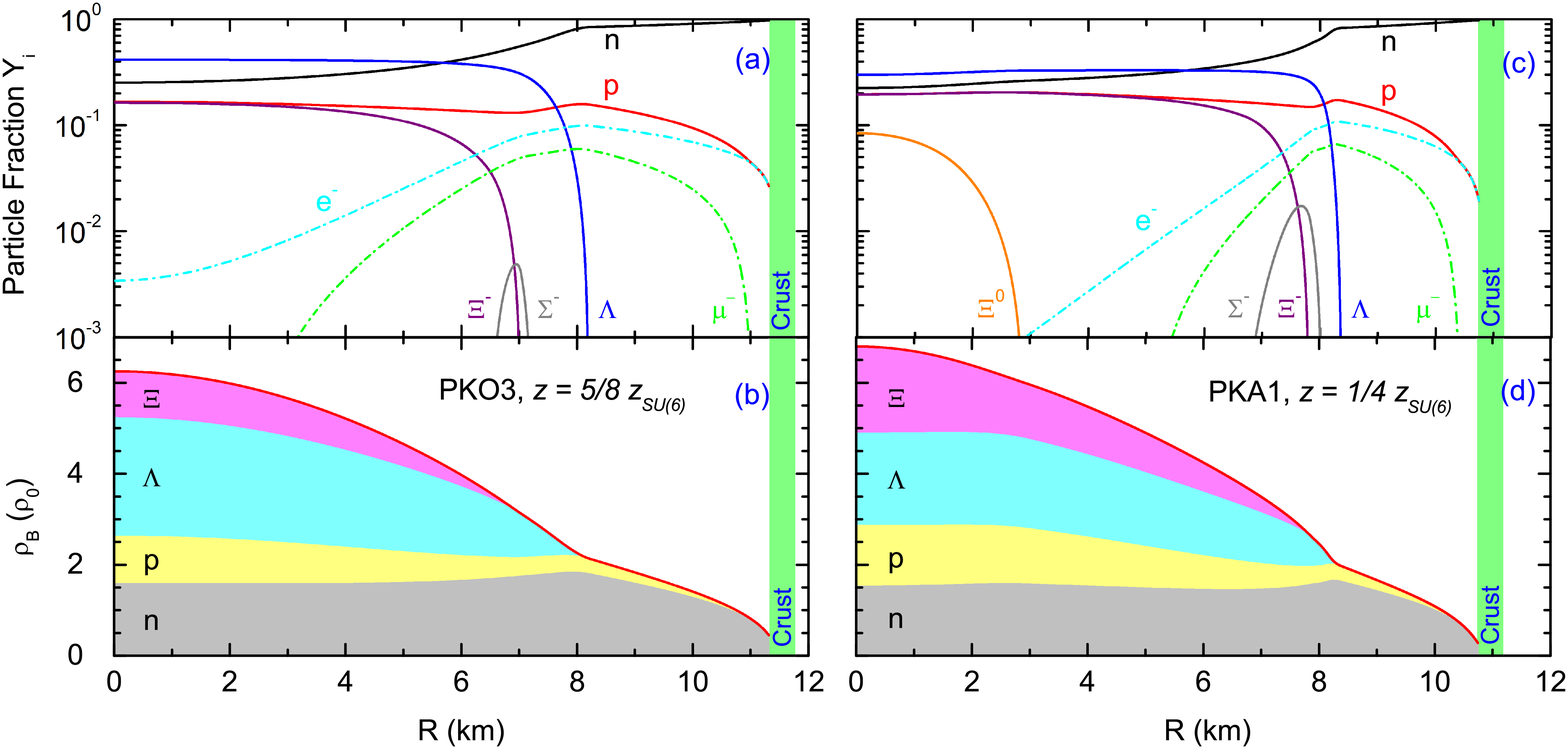}
\fi
\caption{ Particle fractions (a, c) and baryon number density (b, d) of hyperonic matter as a function of circumferential radius in the core of a
2.0$M_\odot$ star, which corresponds to the maximum gravitational mass of the models with the a case SU(3) flavor symmetry parameters as indicated
in the figure.}
\label{fig:Rad_2M}
\end{figure*}
%
\begin{table*}[tb]
\caption{Properties for a 2.0$M_\odot$ configurations. We list the model the parameters specifying the SU(3) parameterization, i.e.,
$z$ and $\alpha_v$, the mass $M_{\text{max}}$ ($M_\odot$), radius $R_{\text{max}}$ (km), central density
$\rho_c$ (fm$^{-3}$) and its strangeness fraction $F_{\text{s}}$ (strangeness per baryon) of the maximum-mass configurations, 
as well as the onset densities of the appearance of hyperons $Y_{\text{o}}$ (fm$^{-3}$).}\setlength{\tabcolsep}{18pt}
\label{tab:hypernuclear star}
\begin{tabular}{ccccccccc}
\hline
  DF   & $z$ & $\alpha_v$ & $M_{\text{max}}$ & $R_{\text{max}}$ & $\rho_c$ & $F_{\text{s}}(\rho_c)$ &
 $Y_{\text{o}}(\Lambda)$ & $Y_{\text{o}}(\Xi)$ \\
\hline
 PKA1  & $\frac{1}{4\sqrt{6}}$  &  1            & 2.006 &  11.096  &  1.088 & 0.286 & 0.316 & 0.428 \\
 PKO1  & $\frac{1}{2\sqrt{6}}$  &  1            & 2.016 &  11.801  &  0.965 & 0.242 & 0.329 & 0.514 \\
 PKO2  & $\frac{1}{2\sqrt{6}}$  &  1            & 2.035 &  11.803  &  0.948 & 0.241 & 0.340 & 0.501 \\
 PKO3  & $\frac{5}{8\sqrt{6}}$  &  1            & 2.005 &  11.822  &  0.957 & 0.248 & 0.327 & 0.482 \\
 PKO3  & $\frac{1}{\sqrt{6}} $  & $\frac{5}{8}$ & 2.019 &  11.731  &  0.973 & 0.242 & 0.332 & 0.526 \\
\hline
\end{tabular}
\end{table*}

Figures~\ref{fig:EoS_2M} (a) and (b) show the EoSs and the corresponding mass-radius relations computed with the four RHF DFs within the SU(3)
symmetric model. The four parametrizations fall into two groups, one corresponding to PKOi ($i=1$-3) and the other to PKA1. The maximum mass
for all these DFs are at about 2.0$M_\odot$, with the radii of this maximal mass star differing by about 1.0~km. Note that the PKA1 
parameterization provides an EoS, which is softer at low density, but becomes stiffer at high density, thus allowing for 2.0$M_\odot$ object. 
The central density of this object is much higher and it is more compact tha its counterparts constructed from PKOi ($i=1$-3) EoSs.

Despite of similarity in the EoS predicted by these DFs, there are large differences in the internal composition of the massive stars. 
This is illustrated in Fig.~\ref{fig:Rad_2M}, which shows particle fractions and baryon densities as a function of the radial coordinate
for a 2$M_\odot$ compact star. As expected, hyperons are concentrated in the inner core (in the region with $r\leq 8$ km) and they become
the dominant component within $r \leq 6$~km, while leptons are concentrated mostly in the outer part of the star ($r \geq 6$ km).
Therefore, our RHF DFs favor a strongly hyperon populated core inside massive star. An interesting feature of PKA1 DF is the appearance
of $\Xi^0$ hyperons in the very inner part of the neutron star inner core, which occurs at the cost of slight reduction of the
$\Lambda$s. It is worthwhile noting that the distribution of nucleonic component in the inner core is almost constant, and has the value
roughly $2.5 \rho_0$.

Finally, we summarize in Table~\ref{tab:hypernuclear star} the relevant properties for a 2.0$M_\odot$ compact star obtained with
different DFs, including the strangeness fraction $F_{\text{s}}$ at central densities $\rho_c$, the onset density of hyperons $Y_{\text{o}}$.
The radii of such stars are seen to be in the range $11.1\sim11.8$ km, the central densities are about $6.5\rho_0$ (see also Fig.~\ref{fig:Rad_2M}),
the strangeness fraction at the center is about 0.25, and the onset density for the dominant hyperon ($\Lambda$) is about $2.2\rho_0$.

\subsubsection{Direct Urca processes}

\begin{figure}[tb]
\centering
\ifpdf
\includegraphics[width = 0.40\textwidth]{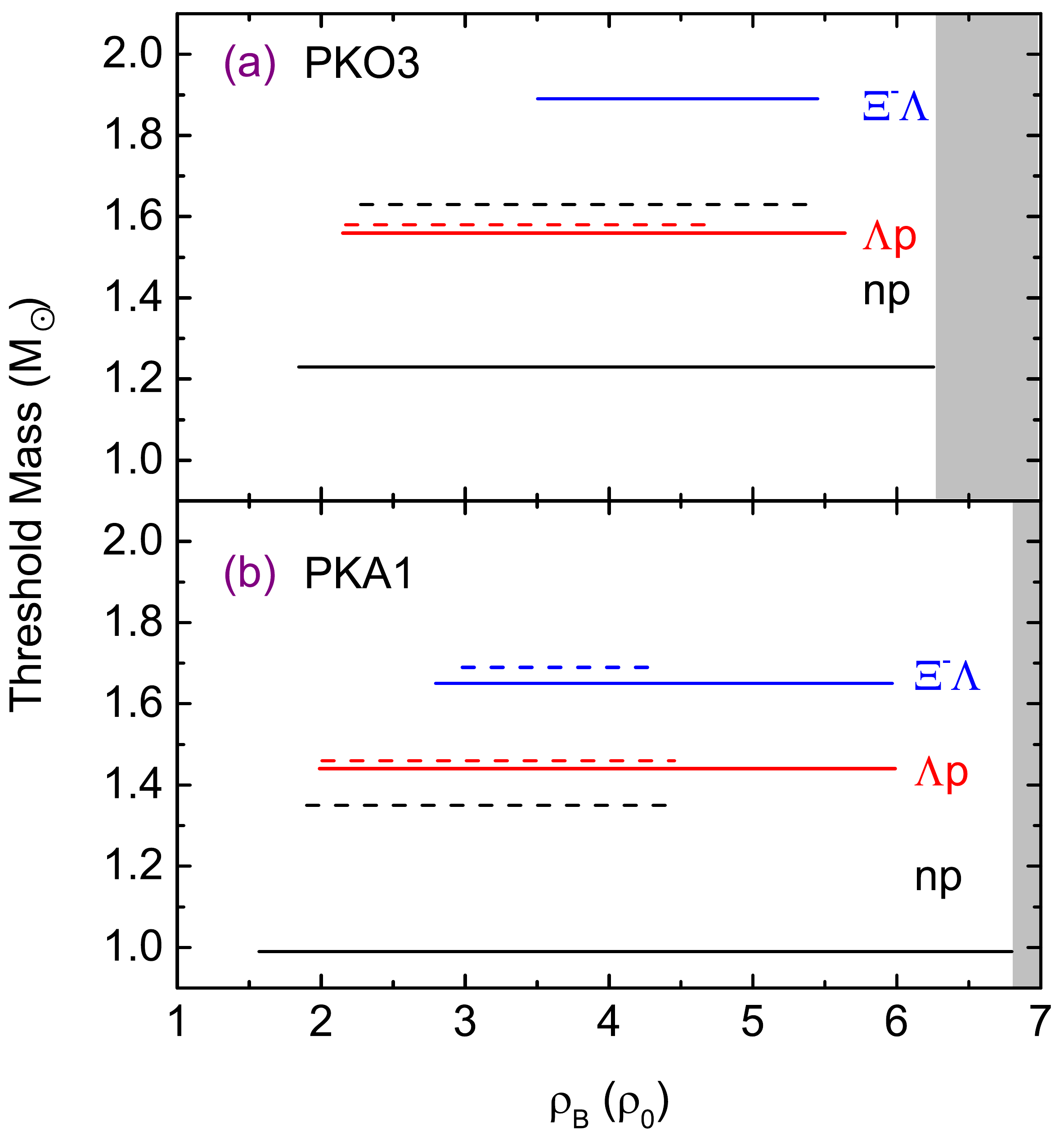}
\else
\includegraphics[width = 0.40\textwidth]{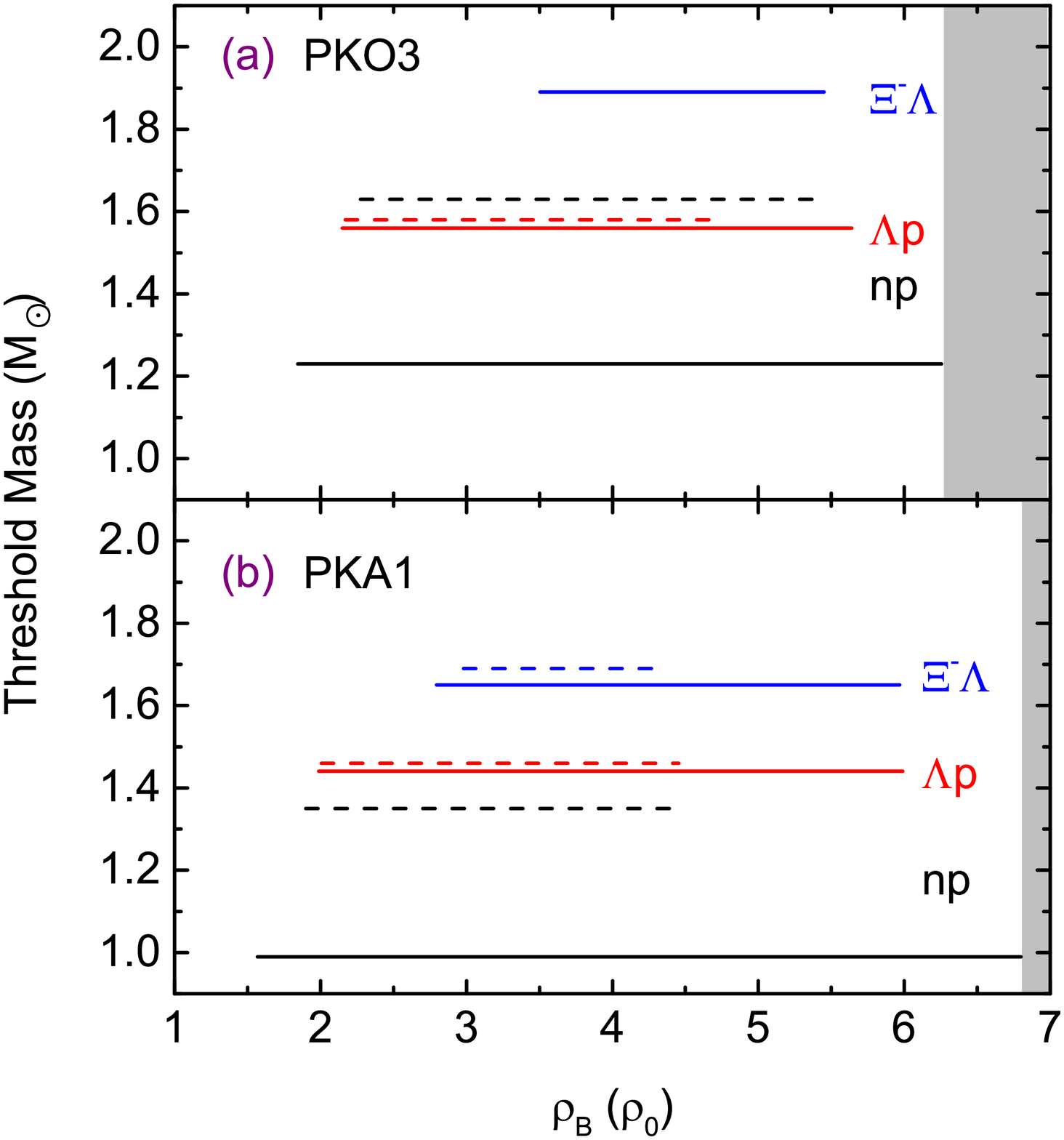}
\fi
\caption{The direct Urca processes in hyperonic stars as a function of the normalized baryonic density, predicted by the two DFs
that lead to a maximum gravitational mass of 2.0$M_\odot$ in the SU(3) flavor symmetric model. The relevant baryons are $n, p, \Lambda$,
and $\Xi^-$. The sold (dashed) lines denote the electron (muon) versions of the processes. The $y$-axis shows the mass of the star 
in which the threshold for a given process is achieved. The shading indicates the region beyond the maximum mass configurations.}
\label{fig:Urca}
\end{figure}

The studies of cooling of a compact star provide important information about their interior composition. The direct Urca process on baryons,
if it is kinematically allowed, is the most powerful mechanism of neutrino emission from the neutron star interior~\cite{Lattimer1991,Page1992}.
Its direct and inverse versions can be written schematically as
\begin{equation}
\label{eq:Urca_process}
B_1 \rightarrow  B_2 + l + \bar{\nu}_l, \quad
B_2 + l \rightarrow  B_1 + \nu_l,
\end{equation}
where $B$ are baryons involved in the weak interaction process, $l$ is a lepton, either an electron or a muon,
and $\nu_l$ is the neutrino associated with lepton $l$. Simultaneous conservation of energy-momentum requires
that the triangle inequality must be fulfilled for the Fermi momenta
\begin{align}
\vert p^F_{B_2} - p^F_l \vert \leq p^F_{B_1} \leq p^F_{B_2} + p^F_l.
\end{align}
It defines the critical density above which the processes~\eqref{eq:Urca_process} are allowed. Once hyperons are
present in matter, various types of hyperon direct Urca process will be allowed, because the above conditions are
fulfilled already for very low fraction of hyperons, of the order of few percent. The hyperonic direct Urca
process can play an important role in the cooling history of a compact star because, they can become operative
already at relatively small proton fraction at which the nucleonic direct Urca process is still
forbidden~\cite{Haensel1994,Schaab1998,Tsuruta2009,Raduta2017}.

Clearly, a direct Urca process can operate in the interior of a star if its central density exceeds the critical
density for this process. If allowed, such a process is operating within a spherical region with a radius corresponding
to the critical density. In Fig.~\ref{fig:Urca} we show the thresholds of the direct Urca processes on nucleons and hyperons
as a function of density calculated for PKO3 and PKA1 parametrizations. The horizontal lines mark the masses of stars
above which the corresponding process operates.

In the case of PKO3 parameterization, shown in Fig.~\ref{fig:Urca}(a), the nucleonic direct Urca process ($np$) is
allowed in the central cores of all stars with $M \gtrsim 1.3M_\odot$. Once a hyperon is present, its concentration
increases rapidly with density, and consequently the critical density for the Urca process generally differs little
from the onset density. For instance, the $\Lambda$ hyperon appears first at density $0.327$~fm$^{-3}$ and the direct
Urca process involving $\Lambda$s occurs if $\rho_B \geq 0.329$~fm$^{-3}$. As a result of such low critical density
for $\Lambda$, the hyperonic process ($\Lambda p$) starts to act in the stars with masses $M \gtrsim 1.5M_\odot$.
The process ($\Xi^- \Lambda$) operates only in massive stars with $M \gtrsim 1.9M_\odot$. In the case of PKA1
parameterization, the general features discussed above are intact, but the critical densities and the corresponding
threshold masses are somewhat smaller, see Fig.~\ref{fig:Urca}(b).

At the same time, the hyperonic processes are quenched above the density 5.5-6.0$\rho_0$ due to the strong suppression of
lepton concentrations, in particular the concentrations of $\mu$-ons, see Fig.~\ref{fig:Rad_2M}. The nucleonic direct 
Urca process which involves only electrons remains intact in a larger range of densities. This implies that in the 
innermost core of a massive star cools predominantly by the nucleonic direct Urca process. One should, however, notice
that the emissivity of the various Urca processes is dependent on the partial concentrations of each baryon and/or 
lepton species, their effective masses which in turn are determined by the EoS as well as their standard model 
weak-couplings constant and the importance of various processes can be ultimately revealed through the cooling
simulations of corresponding models.

\section{Summary and conclusions}\label{section5}

We have provided a first study of the hypernuclear matter in neutron stars with the relativistic Hartree-Fock 
theory using a DFT with density dependent couplings and including the full octet of baryons. This extends 
previous studies of hyperonization in dense matter to the level of the relativistic Hartree-Fock theories 
thus providing a new insight into the role played by the tensor forces, pion-exchanges, space component of vector self-energy, 
among other things, not revealed in the Hartree type approaches. Our work complements previous studies
of the hyperonization problem in the Hartree-Fock theories which were based exclusively on the quark-meson 
coupling model~\cite{Massot2012,Whittenbury2014,Miyatsu2015}.

Two factors contribute to the softening of the EoS in our framework: (a) the Fock terms generically make the 
hyperonic EoS soft, as they could even provide a negative contribution to the pressure; (b) the meson-hyperon
tensor couplings mediate additional attraction among hyperons compared to the models based on Hartree 
self-energies.

We followed the strategy of previous work (see for example ~\cite{Weissenborn2012a,Weissenborn2012b,Colucci2013,Dalen2014})
to (a) use the SU(3) spin-flavor symmetric quark model to tune the hyperonic coupling constants at fixed values of
the nucleonic ones and (b) to vary the depth of hyperonic potential. Within the parameter space of SU(3) 
symmetric quark model we find hypernuclear stellar configurations with masses that span from 1.7$M_\odot$ to
2.2$M_\odot$.  Our assessment of the influence of the uncertainties of the hyperon potentials on the EoS 
and mass-radius relation for different underlying parametrizations of nuclear matter shows that, the variation
of hyperon potentials do not change the maximum mass and the radius of a neutron star significantly due to a 
delicate compensation among the three types of hyperons. In particular, large changes in the $\Xi$ and $\Sigma$
potentials (e.g., 70\% variations around their accepted values) do not produce significant changes in the 
EoS and parameters of stellar configurations. We also find that our results are not sensitive on the
choice of the available nuclear DFs.

We have selected those EoSs which produce compact stars with maximum mass of the order of 2.0$M_\odot$ and
thus can be used in the neutron star phenomenology. For these models we have also estimated the critical 
density and threshold mass for hyperon direct Urca processes. Our analysis shows that once hyperons are present, 
the hyperon direct Urca processes become operative due to the rapid rise in the hyperon population and star 
with a mass $M \gtrsim 1.5 M_\odot$ are very likely to cool via the hyperon direct Urca process. Furthermore,
we point out that the hyperonic direct Urca processes could by quenched above 5.5-$6.0\rho_0$ due to the strong
suppression of lepton concentrations.

In the future it is worthwhile to consider the baryon exchange (transition) processes which may change the 
particle composition, although the effects of these processes were found to be small in a recent work~\cite{Katayama2015}.
The hyperonic coupling constants can be further tuned on hypernuclei instead of using the empirical potential
depth of hyperons in nuclear matter. Another problem is the onset of resonances in dense matter~\cite{Drago2016, Zhuzy2016}
and the interferences among hyperons and $\Delta$ isobars should be investigated as well~\cite{Lijj2018}. 
A natural extension of the present baryonic EoS would be to include a deconfined quark phase at high densities.

J. L. acknowledges the support by the Alexander von Humboldt foundation. A. S. is supported by the Deutsche 
Forschungsgemeinschaft (Grant No. SE 1836/3-2) and by the NewCompStar COST Action MP1304. W. L. is supported by
the National Natural Science Foundation of China under Grant Nos.11375076 and 11675065.


\section*{Appendix A. Functions in the self-energy}
\label{app:A}
%
\begin{table*}[htpb]
\centering
\caption{Functions $A_\alpha$, $B_\alpha$, $C_\alpha$ and $D_\alpha$ in Eq.~(\ref{eq:Fock self-energies}).}
\setlength{\tabcolsep}{20.5pt}
\label{tab:Functions in self-energy}
\centering
\begin{tabular}{ccccc}
\hline
$\alpha_i$   &$A_\alpha$                       & $ B_\alpha$                    & $C_\alpha$                     & $D_\alpha$\\
\hline
$\sigma_S$   & $ g^2_{\sigma B}\Theta_\sigma$ & $  g^2_{\sigma B}\Theta_\sigma$ & $-2g^2_{\sigma B}\Phi_\sigma$  & -\\
$\delta_S$   & $ g^2_{\delta B}\Theta_\delta$ & $  g^2_{\delta B}\Theta_\delta$ & $-2g^2_{\delta B}\Phi_\delta$  & -\\
\\

$\omega_V$   & $2g^2_{\omega B}\Theta_\omega$ & $-4g^2_{\omega B}\Theta_\omega$ & $-4g^2_{\omega B}\Phi_\omega$  & -\\
$\omega_T$   & $-(f_{\omega B}/2M)^2m^2_\omega\Theta_\omega$&$-3(f_{\omega B}/2M)^2m^2_\omega\Theta_\omega$  &
              $4(f_{\omega B}/2M)^2m^2_\omega\Lambda_\omega$& -\\
$\omega_{VT}$&    -     &    -    &    -                           &$12(f_{\omega B} g_{\omega B}/2M)\Omega_\omega$ \\
$\rho_V  $   & $ 2g^2_{\rho B}\Theta_\rho$    & $-4g^2_{\rho B}\Theta_\rho $    & $-4g^2_{\rho B}\Phi_\rho$      & -\\
$\rho_T$     & $-(f_{\rho B}/2M)^2m^2_\rho\Theta_\rho$&$-3(f_{\rho B}/2M)^2m^2_\rho\Theta_\rho$ &
              $4(f_{\rho B}/2M)^2m^2_\rho\Lambda_\rho$& -\\
$\rho_{VT}$  &    -     &    -    &    -                           &$12(f_{\rho B} g_{\rho B}/2M)\Omega_\rho$       \\
\\

$\pi_{PV}$   & $-f^2_{\pi B}\Theta_\pi$       & $-f^2_{\pi B}\Theta_\pi$        & $2f^2_{\pi B}/m^2_\pi\Pi_\pi$  & -\\
\hline
\end{tabular}
\begin{tablenotes}
\item[a] Note. For the strange mesons $\sigma^\ast$ and $\phi$, The functions $A_{\alpha}$, $B_{\alpha}$,
$C_{\alpha}$ and $D_{\alpha}$ for $\sigma^\ast$ coincide with those for $\sigma$ and for $\phi$ coincide with $\omega$.
The index $i$ is specified in the left column, where $S(V)[T]$ stands for the scalar(vector)[tensor] coupling at each
meson-baryon vertex.
\end{tablenotes}
\end{table*}
%
\begin{table*}[tb]
\centering
\caption{The effective interactions PKOi ($i=1$-3) and PKA1 of RHF DFT, where the masses are $M_N = 938.9$~MeV, $m_\omega = 783.0$~MeV,
$m_\rho = 769.0$~MeV, and $m_\pi = 138.0$~MeV.}
\setlength{\tabcolsep}{8pt}
\label{tab:RHF parameters}
\centering
\begin{tabular}{ccccccccccc}
\hline
     & $m_\sigma$ & $g_\sigma$ & $g_\omega$ &  $g_\rho$  &   $f_\pi$  &$\kappa_\rho$& $a_\rho$  &  $a_\pi$   & $a_{\rho_T}$ \\
\hline
PKO1 & 525.769084 &  8.833239  &  10.729933 &  2.434749  &  0.291716  &     -       & 0.076760  &  1.231976  &      -       \\
PKO2 & 534.461766 &  8.920597  &  10.550553 &  2.163268  &      -     &     -       & 0.631605  &      -     &      -       \\
PKO3 & 525.667686 &  8.895635  &  10.802690 &  2.030285  &  0.392931  &     -       & 0.635336  &  0.934122  &      -       \\
PKA1 & 488.227904 &  8.372672  &  11.270457 &  2.118421  &  0.310448  &   3.199491  & 0.544017  &  1.200000  &   0.820583   \\
\hline
     & $a_\sigma$ & $b_\sigma$ & $c_\sigma$ & $d_\sigma$ & $a_\omega$ & $b_\omega$ & $c_\omega$ & $d_\omega$ &   $\rho_0$   \\
\hline
PKO1 &   1.384494 &  1.513190  &   2.296615 &  0.380974  &  1.403347  &  2.008719  &  3.046686  &   0.330770 &     0.152    \\
PKO2 &   1.375772 &  2.064391  &   3.052417 &  0.330459  &  1.451420  &  3.574373  &  5.478373  &   0.246668 &     0.151    \\
PKO3 &   1.244635 &  1.566659  &   2.074581 &  0.400843  &  1.245714  &  1.645754  &  2.177077  &   0.391293 &     0.153    \\
PKA1 &   1.103589 & 16.490109  &  18.278714 &  0.135041  &  1.126166  &  0.108010  &  0.141251  &   1.536183 &     0.160    \\
\hline
\end{tabular}
\end{table*}

The explicit expression for the functions $A_{\alpha}$, $B_{\alpha}$, $C_{\alpha}$ and $D_{\alpha}$
in Eq.~(\ref{eq:Fock self-energies}) are given in Table~\ref{tab:Functions in self-energy}, where,
for compactness, we introduced the following short-hand notations,
\begin{subequations}
\begin{align}
\Theta_{\alpha}(p,p^\prime)  & \equiv  \ln \frac{m^2_{\alpha}+(p+p^\prime)^2}{m^2_{\alpha}+(p-p^\prime)^2},\\
\Phi_{\alpha}(p,p^\prime)    & \equiv  \frac{1}{4 pp^\prime}(p^2+p^{\prime2}+m^2_{\alpha})\Theta_{\alpha}(p,p^\prime)-1,\\
\Pi_{\alpha}(p,p^\prime)     & \equiv  (p^2+p^{\prime2}-\frac{m^2_{\alpha}}{2})\Phi_{\alpha}(p,p^\prime)- pp^\prime\Theta_{\alpha}(p,p^\prime),\\
\Lambda_{\alpha}(p,p^\prime) & \equiv  (p^2+p^{\prime2})\Phi_{\alpha}(p,p^\prime)- pp^\prime\Theta_{\alpha}(p,p^\prime),\\
\Omega_{\alpha}(p,p^\prime)  & \equiv  p\Theta_{\alpha}(p,p^\prime)-2p^\prime\Phi_{\alpha}(p,p^\prime).
\end{align}
\end{subequations}

\section*{Appendix B. Details of nucleonic RHF parameterization}
\label{app:B}

In RHF approach, the explicit density dependence is introduced into the meson-nucleon couplings in the following form
\begin{align}
g_{\alpha N}(\rho_v)=g_{\alpha N}(\rho_0)\eta_{\alpha}(x),
\end{align}
where $x=\rho_B/\rho_0$, $\rho_B$ is the baryonic density, $\rho_0$ is the nuclear saturation density.

In the isoscalar channels, the ansatz $\eta_\alpha$ for $\sigma$ and $\omega$ mesons is given by
\begin{align}
\eta_{\alpha}(x)=a_\alpha\frac{1+b_\alpha(x+d_\alpha)^2}{1+c_\alpha(x+d_\alpha)^2}, \quad \alpha = \sigma, \omega.
\end{align}
This function is subject to the following constraints:
$\eta_{\alpha N}(1)=1$, $\eta^{\prime\prime}_{\alpha N}(0)=0$ and
$\eta^{\prime\prime}_{\sigma N}(1)=\eta^{\prime\prime}_{\omega N}(1)$, which reduce the number of free parameters to three.

In the isovector channels, for simplicity, the density dependence is taken in an exponential form
\begin{align}
f_{\alpha N}(x)&=f_{\alpha N}(\rho_0)e^{-a_\alpha(x-1)}, \quad \alpha = \pi, \rho,
\end{align}
According to the general strategy of relativistic DFT, the masses and the couplings strengths appearing in the RHF
Lagrangian~\eqref{eq:Lagrangian density} have been determined through fits to the masses of reference nuclei and the
bulk properties of symmetric nuclear matter at the saturation. The parameters of the RHF effective interactions PKOi ($i=1$-3) 
and PKA1 are shown in Table~\ref{tab:RHF parameters}.

\section*{Appendix C. SU(3) flavor symmetry model}
\label{app:C}


\begin{table}[tb]
\caption{Values of $R_{\alpha Y} = g_{\alpha Y}/g_{\alpha N}$ and $\kappa_{\alpha Y} = f_{\alpha Y}/g_{\alpha Y}$ for
hyperons within the SU(6) spin-flavor relations.}\setlength{\tabcolsep}{10.4pt}
\label{tab:Couplings in SU6}
\begin{tabular}{ccccc}
\hline
$R\backslash Y$                  & $\Lambda $ & $\Sigma$                  &  $\Xi$                      \\
\hline
 $R_{\sigma Y}$      &    2/3     &    2/3                    &   1/3                     \\
 $R_{\sigma^\ast Y}$      &   -$\sqrt{2}/3$    &     -$\sqrt{2}/3$ &    -2$\sqrt{2}/3$  \\
\\

 $R_{\omega Y}$      &     2/3    &    2/3                    &   1/3                     \\
 $\kappa_{\omega Y}$ &      -1    & $1+2\kappa_{\omega N}$    &  $-2-\kappa_{\omega N}$   \\
 $R_{\phi Y}$        &   -$\sqrt{2}/3$    &     -$\sqrt{2}/3$             &    -2$\sqrt{2}/3$  \\
 $\kappa_{\phi Y}$   & $2+3\kappa_{\omega N}$ & $-2-\kappa_{\omega N}$   &  $1+2\kappa_{\omega N}$   \\
\\

 $R_{\rho Y}  $      &       0    &    2                      &    1                      \\
 $\kappa_{\rho Y} $  &       0    &$-3/5+(2/5)\kappa_{\rho N}$&$-6/5-(1/5)\kappa_{\rho N}$\\
\\

 $f_{\pi Y}   $      &       0    &  2$\alpha_{ps}$           & -($1/2$)$\alpha_{ps}$     \\
\hline
\end{tabular}
\\
Note. $\alpha_{ps}$ = 0.40, see Refs.~\cite{Swart1963, Dover1984}. $\kappa$ denotes the ratio of the tensor to vector couplings of the vector mesons.
\end{table}

The $NY$ and $YY$ interactions are currently not determined due to the lack of sufficiently abundant and accurate experimental data.
This makes understanding the hyperonic sector a long-standing theoretical challenge. One possible approach to this sector is the
use of the symmetries underlying the quark model of hadrons. The SU(3) symmetry in flavor space is commonly regarded as an approximate
symmetry group of strong interaction if one restricts the attention only on three lightest quark flavors (up, down, and strange).

The SU(3) invariant Lagrangian can be constructed using matrix representations for the baryons $B$, and meson nonet (singlet state
$M_1$, and octet state $M_8$). In this work we consider the lowest order baryon octet ($J^P = 1/2^+$). The SU(3) interaction Lagrangian
is a linear combinations of the antisymmetric ($F$-type), symmetric ($D$-type), and singlet ($S$-type) scalars,
\begin{align}
\mathcal{L}_{\text{int}} = &-g_8\sqrt{2}[\alpha\text{Tr}([\bar{B},M_8]B) + (1-\alpha)\text{Tr}(\{\bar{B},M_8\}B)] \nonumber \\
                           &-g_1\frac{1}{\sqrt{3}}\text{Tr}(\bar{B}B)\text{Tr}(M_1).
\end{align}
Here, $g_8$ and $g_1$ denote the meson octet and singlet coupling constant respectively, the parameter $\alpha$, known as $F/(F+D)$
ratio, lies in the range $0\le \alpha \le 1$.

Considering the vector meson sector, the combinations of the unphysical SU(3) singlet, $\vert 1 \rangle$, and octet, $\vert 8
\rangle$, states produces the physical $\omega$ and $\phi$ mesons
\begin{subequations}
\begin{align}
\omega  &= \sin\theta_v \vert 8 \rangle + \cos\theta_v \vert 1 \rangle, \\
\phi    &= \cos\theta_v \vert 8 \rangle - \sin\theta_v \vert 1 \rangle,
\end{align}
\end{subequations}
with $\theta_v$ being the mixing angle. The coupling constants of the physical vector mesons with the baryons
read as follows:
\begin{subequations}
\begin{align}
g_{\omega N}       &= \cos\theta_v g_1 +  \sin\theta_v (4\alpha_v -1)g_8/\sqrt{3}, \\
g_{\omega \Lambda} &= \cos\theta_v g_1 - 2\sin\theta_v (1 -\alpha_v )g_8/\sqrt{3}, \\
g_{\omega \Sigma}  &= \cos\theta_v g_1 + 2\sin\theta_v (1 -\alpha_v )g_8/\sqrt{3}, \\
g_{\omega \Xi}     &= \cos\theta_v g_1 -  \sin\theta_v (1 +2\alpha_v)g_8/\sqrt{3}.
\end{align}
\end{subequations}
One could clearly see that, all possible combinations of the couplings are determined by four parameters.
From these relations one obtains
\begin{subequations}
\begin{align}
\frac{g_{\omega \Lambda}}{g_{\omega N}} &=
\frac{1-\frac{2z}{\sqrt{3}}(1-\alpha_v)\tan \theta_v}{1-\frac{z}{\sqrt{3}}(1-4\alpha_v)\tan \theta_v}, \\
\frac{g_{\omega \Xi}}{g_{\omega N}} &=
\frac{1-\frac{z}{\sqrt{3}}(1+2\alpha_v)\tan \theta_v}{1-\frac{z}{\sqrt{3}}(1-4\alpha_v)\tan \theta_v}, \\
\frac{g_{\omega \Sigma}}{g_{\omega N}} &=
\frac{1+\frac{2z}{\sqrt{3}}(1-\alpha_v)\tan \theta_v}{1-\frac{z}{\sqrt{3}}(1-4\alpha_v)\tan \theta_v}.
\end{align}
\end{subequations}

The corresponding results for the $\phi$-meson couplings follow from those for $\omega$ meson via the replacement
$\cos\theta_v \rightarrow -\sin\theta_v$ and $\sin\theta_v\rightarrow\cos\theta_v$. An additional new term is given by
\begin{align}\label{eq:phiome relation}
\frac{g_{\phi N}}{g_{\omega N}} &= -
\frac{\tan \theta_v + \frac{z}{\sqrt{3}}(1-4\alpha_v)}{1-\frac{z}{\sqrt{3}}(1-4\alpha_v)\tan \theta_v}.
\end{align}

For the isovector meson $\rho$, one has
\begin{subequations}\label{eq:SU3Zr}
\begin{align}
\frac{g_{\rho \Lambda}}{g_{\rho N}} &= 0, \\
\frac{g_{\rho \Xi}}{g_{\rho N}} &= 2\alpha_v-1, \\
\frac{g_{\rho \Sigma}}{g_{\rho N}} &= 2\alpha_v.
\end{align}
\end{subequations}
Note that for $\alpha_v= 0(0.5)$, the couplings $g_{\rho \Sigma} (g_{\rho \Xi})$ are zero, 
although there are no physical reasons for them to be so. Such values of the couplings in 
practice do not affect our results. For example, for the range of  coupling values $0 \le g_{\rho \Xi}\le 1$, 
the remaining meson-$Y$ couplings being fixed to their values, we find that the variations
in the EoS and mass-radius relation are negligible. This is because the $\rho$ couplings depend 
exponentially on density, therefore, the $\rho$ field is largely suppressed at high densities 
important for neutron stars.

The $\phi$ meson taken as pure $\bar{s}s$ state leads to the \textit{ideal} mixing
\begin{align}
\label{eq:idmix}
\theta^{\text{id.}}_v = \tan^{-1}(1/\sqrt{2}).
\end{align}
If one determines the vector meson-baryon couplings on the basis of the assumption that nucleons do not couple
to $\phi$ meson, then
\begin{align}
z \equiv g_8/g_1 = 1/\sqrt{6}.
\end{align}
If we further require the universality assumption for the (electric) $F/(F + D)$ ratio, we have $\alpha_v= 1$, i.e.,
only the $F$-type coupling remains and the coupling constants are related as in the additive quark model.

In the case of the scalar mesons $\sigma$ and $\sigma^\ast$, the coupling are given by expressions entirely analogous
to those of $\omega$ and $\phi$, respectively, with the replacements $\omega\to \sigma$, $\phi\to \sigma^\ast$.
In addition the vector subscripts are changed to scalar ones, i.e., $v\to s$.

Furthermore, there is another type of couplings between the vector mesons and the baryon current $\{8\}\otimes\{8\}$ via
the tensor coupling with coupling constants $f^T$. In order to obtain also relations for the tensor coupling constants,
the corresponding SU(3) relations are in fact applied to the magnetic coupling $G^M$. Using the vector dominance assumption
with ideal mixing, one finds, for example, for the $\omega$ meson
\begin{align}
G^M_{\omega\Lambda} = 0, \quad
G^M_{\omega\Sigma} = &\frac{4}{3}G^M_{\omega N}, \quad
G^M_{\omega\Xi} = -\frac{1}{3}G^M_{\omega N}.
\end{align}
The tensor-to-vector coupling ratio $\kappa$ of a baryon is then given in the static limit $G^M = g^V + f^T$.
We summarize the hyperon-meson coupling ratios $R_{\alpha Y} = g_{\alpha Y}/g_{\alpha N}$ under the
SU(6) spin-flavor model in Table~\ref{tab:Couplings in SU6}, where we show only the coupling constants relevant for our model.
Notice that the effective coupling of the $\rho$ meson to the $\Sigma$ hyperon (isospin $\tau = 1$) is twice that to the nucleon
(isospin $\tau = 1/2$), as required by the symmetries assumed.

If we set $\alpha_v$ to its SU(6) value $\alpha_v=1$ and use ideal mixing~\eqref{eq:idmix} while keeping $z$ as a free
parameter, we obtain
\begin{subequations}\label{eq:SU3Z}
\begin{align}
\frac{g_{\phi N}}{g_{\omega N}}
&= \frac{\sqrt{6}z - 1}{\sqrt{2} + \sqrt{3}z}, \\
\frac{g_{\omega \Lambda}}{g_{\omega N}} = \frac{g_{\omega \Sigma}}{g_{\omega N}}
&= \frac{\sqrt{2}}{\sqrt{2} + \sqrt{3}z}, \\
\frac{g_{\omega \Xi}}{g_{\omega N}}
&= \frac{\sqrt{2} - \sqrt{3}z}{\sqrt{2} + \sqrt{3}z},\\
\frac{g_{\phi \Lambda}}{g_{\omega N}} = \frac{g_{\phi \Sigma}}{g_{\omega N}}
&= -\frac{1}{\sqrt{2} + \sqrt{3}z}, \\
\frac{g_{\phi \Xi}}{g_{\omega N}}
&= -\frac{1 + \sqrt{6}z}{\sqrt{2} + \sqrt{3}z}.
\end{align}
\end{subequations}

Another choice is to set $z$ to its SU(6) value $z=1/\sqrt{6}$ and use the ideal mixing~\eqref{eq:idmix} while keeping $\alpha_v$
as a free parameter instead. As demonstrated in the main body of this work the two alternatives lead to similar results although the
underlying physical assumptions differ.


\bibliographystyle{apsrev}
\bibliography{HNstar_ref.bib}

\end{document}